\documentclass[10pt]{article}
\usepackage{multicol}
\usepackage{simpleConference}
\usepackage{arxiv}
\usepackage[backend=bibtex, style=chem-rsc, autocite=superscript]{biblatex}
\let\tempautocite\autocite
\renewcommand{\autocite}[1]{\tempautocite[][]{#1}{}}
\usepackage[version=3]{mhchem}
\usepackage{bm} 
\usepackage{float}
\usepackage{fancyhdr} 
\usepackage{comment}

\usepackage[utf8]{inputenc} 
\usepackage[T1]{fontenc}    
\usepackage{hyperref}       
\usepackage{url}            
\usepackage{booktabs}       
\usepackage{amsfonts}       
\usepackage{nicefrac}       
\usepackage{microtype}      
\usepackage{lipsum}		
\usepackage{graphicx}
\usepackage{float}
\usepackage{subfig}
\usepackage{caption}
\usepackage{xcolor}

\usepackage{braket} 
\usepackage{amsmath}
\usepackage{mathtools}
\usepackage{amssymb}
\usepackage{subfig} 
\usepackage{tikz}
\usepackage{pgfplots}
\pgfplotsset{colormap={RdBu}{rgb=(0,0,1) rgb255=(1,0,0)}}
\usetikzlibrary{fadings}
\usetikzlibrary{calc,patterns,angles,quotes}
\tikzset{near start abs/.style={xshift=1cm}}
\makeatletter
\tikzset{
  @arc through/.style 2 args={
    to path={
      \pgfextra
        \pgfextract@process\pgf@tostart{\tikz@scan@one@point\pgfutil@firstofone(\tikztostart)\relax}%
        \pgfextract@process\pgf@tothrough{\tikz@scan@one@point\pgfutil@firstofone#1}%
        \pgfextract@process\pgf@totarget{\tikz@scan@one@point\pgfutil@firstofone(\tikztotarget)\relax}%
        \pgfextract@process\pgf@topointMidA{\pgfpointlineattime{.5}{\pgf@tostart}{\pgf@tothrough}}%
        \pgfextract@process\pgf@topointMidB{\pgfpointlineattime{.5}{\pgf@totarget}{\pgf@tothrough}}%
        \pgfextract@process\pgf@tocenter{%
          \pgfpointintersectionoflines{\pgf@topointMidA}
            {\pgfmathrotatepointaround{\pgf@tothrough}{\pgf@topointMidA}{90}}
            {\pgf@topointMidB}{\pgfmathrotatepointaround{\pgf@tothrough}{\pgf@topointMidB}{90}}}%
        \pgfcoordinate{arc through center}{\pgf@tocenter}%
        \pgfpointdiff{\pgf@tocenter}{\pgf@tostart}%
        \pgfmathveclen@{\pgfmath@tonumber\pgf@x}{\pgfmath@tonumber\pgf@y}%
        \edef\pgf@toradius{\pgfmathresult pt}
        \pgfmathanglebetweenpoints{\pgf@tocenter}{\pgf@tostart}%
        \let\pgf@tostartangle\pgfmathresult
        \pgfmathanglebetweenpoints{\pgf@tocenter}{\pgf@totarget}%
        \let\pgf@toendangle\pgfmathresult
  
      \ifdim\pgf@tostartangle pt>\pgf@toendangle pt\relax
          \pgfmathsetmacro\pgf@tostartangle{\pgf@tostartangle-360}%
        \fi
        #2%
          \pgfmathsetmacro\pgf@toendangle{\pgf@toendangle-360}%
        \fi
      \endpgfextra
      arc [radius=+\pgf@toradius, start angle=\pgf@tostartangle, end angle=\pgf@toendangle] \tikztonodes
    }},
  arc through ccw/.style={@arc through={#1}{\iffalse}},
  arc through cw/.style={@arc through={#1}{\iftrue}},
clabel/.style args={#1 between #2 and #3}{append after command={
(\tikzlastnode.#2) -- (\tikzlastnode.#3) 
node[pos=0.5,rectangle,rounded corners,draw,font=\sffamily,fill=black,inner sep=0.8pt,very thick]{\color{white}\small \textbf{#1}}
}},
container1/.style={draw, rectangle,dashed,inner sep=0.48cm, rounded
corners,fill=blue!35,minimum height=4cm,  very thick},
container2/.style={draw, rectangle,dashed,inner sep=0.48cm, rounded
corners,fill=orange!40,minimum height=4cm, very thick},
container3/.style={draw, rectangle,dashed,inner sep=0.48cm, rounded
corners,fill=blue!35,minimum height=4cm, very thick}
}
\makeatother
\usetikzlibrary{decorations.markings}
\usetikzlibrary{shapes.geometric}
\usepackage{circuitikz}
\usetikzlibrary{shapes,arrows.meta,calc,fit,backgrounds,shapes.multipart,positioning,arrows}
\tikzstyle{flowbox} = [%
    minimum width=3cm,%
    minimum height=1cm,%
    text width=3cm,%
    text centered,%
    draw=black,%
]
\tikzstyle{startstop} = [%
    very thick,
    flowbox,%
    diamond,%
    aspect=2,
    rounded corners,%
    fill=yellow!30%
]
\tikzstyle{process} = [%
    very thick,
    flowbox,%
    rectangle,%
    rounded corners,%
    fill=yellow!30%
]
\tikzstyle{process2} = [%
    process,
    fill=gray!15,%
    rounded corners%
]
\tikzstyle{process3} = [%
    very thick,
    flowbox,%
    rectangle,%
    fill=gray!15,%
    rounded corners%
]
\tikzstyle{marrow} = [%
   very thick,%
   scale=2,
   decoration={markings,mark=at position 0.5 with {\arrow{Stealth[scale=1.5]}}},
   postaction={decorate}
]

\newcommand\blfootnote[1]{%
  \begingroup
  \renewcommand\thefootnote{}\footnotetext{#1}%
  \addtocounter{footnote}{-1}%
  \endgroup
}
\usepackage{fontawesome5}
\usepackage{ulem}
\def\dout{\bgroup
 \markoverwith{\lower-0.4ex\hbox
 {\kern-.03em\vbox{\hrule width.2em\kern0.25ex\hrule}\kern-.03em}}%
 \ULon}
\MakeRobust\dout
\usepackage[perpage,hang,bottom]{footmisc}
\makeatletter
\newcommand*{\shifttext}[2]{%
  \settowidth{\@tempdima}{#2}%
  \makebox[\@tempdima]{\hspace*{#1}#2}%
}
\makeatother
\newcommand{\darr}{\shifttext{0.1ex}{$\downarrow$}}
\renewcommand{\thefootnote}{\ifcase\value{footnote}\or \darr  \or \sout{\darr} \or \dout{\darr} \fi}
\usepackage{perpage} 
\MakePerPage{footnote} 

\date{January 10,  2023}
\begin{document}

\title{Global descriptors of a water molecule for machine learning of potential energy surfaces}

\author{Fabio E. A.~Albertani\textsuperscript{$\ast \dagger$}\ , \ Alex J. W.~Thom\textsuperscript{$\ast$}}
\thispagestyle{empty}

  	\maketitle
	\begin{abstract}
Machine learning of multi-dimensional potential energy surfaces,  from purely \textit{ab initio} datasets,  has seen substantial progress in the past years.  Gaussian processes,  a popular regression method,  have been very successful at producing realistic potential energy surfaces from sparse datasets allowing to reduce the computational cost of highly accurate models.  However,  there are many choices one has to take in the design of the kernel and the feature space which can substantially change the performance and the characteristic of the model.  In this study we explore different Gaussian processes set ups and,  more specifically,  how the permutational invariance of the target surface is controlled through the feature space selection.
	\end{abstract}
	\blfootnote{\textsuperscript{$\ast$}\ Yusuf Hamied Department of Chemistry,  University of Cambridge,  Cambridge,  Lensfield Road,  CB2 1EW}
	\blfootnote{\textsuperscript{$\dagger$}\ fa381@cam.ac.uk}
	\setcounter{footnote}{0}
\begin{multicols}{2}

\section{Introduction}
Machine learning (ML) has,  as in many fields of science,  changed the tools we use to predict accurate multi-dimensional potential energy surfaces (PESs) from electronic structure data.  Evaluating accurate energies of flexible molecules,  that are required for quantum dynamics methods,  at the accuracy of post Hartree--Fock electronic structure methods is often too computationally demanding for on the fly methods and has consequently very restricted use\autocite{Deumens2001,Worth2003}.  Using ML methods gives us the ability to use sparser datasets and infer accurate energies in a universal approach since target function do not need to have a ``predefined'' form.  As opposed to analytical fits where one requires an \textit{a priori} knowledge of the function\autocite{Jordan1995, Moyano2005,Chakraborty2006,Li2013,Palma2015},  ML have the ability to learn general unknown functions\autocite{Sacks1989}.
\par
Amongst the various ML methods,  Gaussian processes regression (GP) provides a solid framework for learning continuous functions,  such as PESs\autocite{Bartok2015,Das2015,Cresswell2016,Cui2016,Uteva2017,Kolb2017,Uteva2018,Dragoni2018, Dai2020}.  GPs are easier to implement than other ML techniques like neural networks\autocite{Handley2009} (NN) and also provide an error estimate of the predicted function which can have useful properties in applications to chemistry.  Since the pattern of the target function is heavily dependent on the feature space,  it is important to assess the impact that the choice of the coordinate system has on the ability of a GP to build a latent function,  \textit{i.e.} the predicted surface generated by the GP,  with accurate properties. 
\par
A key aspect of setting up GP is to define a feature space for learning onto which we project the training set.  A few aspects of our target global PES should come into consideration\autocite{Peyerimhoff1985}: physical invariances (translational,  rotational and permutational),  dimensionality (since kernels will become more complex and harder to optimise for large feature spaces) and,  optionally,  chemical intuition. The latter has a direct relation to the complexity of the pattern of the target function which,  if simple along feature dimensions,  can improve the performance of the latent function.  Expressing the molecular geometry using appropriate descriptors is then essential to produce successful ML models and it should be able to create physically invariant PESs. 
\par
In this paper,  we focus on feature spaces derived from global descriptors to reproduce symmetry in the PES.  Translational and rotational invariances are often easily accounted for,  while permutational invariances are harder to handle\autocite{Wang2008, Braams2009, Conte2015, Jiang2016}.  There has been substantial progress in the last years in developing permutationally invariant fitting bases which are based on symmetrisation of monomials to construct polynomials of interatomic distances: the so-called PIP (permutational invariant polynomials) approach\autocite{Xie2009,Braams2009, Qu2018,Babin2013,Babin2014} which is usually used in conjunction with NN approaches,  but have also been used in conjunction with GPs\autocite{Qu2018a,Qu2018b}. 
\par
In ML,  one needs to consider a ``feature space'' which defines the input coordinates of the model as well as,  in most cases,  a description of the geometry of the molecule.  Global descriptors of the molecular geometry are usually simple to derive but have no transferability.  On the other hand,  local atomic descriptors are harder to derive but are often used in ML learning as demonstrated by NN approaches of Behler \textit{et al}\autocite{Behler2014,Behler2017},  SOAP approaches by Cs\`{a}nyi \textit{et al}\autocite{Bartok2013, Bartok2015} as well as others\autocite{Langer2003,Behler2016}.  Local descriptors are well studied and successful but we will not consider them here since we only consider small molecular systems that do not suffer from global descriptors (they are quite easy to define for small systems).
\par
Performance of models can also be obtained by accounting for the physical invariance,  within the training data itself,  by adding equivalent data points.  An example of this is Thompson \textit{et al}\autocite{Thompson1998} where switching labels of the hydrogens for the H + CH$_4 \to$ H$_2$ + CH$_3$ reaction PES,  a large training set is obtained to improve the predicted energies.  The training data ``augmentation'' can be done for any non-permutationally variant feature spaces and has two advantages: as ML methods tend to monotonically improve with respect to training set size it allows us to increase the performance with a small computational cost and one can use feature spaces that are not too abstract.
\par
With large sizes of training sets often used in literature,  feature space are often compared in terms of numerical accuracy of the resulting method and,  often,  have a rather obvious choice given the specific target functions that is considered.  There is however importance in understanding how changing the feature space and the training set can affect the GP performance through the change in the log-marginal likelihood space.  This is especially true as GP are known to work well for sparser sets and often used as they scale favourably with the size of the training set\autocite{Loeppky2009}. 

\section{Gaussian Processes\label{sec:GAP}}
A Gaussian process is a machine learning regression method and is defined as \textit{a collection of random variables, any finite number of which have a joint Gaussian distribution}\autocite{Book:Rasmussen2005}.  An essential part of a GP model is its kernel function and its feature space which is the input space of the GP.
\par
Kernel functions associate a numerical value to the ``similarity'' of any two given points and are often dependent on optimisable hyperparameters to make them flexible.  For example,  a general kernel evaluation from the Mat\'{e}rn kernel class\footnote{To allow further flexibility,  multiply it by a constant kernel (CK) and sum a White Kernel (WK) noise. } for two vectors over the feature space,  $\mathbf{X}$ and $\mathbf{X}'$,  is given by
\begin{equation}
\mathrm{K}(\mathbf{X}, \mathbf{X}') = \sigma^2\ \frac{2^{1-\nu}}{\Gamma(\nu)}\Bigg(\sqrt{2\nu}\frac{d}{\rho}\Bigg)^\nu K_\nu\Bigg(\sqrt{2\nu}\frac{d}{\rho}\Bigg) + \lambda^2
\label{eq:covariance-Mat\'{e}rn}
\end{equation}
where $\Gamma$ is the gamma function, $K_\nu$ is the modified Bessel function of the second kind of degree $\nu$, $\rho$ are length scales and $d$ is the Euclidean distance in feature space $|\mathbf{X}-\mathbf{X}'|$.  The $\rho$ hyperparameter (each feature dimension would have a specific $\rho$ when using an anisotropic kernel) is optimised by the GP while the $\nu$ parameter is not optimised and defines the smoothness of the kernel\autocite{Book:Rasmussen2005}.  At the infinitely smooth limit of the Mat\'{e}rn kernel,  when $\nu \to \infty$,  one obtains the radial basis function (RBF) kernel:
\begin{equation}
\mathrm{K}(\mathbf{X},\mathbf{X}') = \sigma^2 \  \mathrm{exp} \big( -\frac{|\mathbf{X} - \mathbf{X}'|}{\rho^2} \big) 
\label{eq:covariance-rbf}
\end{equation}
To produce at least twice-differentiable latent functions (this is of particular interest for PES modelling since twice-differentiable ensures continuous atomistic forces and Hessians which is expected from the true physics of the problem),  we will consider Mat\'{e}rn ($\nu=2.5$) and RBF kernels as the two extremes of the Mat\'{e}rn class.
\par
At a set of query point,  forming a matrix $\mathbf{X}_p$ of size $N_p \times N_{\mathrm{features}}$,  a GP model predicts a Gaussian distribution with a mean (sometimes called the latent function),  here denoted $y(\mathbf{X}_p)$,  and a variance,  here denoted $\Delta(\mathbf{X}_p)$,  which is associated to the model confidence.  For a set of prediction points,  $\mathbf{X}_p$,  the predicted distribution are given by\autocite{Book:Rasmussen2005}:
\begin{equation}
\begin{gathered}
y (\mathbf{X}_p) = \mathbf{K}_{pt} \ \mathbf{K}^{-1}_{tt} \ \mathbf{y} \\
\Delta (\mathbf{X}_p) = \mathbf{K}_{pp} \ - \mathbf{K}_{pt} \ \mathbf{K}_{tt}^{-1} \ \mathbf{K}_{tp}
\label{eq:gp-maineq}
\end{gathered}
\end{equation}
where the kernel matrices are subscripted with the matrices they evaluate ($p$ for query points and $t$ for training) and the $ij$\textsuperscript{th} element of the matrix $\mathbf{K}_{nm}$ is given by $\mathrm{K}(\mathbf{X}_{n,i}, \mathbf{X}_{m,j})$.   A common meter to define the confidence in a model prediction,  used in the ML community,  is the $\Delta_{95\%}$ confidence interval which is given as $y \pm 2 \Delta$.
\par
Optimising a GP is performed by varying the hyperparameters of its kernel.  One can design different loss and regularisation functions or take a Bayesian approach and optimise the hyperparameters by maximising (or more commonly by minimising the negative) the log-marginal likelihood (LML) defined as\autocite{Book:Rasmussen2005}
\begin{equation}
\mathrm{LML} = -\frac{1}{2} \mathbf{y}^{\mathrm{T}} \mathbf{K}_{tt}^{-1} \mathbf{y} - \frac{1}{2} \mathrm{log} |\mathbf{K}_{tt}| - \frac{n}{2} \mathrm{log} (2\pi)
\label{eq:lml}
\end{equation}
where $\mathbf{K}_{tt}$ is the covariance matrix of the training set to itself.  The terms on the LHS of equation \ref{eq:lml} can be understood as a fit,  a regularisation and a normalisation term respectively\autocite{Book:Rasmussen2005}.
\par
One can also design,  outside of the Bayesian approach,  a series of loss and regularisation functions to optimise the kernel hyperparameters.  These would produce hypersurfaces  which,  like the LML,  could be studied with respect to changes in feature space and training data.  However,  we will not cover these alternative surfaces in this work.

\section{Methodology}
A initial set of water geometries was sampled from the Boltzmann distribution using a Metropolis--Hastings scheme at a temperature that allowed samples up to 0.3 mHa above the equilibrium energy.  Agglomerative clustering\autocite{Nielsen2016} is used to segregate data points with respect to their euclidean distance and,  finally,  a training set of 24 geometries calculated at UHF/aug-cc-pVDZ using the Q-Chem software\autocite{QChem} is created by sampling each cluster at random. 
\par
GPs using Mat\'{e}rn ($\nu=2.5$) kernels,  scaled by a CK and with an added WK,  and different feature spaces are then trained with the same training set using the GMIN suite\autocite{GMIN,OPTIM,PATHSAMPLE} and visualised using disconnectivity graphs\autocite{disconnectionDPS,Becker1997,Wales1998}: given the relatively low dimensionality of the LML,  5D with our choice of kernel,  we pick very large bounds for the hyperparameters with length scales and amplitude ranging from $10^{5}$ to $10^{-5}$ and the noise ranging from unity to $10^{-5}$.  Steps are computationally quite cheap and a full exploration of the landscape is ensured.  Each minimum is then associated to a model and,  unless specified,  we use the lowest minimum on the LML as the best model.
\par
The performance of the latent functions resulting from the differently trained GPs are assessed with the mean absolute error (MAE) on two distinct Boltzmann distributed testing sets also calculated at UHF/aug-cc-pVDZ sampled from a Metropolis--Hastings scheme at different temperatures.  The first testing set,  the ``low energy'' set,  consists of 500 geometries near equilibrium with no data at higher energies than training data while the second testing set,  the ``high energy'' set,  consists of 500 different geometries stretching further away and with some data in the extrapolation regime. 
\par
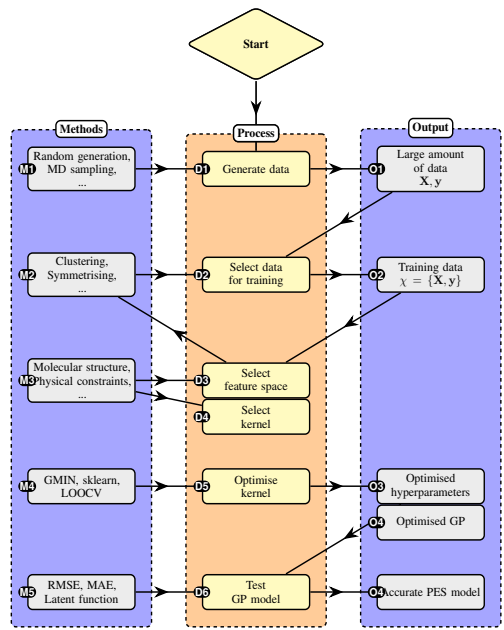
\begin{figure}[H]
	\centering
\resizebox{0.4\textwidth}{0.5\textwidth}{
    \begin{tikzpicture}[node distance=1cm and 2cm, scale=0.45]
        \node (start) [startstop] { Start};
        \node (10) [process, below=2cm of start, clabel=D1 between south west and north west] {Generate data};
        \node (20) [process, below=2cm of 10, clabel=D2 between south west and north west] {Select data\\ for training};
        \node (30) [process, below=2cm of 20, clabel=D3 between south west and north west] {Select\\ feature space};
        \node (30bis) [process, below=0cm of 30,clabel=D4 between south west and north west] {Select\\ kernel};
        \node (40) [process, below=2cm of 30,clabel=D5 between south west and north west] {Optimise \\ kernel};
        \node (50) [process, below=2cm of 40,clabel=D6 between south west and north west] {Test \\ GP model};

        \node (11) [process2, left=of 10,clabel=M1 between south west and north west]  {Random generation, \\ MD sampling, \\ ...};
        \node (22) [process2, left=of 20,clabel=M2 between south west and north west]  {Clustering,\\ Symmetrising, \\ ...};
        \node (33) [process2, left=of 30,clabel=M3 between south west and north west]  {Molecular structure, \\ Physical constraints, \\ ...};
        \node (44) [process2, left=of 40,clabel=M4 between south west and north west]  {GMIN, sklearn, \\ LOOCV};
        \node (55) [process2, left=of 50,clabel=M5 between south west and north west]  {MAE,  MAE, \\ Latent function};

        \node (111) [process3, right=of 10,clabel=O1 between south west and north west] {Large amount of data \\ $ \mathbf{X},  \mathbf{y} $};
        \node (222) [process3, right=of 20,clabel=O2 between south west and north west] {Training data \\ $\chi = \{ \mathbf{X}, \mathbf{y} \}$};
        \node (444) [process3, right=of 40,clabel=O3 between south west and north west] {Optimised \\ hyperparameters};
        \node (444bis) [process3, below=0cm of 444,clabel=O4 between south west and north west] {Optimised GP};
        \node (555) [process3, right=of 50,clabel=O4 between south west and north west] {Accurate PES model};

\node [container1, fit=(11) (22) (33) (44) (55), minimum height=12cm] (METHOD) {};
\node at (METHOD.north) [fill=white,draw,rounded corners, very thick] {\textbf{Methods}};
\node [container2, fit=(10) (20) (30) (30bis) (40) (50), minimum height=12cm] (PROCESS) {};
\node at (PROCESS.north) [fill=white,draw,rounded corners, very thick] {\textbf{Process}};
\node [container3, fit=(111) (222) (444) (555), minimum height=12cm] (OUTPUT) {};
\node at (OUTPUT.north) [fill=white,draw,rounded corners, very thick] {\textbf{Output}};

	\draw (start) edge[marrow] (10);
	\draw[thick, scale=2, decoration={markings,mark=at position 0.2 with {\arrow{>}}}]   (start) -- (10);
	\draw (111) edge[marrow] (20);
	\draw (222) edge[marrow] (30);
	\draw (444) edge[marrow] (50);

        \draw (11) edge[marrow] (10);
        \draw (22) edge[marrow] (20);
        \draw (33) edge[marrow] (30);
        \draw (33) edge[marrow] (30bis);
        \draw (44) edge[marrow] (40);
        \draw (55) edge[marrow] (50);

        \draw (10) edge[marrow] (111);
        \draw (20) edge[marrow] (222);
        \draw (40) edge[marrow] (444);
        \draw (50) edge[marrow] (555);

        \draw (30) edge[marrow] (22);

        \node (start) [startstop] {\textbf{Start}};
        \node (10) [process, below=2cm of start, clabel=D1 between south west and north west] {Generate data};
        \node (20) [process, below=2cm of 10, clabel=D2 between south west and north west] {Select data\\ for training};
        \node (30) [process, below=2cm of 20, clabel=D3 between south west and north west] {Select\\ feature space};
        \node (30bis) [process, below=0cm of 30,clabel=D4 between south west and north west] {Select\\ kernel};
        \node (40) [process, below=2cm of 30,clabel=D5 between south west and north west] {Optimise\\ kernel};
        \node (50) [process, below=2cm of 40,clabel=D6 between south west and north west] {Test \\ GP model};
\node at (PROCESS.north) [fill=white,draw,rounded corners, very thick] {\textbf{Process}};

        \node (11) [process2, left=of 10,clabel=M1 between south west and north west]  {Random generation, \\ MD sampling, \\ ...};
        \node (22) [process2, left=of 20,clabel=M2 between south west and north west]  {Clustering,\\ Symmetrising, \\ ...};
        \node (33) [process2, left=of 30,clabel=M3 between south west and north west]  {Molecular structure, \\ Physical constraints, \\ ...};
        \node (44) [process2, left=of 40,clabel=M4 between south west and north west]  {GMIN, sklearn, \\ LOOCV};
        \node (55) [process2, left=of 50,clabel=M5 between south west and north west]  {RMSE,  MAE, \\ Latent function};

        \node (111) [process3, right=of 10,clabel=O1 between south west and north west] {Large amount of data \\ $ \mathbf{X},  \mathbf{y} $};
        \node (222) [process3, right=of 20,clabel=O2 between south west and north west] {Training data \\ $\chi = \{ \mathbf{X}, \mathbf{y} \}$};
        \node (444) [process3, right=of 40,clabel=O3 between south west and north west] {Optimised \\ hyperparameters};
        \node (444bis) [process3, below=0cm of 444,clabel=O4 between south west and north west] {Optimised GP};
        \node (555) [process3, right=of 50,clabel=O4 between south west and north west] {Accurate PES model};
    \end{tikzpicture}
}
\vspace{0.4cm}
\caption[Flowchart of GP setup for PES learning.]{Flowchart that schematically explains the different levels of the GP setups where choices of methods and GP kernels, feature space and so on.  We define "M" as methods,  "D" as decisions and "O" as outputs.}
\label{fig:gp-flowchart}
\end{figure}

In order to represent the steps and decisions one has to take when learning data with GPs,  we define a chart in figure \ref{fig:gp-flowchart}.  As the methodology above states,  some of the decisions are not changed in this paper: this is the case for D1 and D2\footnote{Although we do explore some of the methods in M2 which do affect what the training data is.} since the training data is fixed,  and D5 and D6 since we only use LML and MAEs as means of selecting hyperparameters and test our models.  However,  the D3 and D4 are explored with different feature space and kernels tested. 

\section{Results}
\begin{figure}[H]
\vspace{0.4cm}
\centering
\begin{tikzpicture}[scale=0.5]
\node[inner sep=0pt] (graph) at (0,-3) {\includegraphics[width=0.25\textwidth]{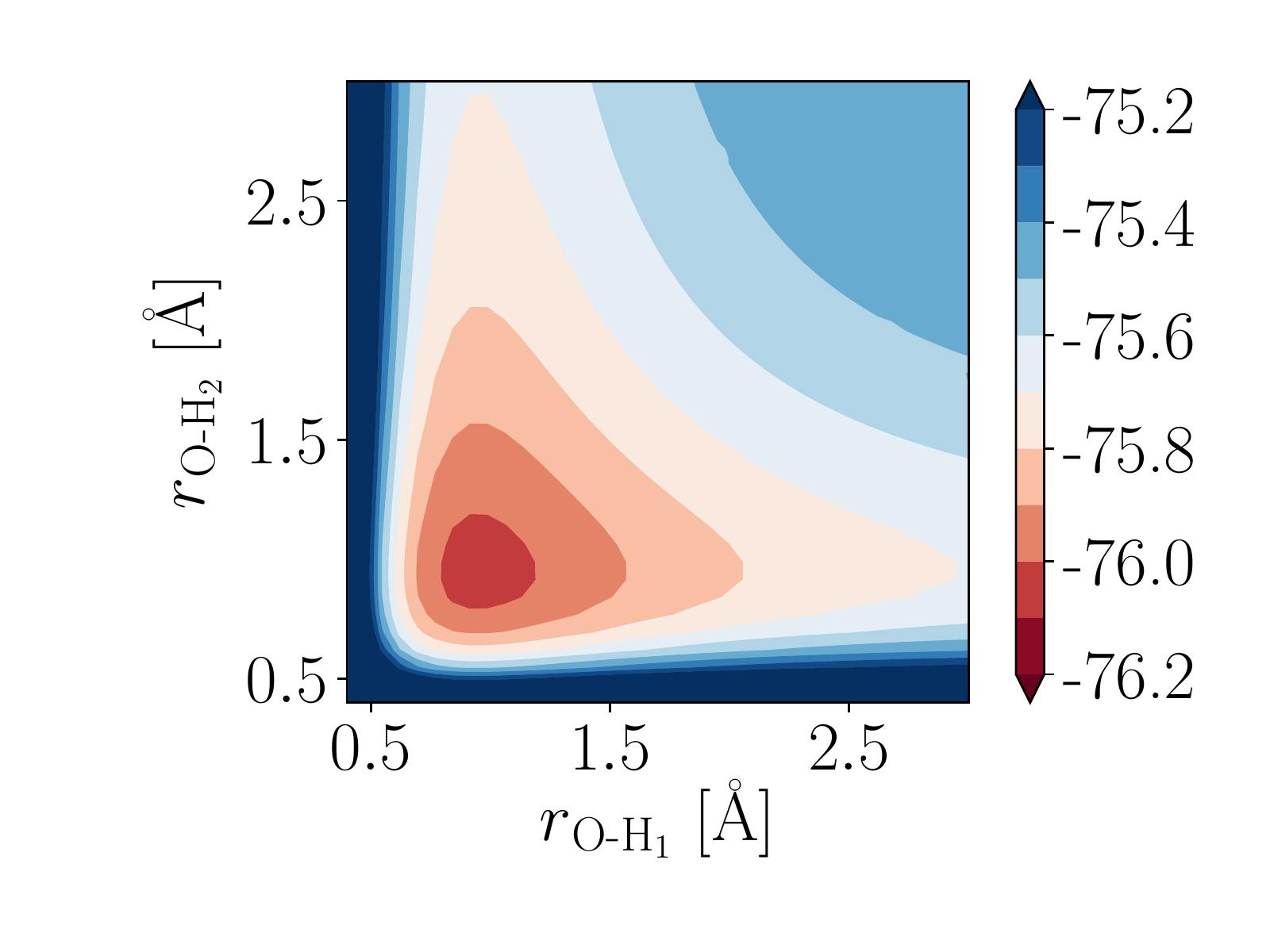}};
    \node[shape=circle,fill=red, scale=1.0,label=above:{O}] (a) at (7, -2) {};
    \node[shape=circle,fill=gray, scale=0.5,label=left:{H$_1$}]  (b) at (6, -3) {};
    \node[shape=circle,fill=gray, scale=0.5,label=right:{H$_2$}]  (c) at (8, -3) {};
    \foreach \from / \to in { a/b, a/c}
        \draw [-, very thick] (\from) -- (\to);
\end{tikzpicture}
\vspace{0.4cm}
\caption[True PES of the water at UHF/aug-cc-pVDZ.]{The true surface of the PES at UHF/aug-cc-pVDZ of a single water molecule in the gas phase.  One can see the rather simple PES spanned along the two O-H bond stretches,  here plotted between 0.4 and 3.0 \AA.  A water molecule is drawn on the right to show the labelling which is used in later sections. }
\label{fig:h2o-truesurface}
\end{figure}
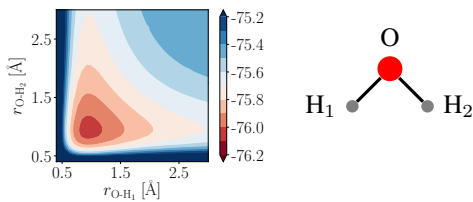

The function we are trying to model with a GP is the true UHF/aug-cc-pVDZ surface,  shown in figure \ref{fig:h2o-truesurface}.  The PES is rather simple and exhibits a single minimum.  The projection of the surface along the stretches of the two O-H bonds is the most interesting as it is symmetric due to the permutational invariance of the water molecule.  The ability of a GP model to replicate this symmetry is easily seen on the plot of the latent function. 

Obtaining accurate models for such a simple surface is not a hard task with methods that are available today and one can produce,  using semi-empirical models,  sub-$\mu$Ha accuracy water potentials\autocite{Mizus2018}.  Closer to ML applications,  sub-0.1 eV ($0.1\ \mathrm{eV} \approx 4\ \mathrm{mHa}$) are commonly reported\autocite{Abbott2019,cheng2019universal} often with rather large training sets (equivalent to $>6$ grid points per degree of freedom\footnote{In this study we use 24 geometries which equate to less than 3 grid points per degree of freedom.}) and with allowed stretches of around 0.2 \AA.  More recently,  sub-mHa ML potentials with around 100 training geometries for a single water molecule have also been reported using novel method in quantum computing\autocite{Schumacher2022}.
\par
In this study,  we do not aim to produce PES models which are more accurate than the current best published ML potentials.  We are instead interested in the Bayesian optimisation of GP models for sparse and high energy data.  Our errors are rather large but are only compared to one another to measure the ability of the various GP schemes at reproducing target patterns.

\subsection{Internuclear distances}

The first feature space we consider is the set of internuclear distances (ID), given by $\mathrm{X} = [ r_{\mathrm{O-H}_1},  r_{\mathrm{O-H}_2}, r_{\mathrm{H}_1-\mathrm{H}_1} ]$.  The resulting latent function has a large error on the testing sets with 8.0 mHa and 37.0 mHa. The GP latent function misses the symmetry of the PES with very different length scales for the two bonds which are permutationally invariant.
\begin{figure}[H]
\vspace{0.4cm}
\centering
	\begin{tikzpicture}[scale=0.5]
	\node[inner sep=0pt, label=above:{\small MAE: 8.0/38.1 mHa}] (graph1) at (0,-2) {\includegraphics[width=0.25\textwidth]{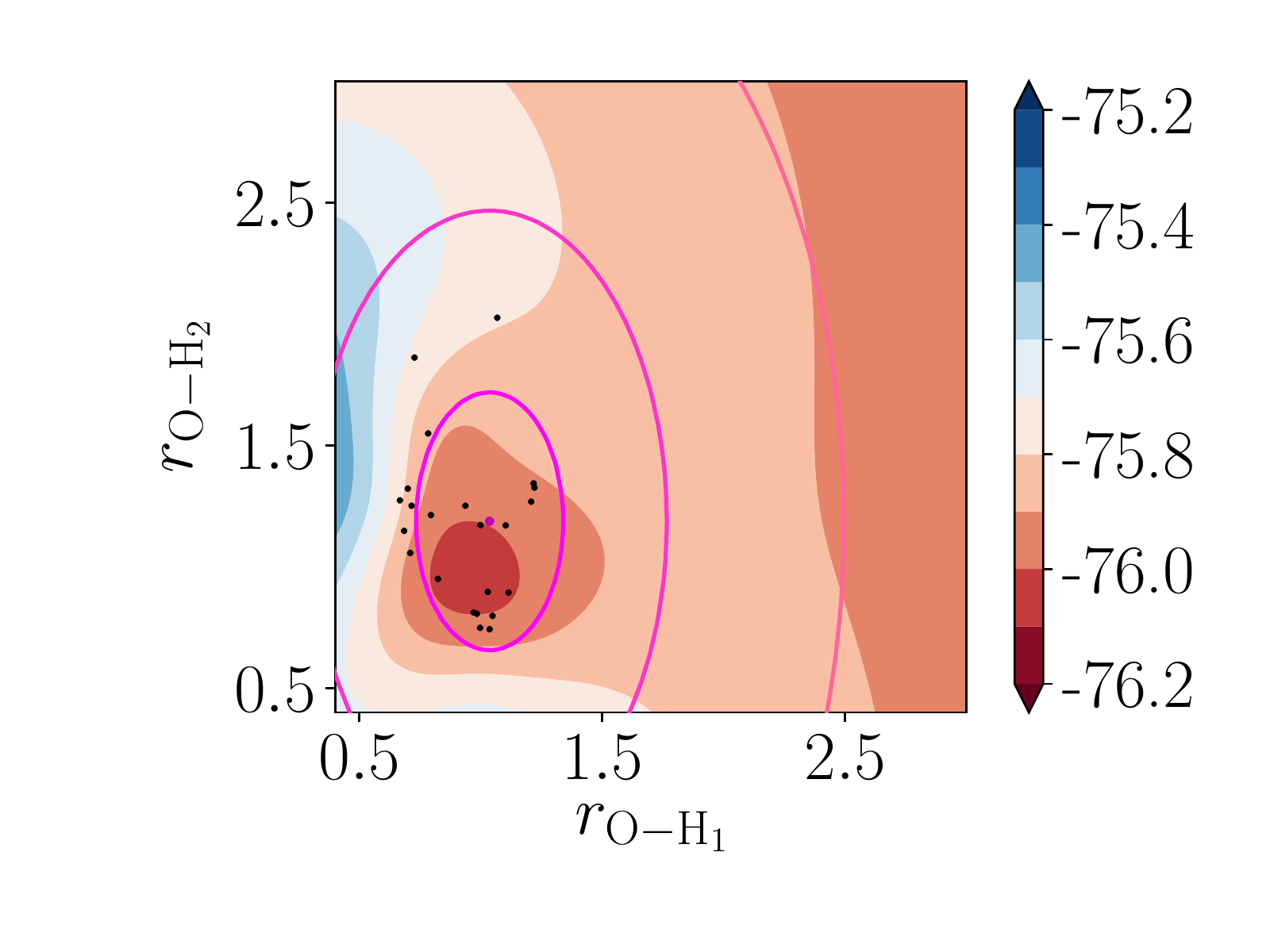}};
	\node[inner sep=0pt] (graph2) at (7,-1.8) {\includegraphics[width=0.09\textwidth,trim = 3.7cm -5cm 3.7cm 0cm,clip]{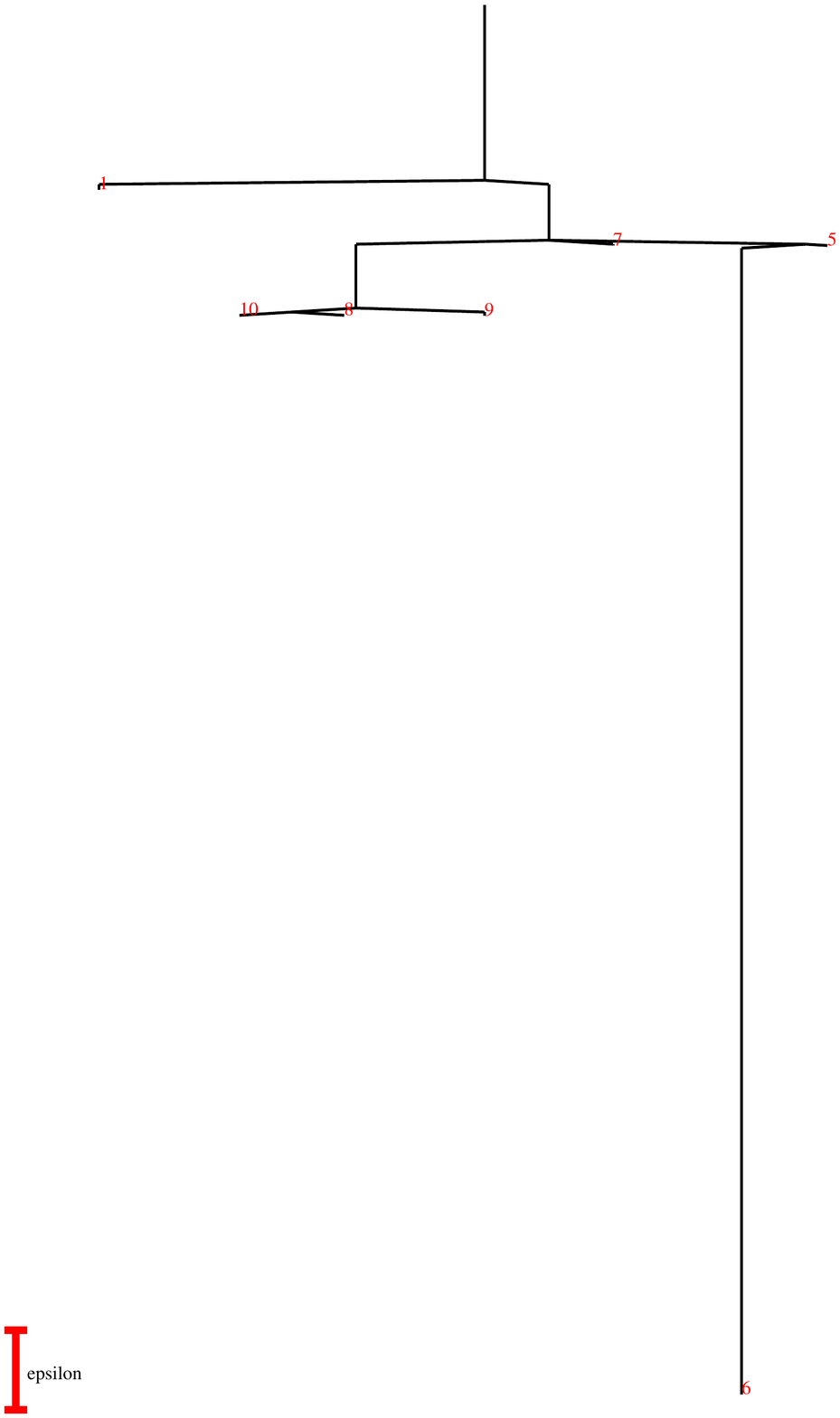}};
	\end{tikzpicture}
	\vspace{0.4cm}
	\caption{Predicted PES by a GP trained on the ID feature space alongside the disconnectivity graph of the LML function.  Black dots represent training data and magenta contours are isovalue of the covariance function for one training point selected at random. }
	\label{fig:id_feature_space}
\end{figure}

A way to understand the ability of the GP to predict the surface from the sparse data,  is to see how far a data point ``influences'' the prediction (this is controlled by the length scale hyperparameter).  In figure \ref{fig:id_feature_space},  we plot magenta contours which represent isovalues of the covariance function from a selected point.  The further they stretch the more ``information'' is carried over from the training data to the GP model.
\par
A common projection of the internuclear distances used in ML-PESs,  is the Morse-transformation defined as
\begin{equation}
\tilde{\mathrm{X}}_i = \mathrm{exp}(-(\mathrm{X}_i - \mathrm{X}_0)/\alpha) \qquad \mathrm{for} \qquad i=1,2,3 
\end{equation}
where $\alpha$ and $\mathrm{X}_0$ are Morse parameters that control the projection.  The choice of the latter can be derived from chemistry or,  more generally,  can be defined for all internuclear distances as a set of optimal parameters.  We have,  in a different study\autocite{Fabiochap5},  addressed the way one can find optimal parameters.
\par
The PES projected on this feature space,  denoted MID here,  is ``simplified'': the bond stretches are expanded in the steep,  nuclear repulsion,  regions of the PES at short bond lengths and are contracted in the slowly changing regions towards bond dissociation.  The MID feature space,  here projected using $\alpha=2.0$ and $\mathrm{X}_0=0.0$, does not improve on the ID feature space with similar large errors on the testing sets.  The MID latent function also misses the symmetry in the latent function. 
\begin{figure}[H]
	\centering
	\vspace{0.4cm}
\centering
	\begin{tikzpicture}[scale=0.5]
	\node[inner sep=0pt, label=above:{\small MAE: 7.7/37.9 mHa}] (graph1) at (0,-2) {\includegraphics[width=0.25\textwidth]{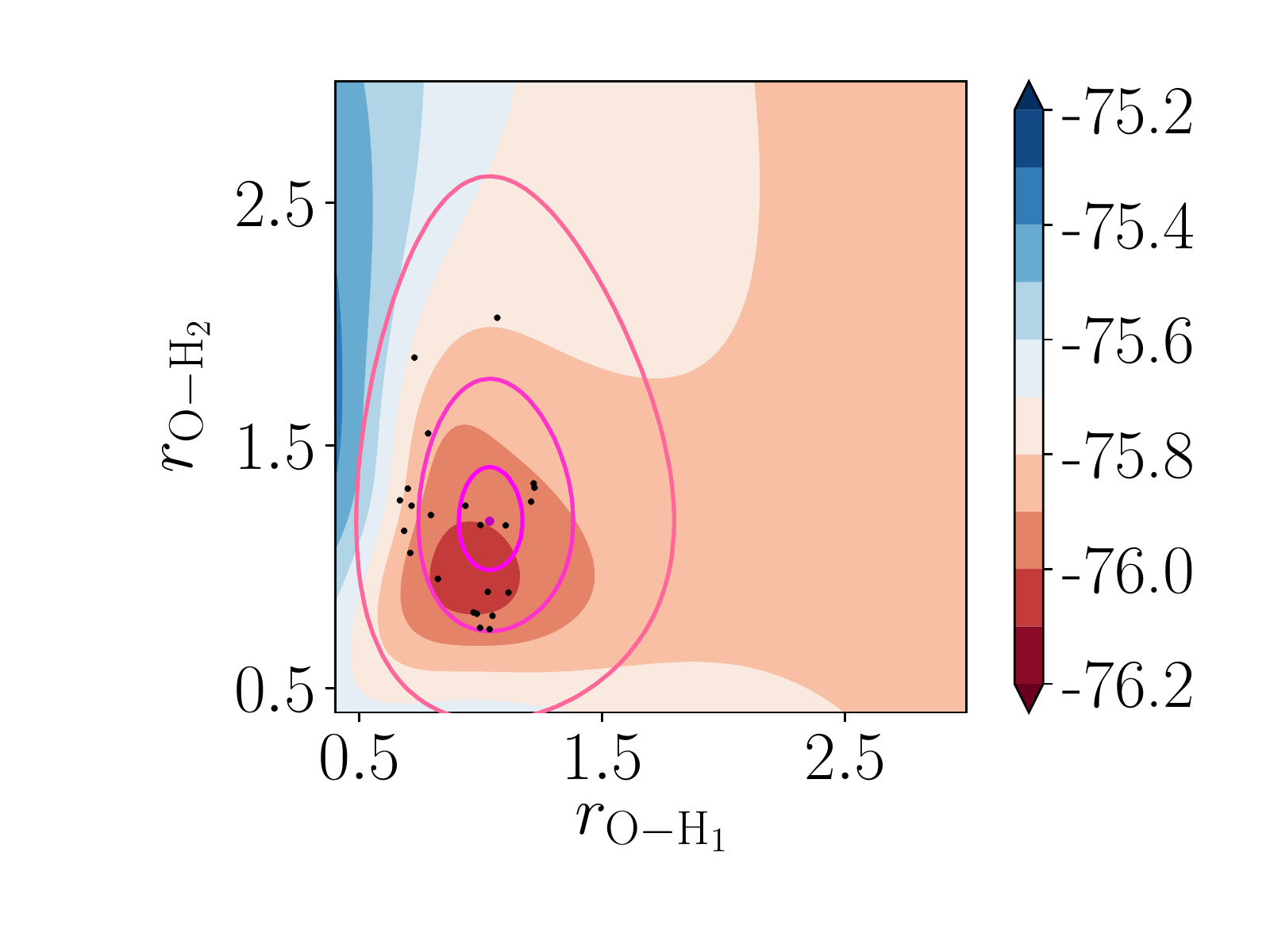}};
	\node[inner sep=0pt] (graph2) at (7,-1.8) {\includegraphics[width=0.09\textwidth,trim = 3.7cm -5cm 3.7cm 0cm,clip]{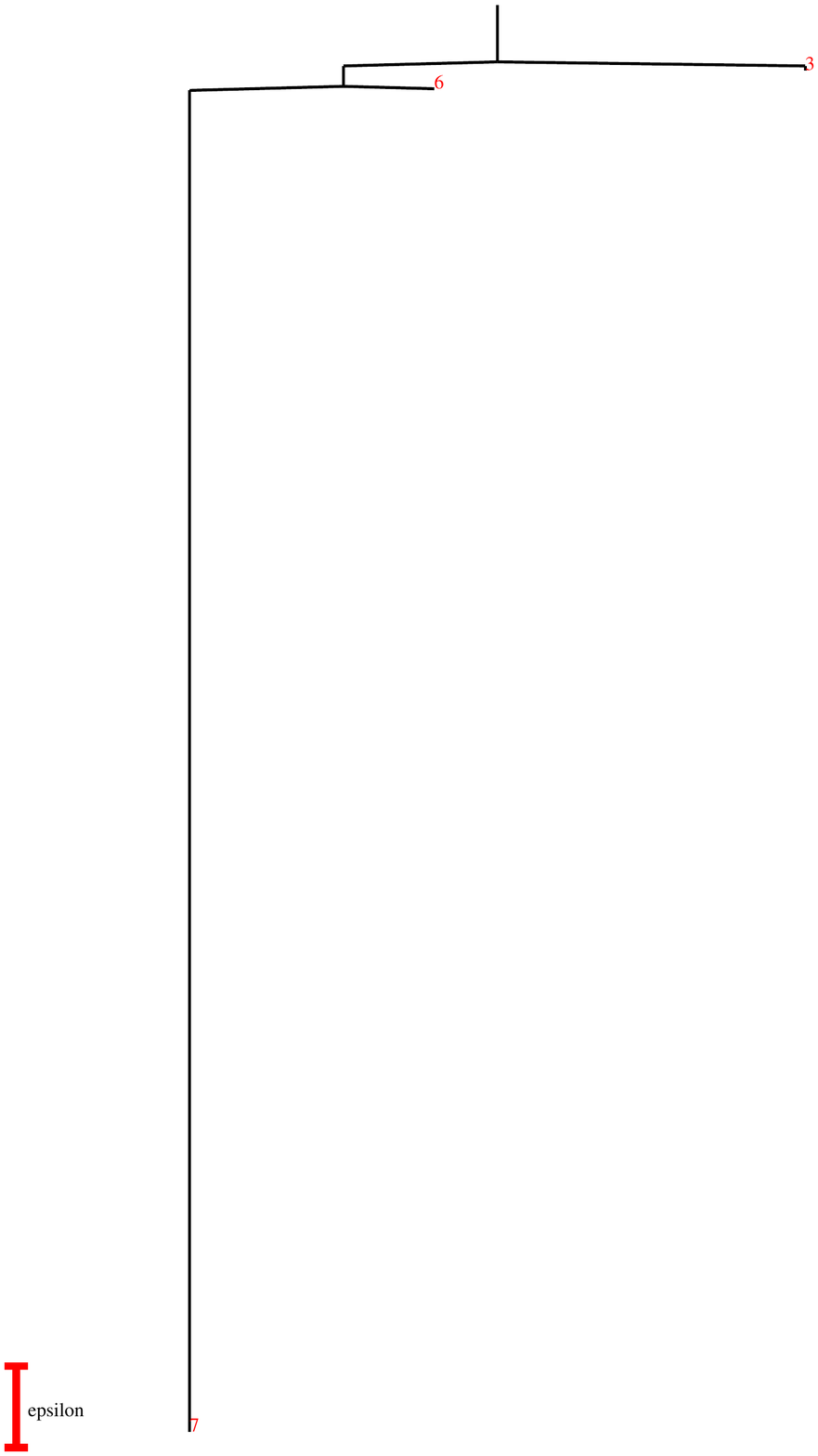}};
	\end{tikzpicture}
	\vspace{0.4cm}
	\caption{Predicted PES by a GP trained on the MID feature space alongside the disconnectivity graph of the LML function.  Black dots represent training data and magenta contours are isovalue of the covariance function. }
	\label{fig:mid_feature_space}
\end{figure}
Simple feature space do not easily reproduce the symmetry of the PES with very sparse sets.  One needs to consider baking the symmetry into the feature space or into the training data to improve the modelling.

\subsection{Fundamental invariants}

We now consider the enforcement of symmetry using mathematical considerations,  and start by projecting the training set on the coefficients of the fundamental invariant polynomials (FI feature space) obtained from so called primary and secondary invariants\autocite{Xie2009}.  Fundamental invariants are the minimal basis that spans the PIPs\autocite{Shao2016,Opalka2013} and they can be obtained using algebra software like Singular\autocite{Singular} which provide,  for a triatomic specifically,  an alternative 3D feature space for learning.  The feature space being completely invariant w.r.t.  same atom permutations,  we are guaranteed to produce a symmetrical PES.  However,  despite describing a correctly permutationally invariant surface,  the FI-based model is about as performant as the ID-based one with MAEs of 8.0 / 38.1 mHa and of 7.4/ 37.2 mHa,  respectively,  given for both low and high energy testing sets mentioned in the previous section.

\begin{figure}[H]
\vspace{0.4cm}
\centering
	\begin{tikzpicture}[scale=0.5]
	\node[inner sep=0pt,label=above:{\small MAE: 7.4/37.2 mHa}] (graph1) at (0,-2) {\includegraphics[width=0.25\textwidth]{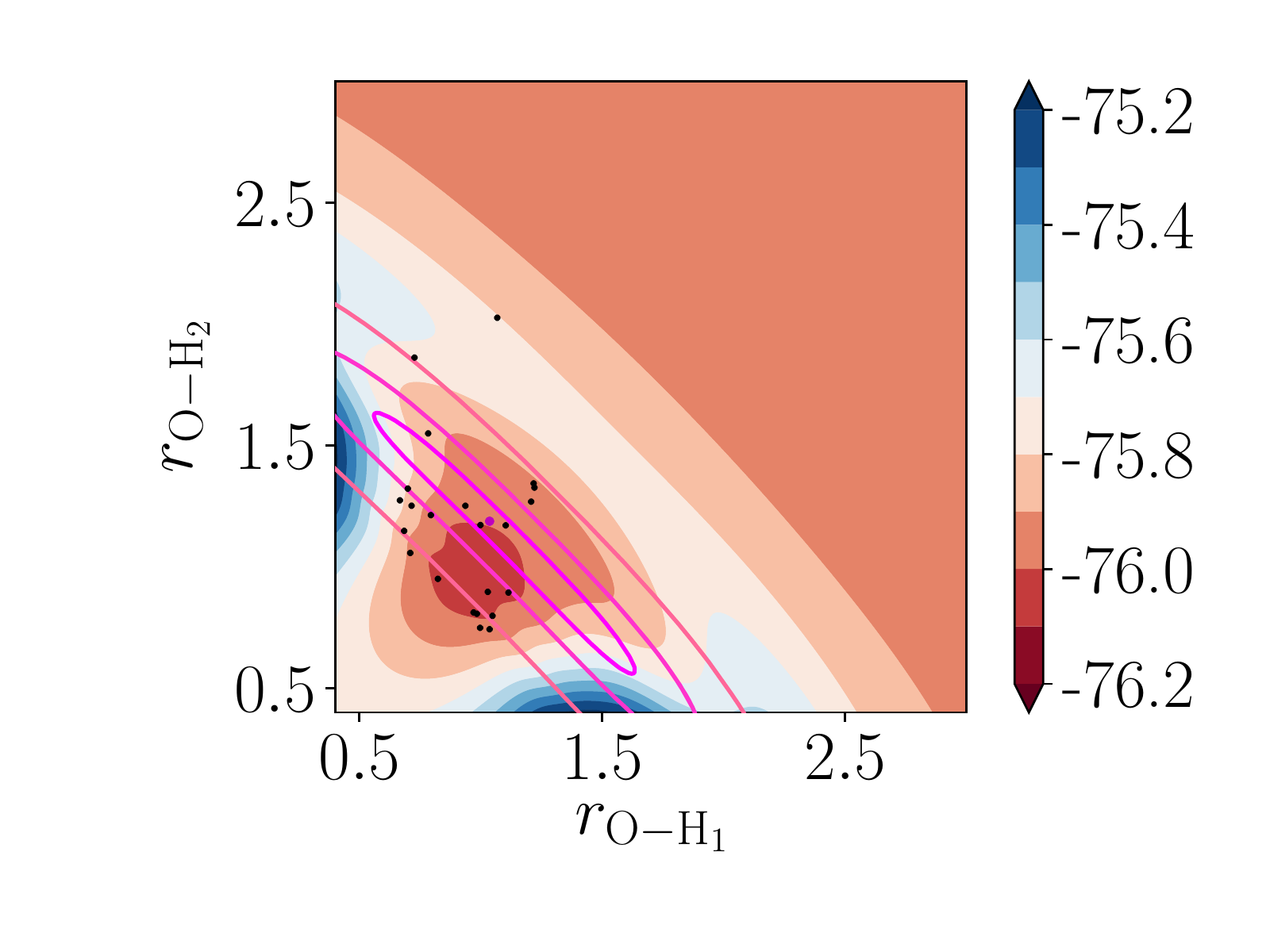}};
	\node[inner sep=0pt] (graph2) at (7,-1.8) {\includegraphics[width=0.09\textwidth,trim = 3.7cm -5cm 3.7cm 0cm,clip]{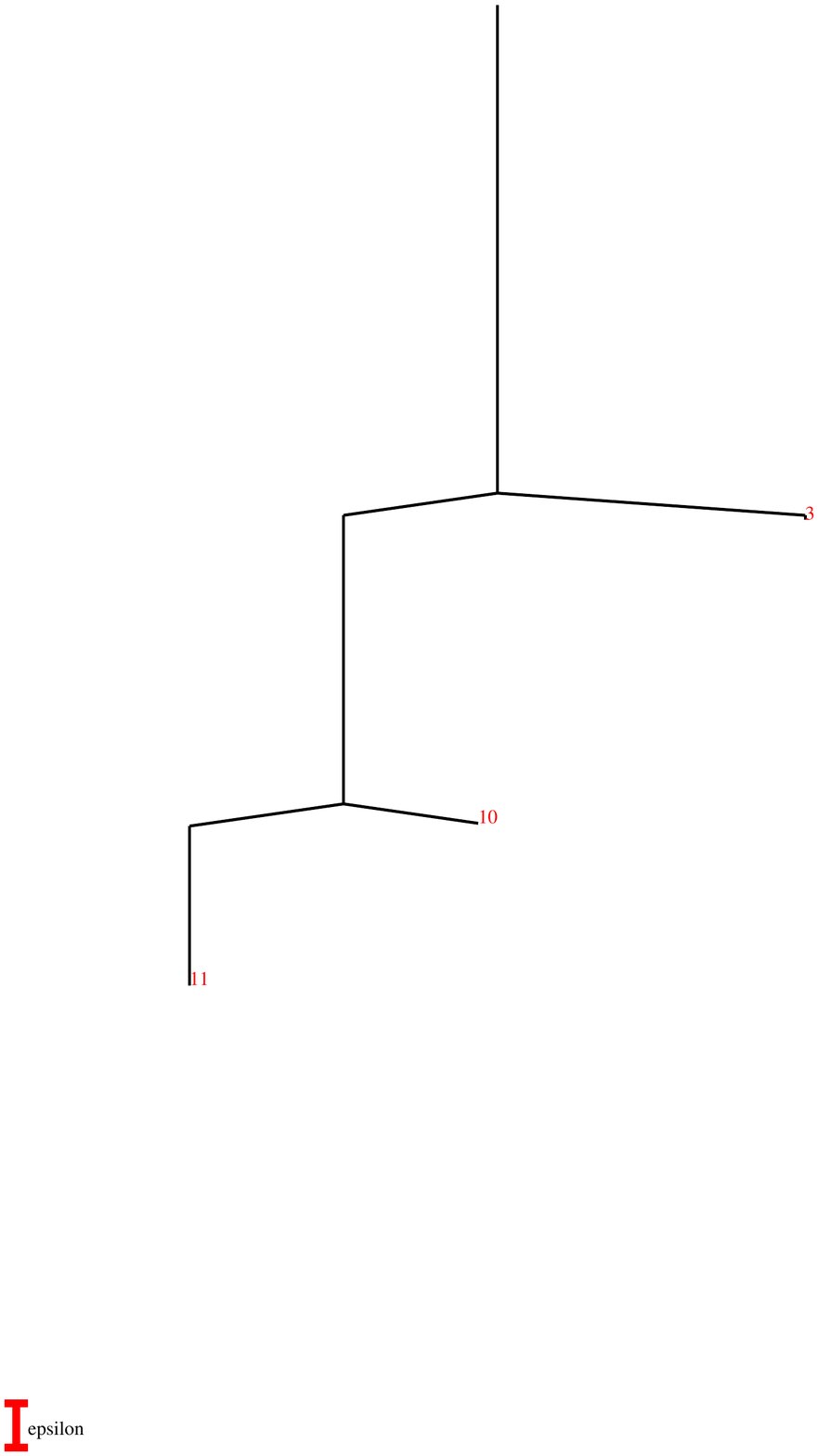}};
	\end{tikzpicture}
	\vspace{0.4cm}
	\caption{Predicted PES by a GP trained on the FI feature space alongside the disconnectivity graph of the LML function.  Black dots represent training data and magenta contours are isovalue of the covariance function. }
	\label{fig:fi_feature_space}
\end{figure}

The main reason for FIs to not perform well is that the space where data was sampled is greatly compressed and,  despite having a potentially good description of the region of feature space near data where the model is learned.  This is seen in the projected FI ellipse that takes a crescent shape in the O-H bond length space giving a quite bad range of \textit{influence} along the $x=y$ direction despite a good range on the perpendicular direction.  Gaussian process optimise compressed sets badly and typically go into an \textit{over fitting} regime making an overall PES which only describes the energy well near the sampled data.  Given the different power of the polynomials of the FI feature dimensions,  one could also consider to make them ``power-consistent'' by taking the $n^{\mathrm{th}}$ root of the polynomial: for example for the FI of AB$_2$ molecules one would take the square root of $f_1=\mathrm{X}_{\mathrm{O-H_1}}^2 + \mathrm{X}_{\mathrm{O-H_2}}^2$.  This does not however improve the performance of the latent function for this training set. 
\par
For sparse training sets,  the FI feature dimensions seem to have a surface that is not well-fitted by the latent function of the Gaussian process.  Moreover, the number of FIs greatly increases with the number of atoms: for AB$_3$ type molecule the 6 MID feature space would have 6 feature dimensions against 15 for the FI feature space and,  for even larger molecules of type A$_2$B$_4$ one would need 122 dimensions in the FI space against only 15 features for the MID space (this would be the case for a water dimer for example).  


\subsection{Normal modes}
Fitting global PESs to normal modes appears to be a sensible choice as they describe,  to a truncated Taylor approximation,  a ``simpler'' surface around the equilibrium geometry\autocite{Carter1997,Yagi2003,Rauhut2006}.  Solving the eigenvalue problem of the mass-weighted Hessian around a symmetrical geometry (C$_{2v}$  for the water molecule) yields Cartesian displacement vectors that describe motions of the molecule which correspond to irreducible representations.  This allows us to construct a feature space with symmetry from a chemical perspective rather than a mathematical one (as done for FIs).
\par
We first define the mass-weighted Hessian of the energy $E$,  as
\begin{equation}
\mathbf{H}_{\mathrm{MW}} = \mathbf{M}
\begin{bmatrix}
\frac{\partial^2 E}{\partial  \xi_1 \partial  \xi_1} &\frac{\partial^2 E}{\partial  \xi_1 \partial  \xi_2} & \ldots & \frac{\partial^2 E}{\partial  \xi_1 \partial  \xi_{3N}} \\[0.3cm]
\frac{\partial^2 E}{\partial  \xi_2 \partial  \xi_1} &\frac{\partial^2 E}{\partial  \xi_2 \partial  \xi_2} & \ldots & \frac{\partial^2 E}{\partial  \xi_2 \partial  \xi_{3N}} \\[0.3cm]
\vdots & \vdots & \ddots & \vdots \\[0.3cm]
\frac{\partial^2 E}{\partial  \xi_{3N} \partial  \xi_1} &\frac{\partial^2 E}{\partial  \xi_{3N} \partial  \xi_2} & \ldots & \frac{\partial^2 E}{\partial  \xi_N \partial  \xi_{3N}} \\
\end{bmatrix} \mathbf{M}
\label{eq:mwH}
\end{equation}
where $\mathbf{M}$ is a diagonal matrix with the diagonal terms given by M$_{ii} = m_i^{1/2}$ and $\xi_i$ are the Cartesian coordinates labelled as $\xi_1,\xi_2,\xi_3, \xi_4 \ldots \xi_{3N} = \Delta x_1,  \Delta y_1,  \Delta z_1,  \Delta x_2 \ldots \Delta z_N$ where one labels the $N$ atoms for the $\Delta$s and the $3N$ coordinates for the $\xi$.  An important step to obtain a feature space,  is to project the translation and rotations out of the Hessian.  This can be done by generating the translation and rotation mass-weighted vectors and then performing a Schmidt orthogonalisation to produce $3N-6$ mass-weighted vectors\footnote{One would only produce $3N-5$ vectors for a linear molecule.} which are orthogonal to the translations and rotations.  The latter form the matrix $\mathbf{D}$,  and considering the eigenvalue problem of the projected mass-weighted Hessian, 
\begin{equation}
\mathbf{D}^{\dagger} \mathbf{H}_{\mathrm{MW}} \mathbf{D} \ \mathbf{v}_i = \lambda_i \ \mathbf{v}_i
\label{eq:mweig}
\end{equation}
one gets the eigenvectors,  $\mathbf{v}_i$,  which we will call ``normal modes'' and the eigenvalues,  $\lambda_i$,  which give the normal modes vibrational frequencies.  The NM coefficients that one uses to create a feature space are the projection of the Cartesian displacements on the $3N-6$ normal modes:
\begin{equation}
\begin{bmatrix}
v_1\\
v_2\\
\vdots\\
v_{3N-6}\\
\end{bmatrix} = \mathbf{U}^{-1} \  \mathbf{D}^{\dagger} \  \mathbf{M} \ \Delta \mathbf{q}
\label{eq:proj-nm}
\end{equation}
where $\Delta \mathbf{q} = \mathbf{q} - \mathbf{q}_{\mathrm{eq}}$ is a $3N$ vector of the Cartesian displacements from the equilibrium geometry,  $\mathbf{U}^{-1}$ is the inverse of the matrix whose columns are the eigenvectors of the projected mass-weighted Hessian of equation \ref{eq:mweig}. 
\begin{figure}[H]
\vspace{0.4cm}
\centering
\begin{tikzpicture}[scale=0.7]
    \node[shape=circle,fill=red, scale=0.5, label=above:{$v_2$}] (a) at (1.6, 0) {};
    \node[shape=circle,fill=black,scale=0.25]  (b) at (1.1, -0.5) {};
    \node[shape=circle,fill=black,scale=0.25]  (c) at (2.1, -0.5) {};

    \node[shape=circle,fill=black, scale=0.05]  (d) at (0.75, -0.85) {};
    \node[shape=circle,fill=black, scale=0.05]  (e) at (2.45, -0.85) {};

    \foreach \from / \to in { a/b, a/c}
        \draw [-,thick] (\from) -- (\to);
    \foreach \from / \to in { b/d, c/e}
        \draw [->,blue,thick] (\from) -- (\to);

    \node[shape=circle,fill=red, scale=0.5, label=above:{$v_3$}] (a) at (5.3, 0) {};
    \node[shape=circle,fill=black,scale=0.25]  (b) at (4.8, -0.5) {};
    \node[shape=circle,fill=black,scale=0.25]  (c) at (5.8, -0.5) {};

    \node[shape=circle,fill=black, scale=0.05]  (d) at (4.45, -0.85) {};
    \node[shape=circle,fill=black, scale=0.05]  (e) at (5.45, -0.15) {};

    \foreach \from / \to in { a/b, a/c}
        \draw [-,thick] (\from) -- (\to);
    \foreach \from / \to in { b/d, c/e}
        \draw [->,blue,thick] (\from) -- (\to);

    \node[shape=circle,fill=red, scale=0.5, label=above:{$v_1$}] (a) at (8.6, 0) {};
    \node[shape=circle,fill=black,scale=0.25]  (b) at (8.1, -0.5) {};
    \node[shape=circle,fill=black,scale=0.25]  (c) at (9.1, -0.5) {};

    \node[shape=circle,fill=black, scale=0.05]  (d) at (8.45, -0.85) {};
    \node[shape=circle,fill=black, scale=0.05]  (e) at (8.75, -0.85) {};

    \foreach \from / \to in { a/b, a/c}
        \draw [-,thick] (\from) -- (\to);
    \foreach \from / \to in { b/d, c/e}
        \draw [->,blue,thick] (\from) -- (\to);
        
    \node[inner sep=0pt] (a) at (1.5, -3) {\includegraphics[width=0.12\textwidth,trim=0cm 2.2cm 0cm 0cm, clip]{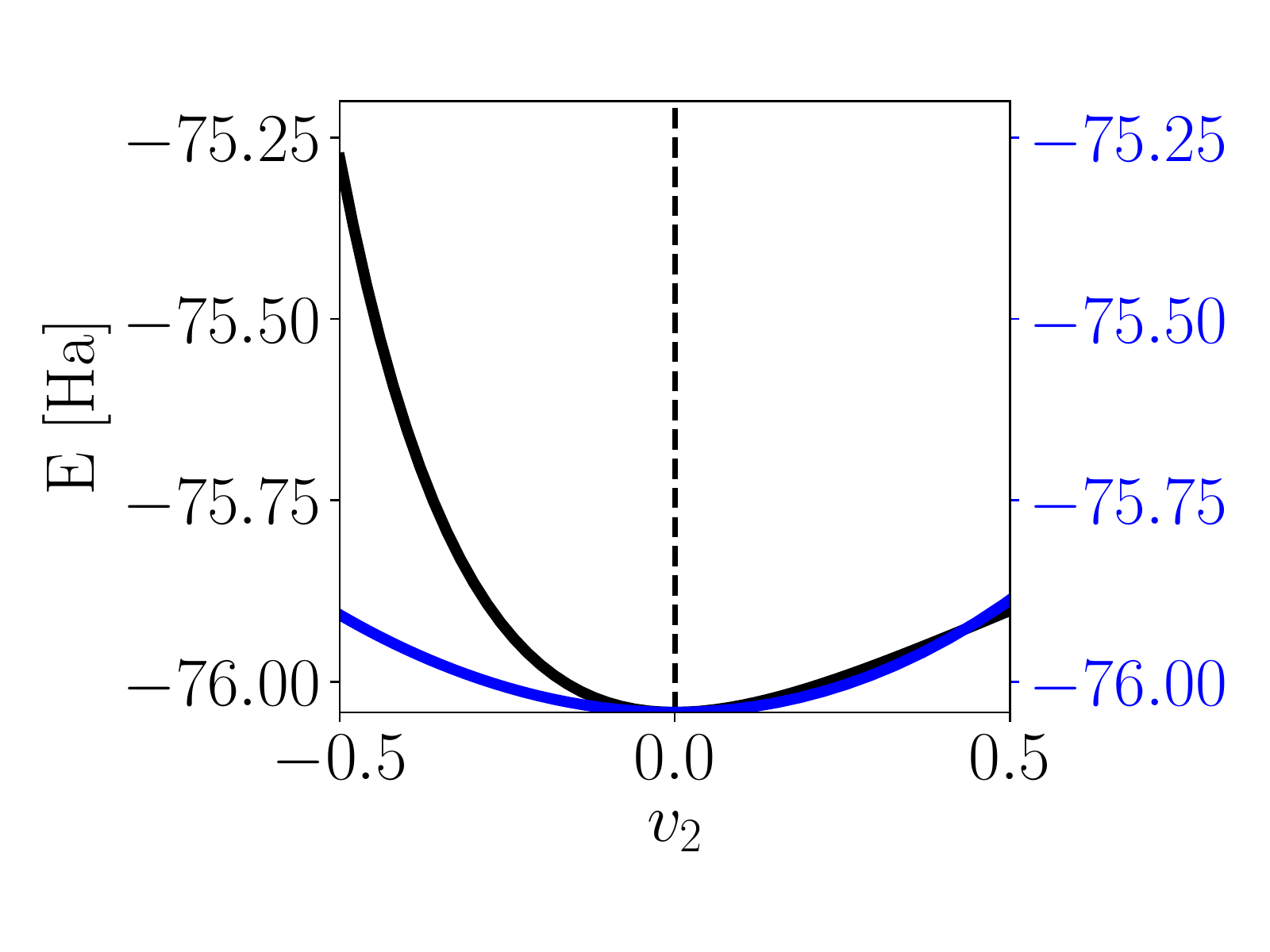}};
    \node[inner sep=0pt] (a) at (5, -3) {\includegraphics[width=0.12\textwidth,trim=0cm 2.2cm 0cm 0cm ,clip]{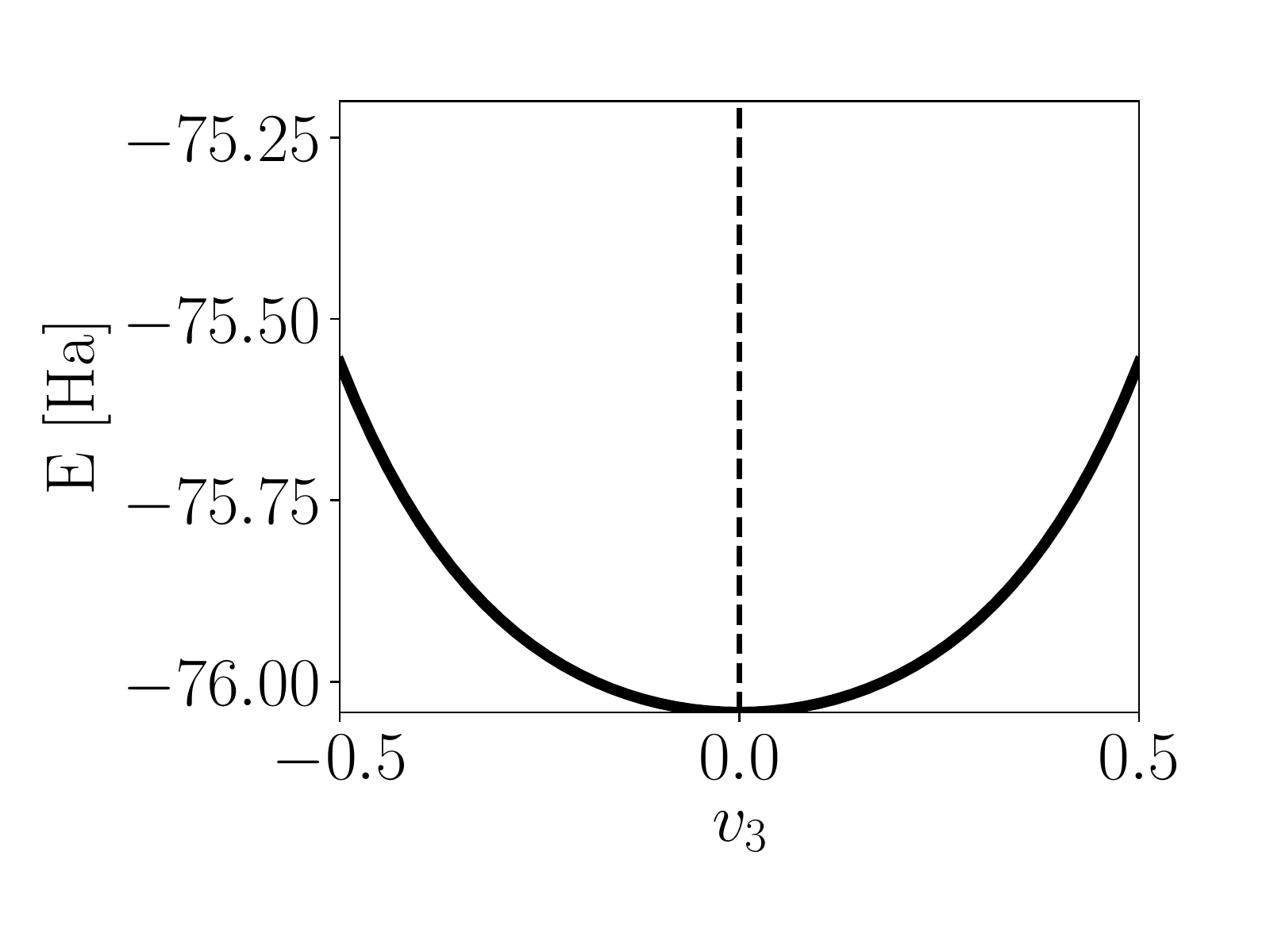}};
    \node[inner sep=0pt] (a) at (8.5, -3) {\includegraphics[width=0.12\textwidth,trim=0cm 2.2cm 0cm 0cm, clip]{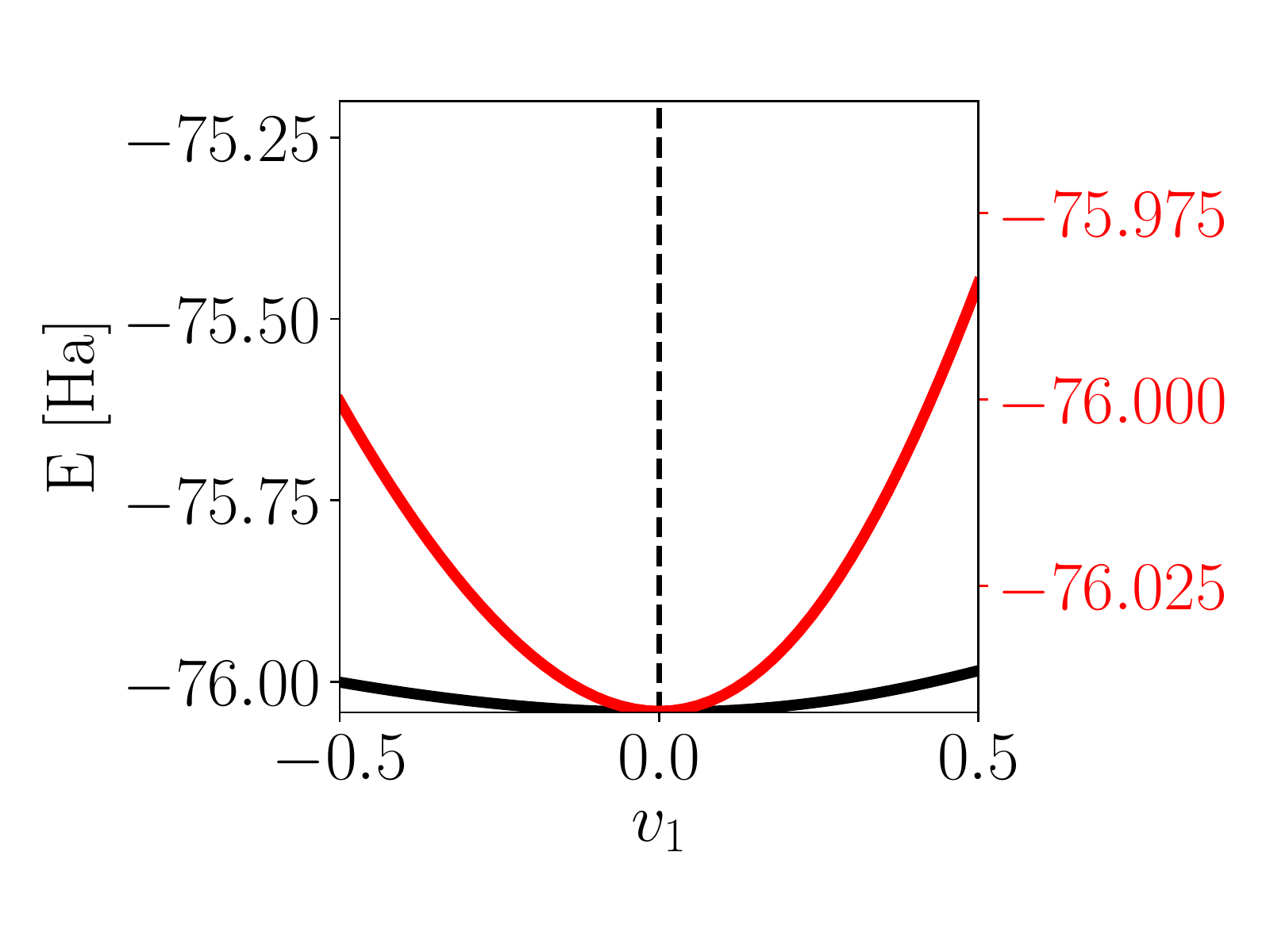}};
\end{tikzpicture}
\vspace{0.6cm}
\caption{Normal modes of vibration of the water molecule with their respective energy curves shown by the black lines.  The bending mode $v_1$ is shown again on a different scale with a red line as it has a much flatter energy curve.  Finally,  the blue energy profile for the $v_2$ normal mode is the Morse-transformed normal mode which is discussed further below (the equilibrium geometry,  when Morse transformed,  has a coefficient of 1 instead of 0 and the blue line is hence also shifted).}
\label{fig:nm}
\end{figure}
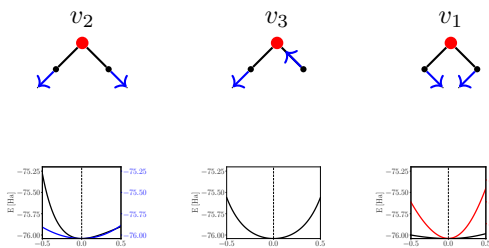

\par
\begin{figure}[H]
\vspace{0.4cm}
\centering
	\begin{tikzpicture}[scale=0.5]
	\node[rotate=90] (a) at (-2,-1.5) {\footnotesize MAE: 12.7/33.1 mHa};
	\node[inner sep=0pt] (graph1) at (2,-2) {\includegraphics[width=0.22\textwidth]{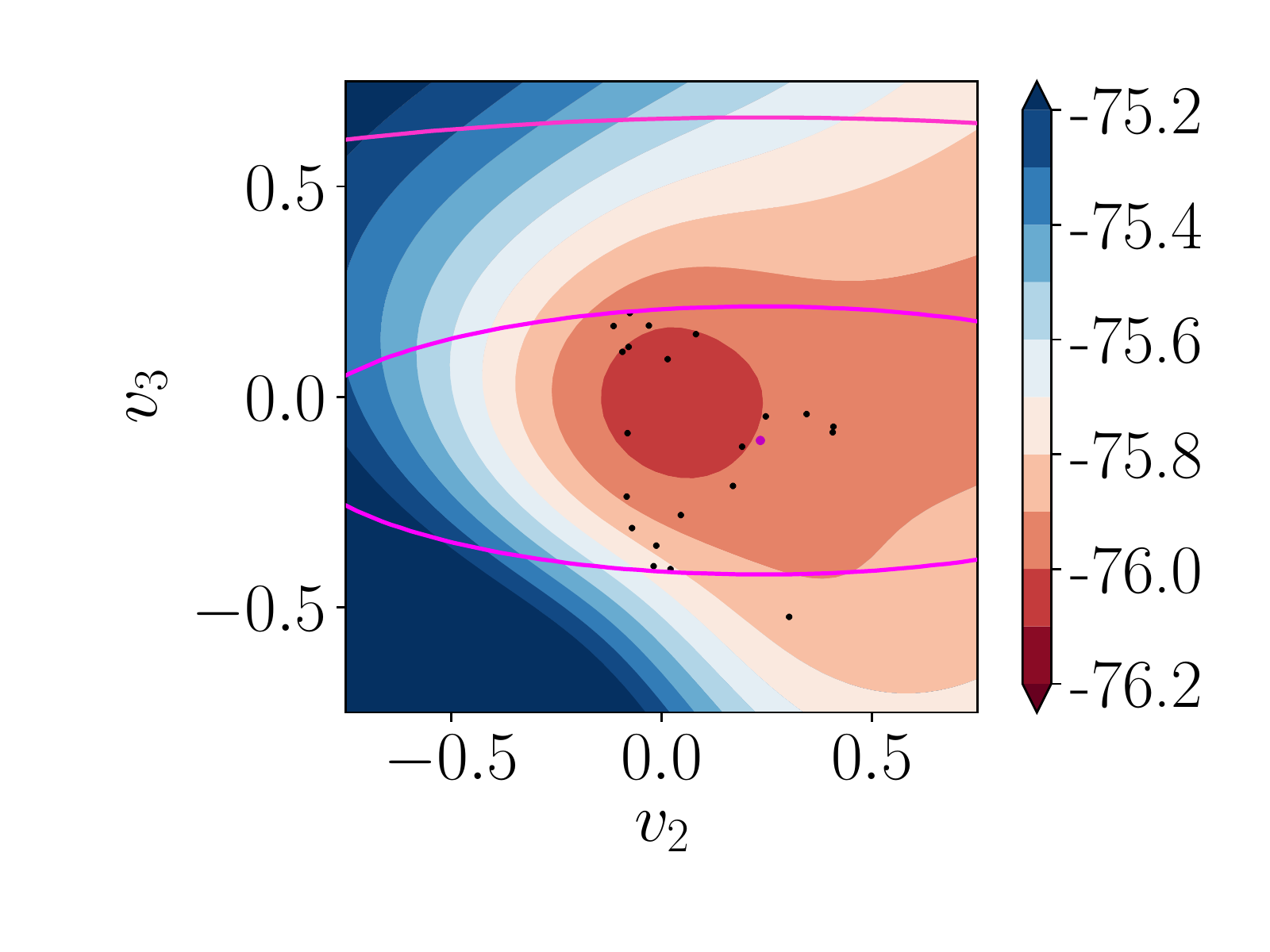}};
	\node[inner sep=0pt] (graph2) at (9,-2) {\includegraphics[width=0.22\textwidth]{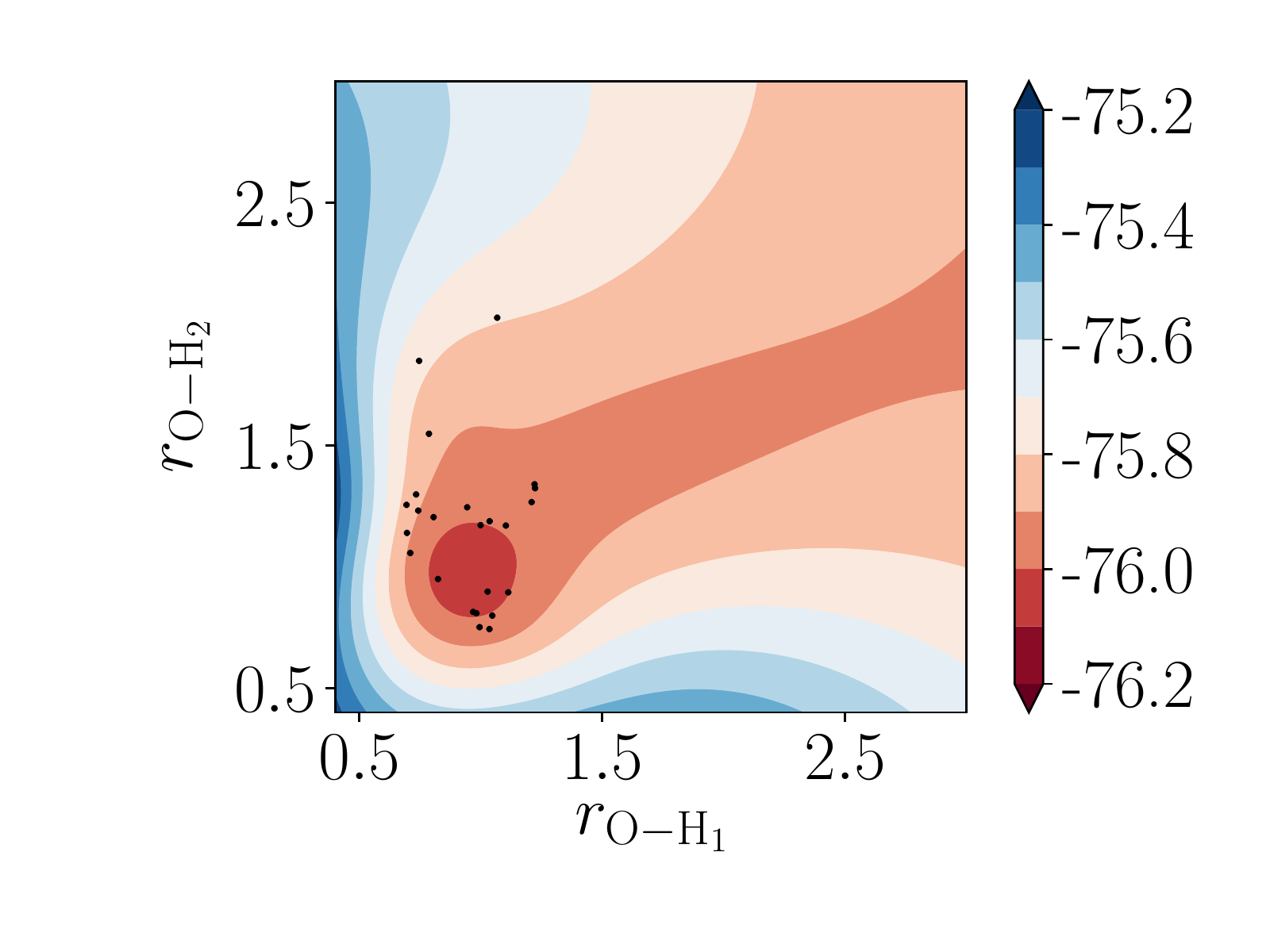}};
	\end{tikzpicture}
	\vspace{0.4cm}
	\caption{Latent functions trained on normal mode coefficients.  Since unlike bond lengths,  MID and FI feature space all share 2 dimensions which can be projected on one another each normal mode has contribution from all degrees of freedom,  one cannot associate the plot along $v_1/v_2$ to a plot along $r$\textsubscript{O-H$_1$}/$r$\textsubscript{O-H$_1$}.  For this reason it is hard to truly understand how the normal mode feature space affects the latent function at a long range where it seems to lack accuracy. }
	\label{fig:nm-pes}
\end{figure}

The NM trained GP produces a non-symmetrical PES model with a longer range description than ID and MID models,  as seen on figure \ref{fig:nm-pes}. GP trained on normal modes coefficients show latent functions that seem to agree with the general concept of normal modes: each length scale is optimised to values that are proportional to the eigenvalue of each normal mode.  Displacements along $v_2$ produce ``symmetrical'' results in figure \ref{fig:nm-pes}(a) despite the symmetrised training set only ensuring that the $v_2$ feature dimension is symmetrical.  The absence of symmetry along the $v_1$ feature dimension is not relevant as moving along it moves the entire water geometry in a symmetrical manner and thus is, by definition,  symmetrical. 

In terms of accuracy,  the NM projection performs as badly,  if not worse,  to the previous feature spaces for the low energy testing set with a MAE of 12.68 mHa.  For the higher energy Boltzmann distribution,  the NM feature space performs better with an MAE of 33.06 mHa,  showing the longer range description,  but is still inaccurate.  Although it seem counter-intuitive that NMs,  a local descriptor of the molecular geometry,  only improves the MAE of the previous training sets at long range one has to remember that the performance at long range is an extrapolation problem that is mostly affected by the length scale hyperparameters of the Gaussian process which allows those testing points to be predicted ``using'' the training data (a short length scale would prevent that and predict for any extrapolated point to be equal to the mean of the training set).
\par
The Morse transformation of the normal modes (MNM),  see figure \ref{fig:nm},  does not improve the model and still produces large MAEs of 5.5/20.8 mHa on the testing sets.  The GP latent function trained on the MNM feature space are not visually different from the ones shown in figure \ref{fig:nm-pes} and are not represented here.

\subsection{Non-local normal modes}
Normal modes are by definition a \textit{local} simplification of the PES: the Taylor expansion truncated at the second order becomes less accurate as we move away from equilibrium.  In terms of feature space design,  it is not important that the true surface deviates from the harmonic approximation since we do not simply fit a quadratic curve but we do learn the true value of the PES.  It is however important,  in keeping with the idea of learning simple surface,  to understand that normal modes being local do provide a PES that can be rather complex when projected onto the normal modes away from equilibrium.  One can define the latter as ``local'',  although this is a redundant label given their definition from equation \ref{eq:proj-nm},  and one can wonder if there was a definition of ``general'' normal modes that would require to specify that these were indeed the sub-family of local normal modes.
\par
For the water molecule we consider,  the breakdown of the simplicity of the PES along the normal modes can be seen with the bending mode of the water molecule.  As the angle changes,  moving along the symmetrical stretching of the water molecule become less and less quadratic and the multidimensional surface along those stretching is non-trivial.  Taking this further one could imagine a molecule that has a low barrier between two equilibria for which one wants to build a PES: which normal modes are more relevant for the overall PES ? It is hard to argue in general that both local normal modes of the equilibria are equivalent.  Moreover,  how relevant are the normal modes of the equilibria around the TS? We try to address this by considering a feature space that uses a general expression of normal modes,  generated on the fly,  on which to project the training data. 
\par
Since the position of a data point is expressed by a projection which itself depends on the position,  $\Delta \mathbf{q}$ which we will assume is projected onto a one dimensional variable $x$,  a general normal mode projection would modify equation \ref{eq:proj-nm} as
\begin{equation}
\begin{aligned}
&\mathbf{v}      = \mathbf{U}^{-1} \  \mathbf{D}^{\dagger} \ \mathbf{M} \Delta \mathbf{q} \\
&\mathbf{w} = \mathbf{U}^{-1}(x) \  \mathbf{D}^{\dagger}(x) \ \mathbf{M} \Delta \mathbf{q}
\end{aligned}
\label{eq:eq23}
\end{equation}
where,  to reiterate,  $\Delta \mathbf{q}$ is the 3$N$ vector of the Cartesian displacement from the equilibrium\footnote{In the case of multiple equilibria we assume that the functions should project onto the same curvilinear space and that it would not matter which equilibria is taken as the one to build the $\Delta \mathbf{q}$.}.  In the definition of equation \ref{eq:eq23},  the projection matrices $\mathbf{U}^{-1}$ and $\mathbf{D}^{\dagger}$ depend on the variable $x$ which is some function of the molecular geometry.
\par
One obvious problem with such a projection is that it is a arduous task to build a projection back to the Cartesian space (as it is the case for FI projections).  This would only affect some applications and creating a feature space onto which a GP can learn is still possible.
\par
These new normal modes effectively remove a DOF since the bending normal mode,  $\mathbf{w}_1$, will always project onto zero as it is ``centred'' at all times,  \textit{i.e.} the amount by which it moves from the original geometry,  is completely absorbed into the $x$ variable.  This is easily solved by swapping the fixed bending mode coefficient by the coordinate and use the $[ x, w_2,  w_3 ]$,   where $w_i$ is the coefficient of the projection onto the $\mathbf{w}_i$ non-local normal mode,  coordinates for learning.  This can be seen as a general approach to use one DOF for evaluating the remaining $3N-7$ feature dimensions as general normal modes and thus forming a complete $3N-6$ dimensions space on which to learn.

\begin{figure}[H]
\vspace{0.4cm}
\centering
	\begin{tikzpicture}[scale=0.5]
	\node[rotate=90] (a) at (-2,-1.5) {\footnotesize MAE: 8.3/38.0 mHa};
	\node[inner sep=0pt] (graph1) at (2,-2) {\includegraphics[width=0.22\textwidth]{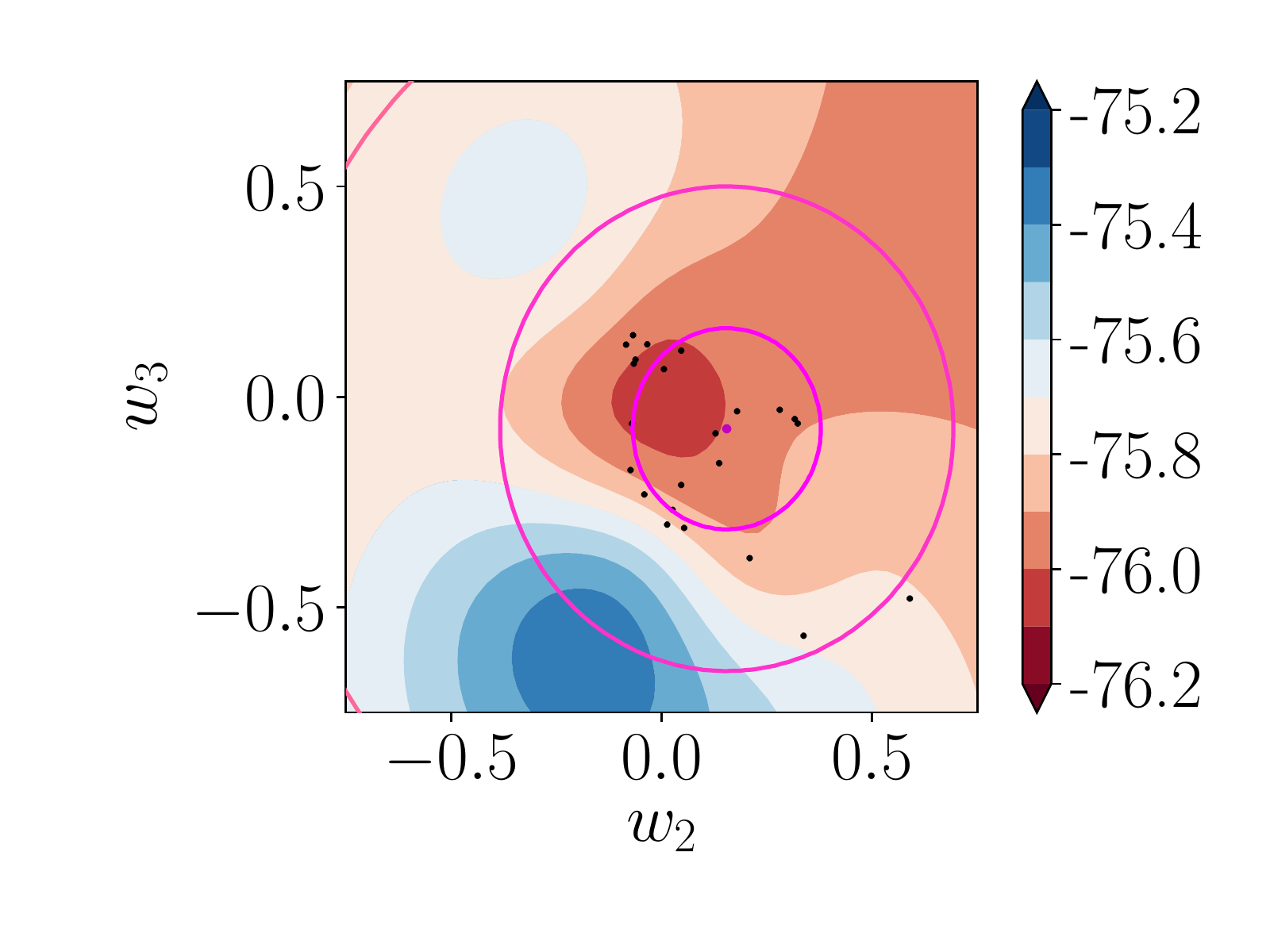}};
	\node[inner sep=0pt] (graph2) at (9,-2) {\includegraphics[width=0.22\textwidth]{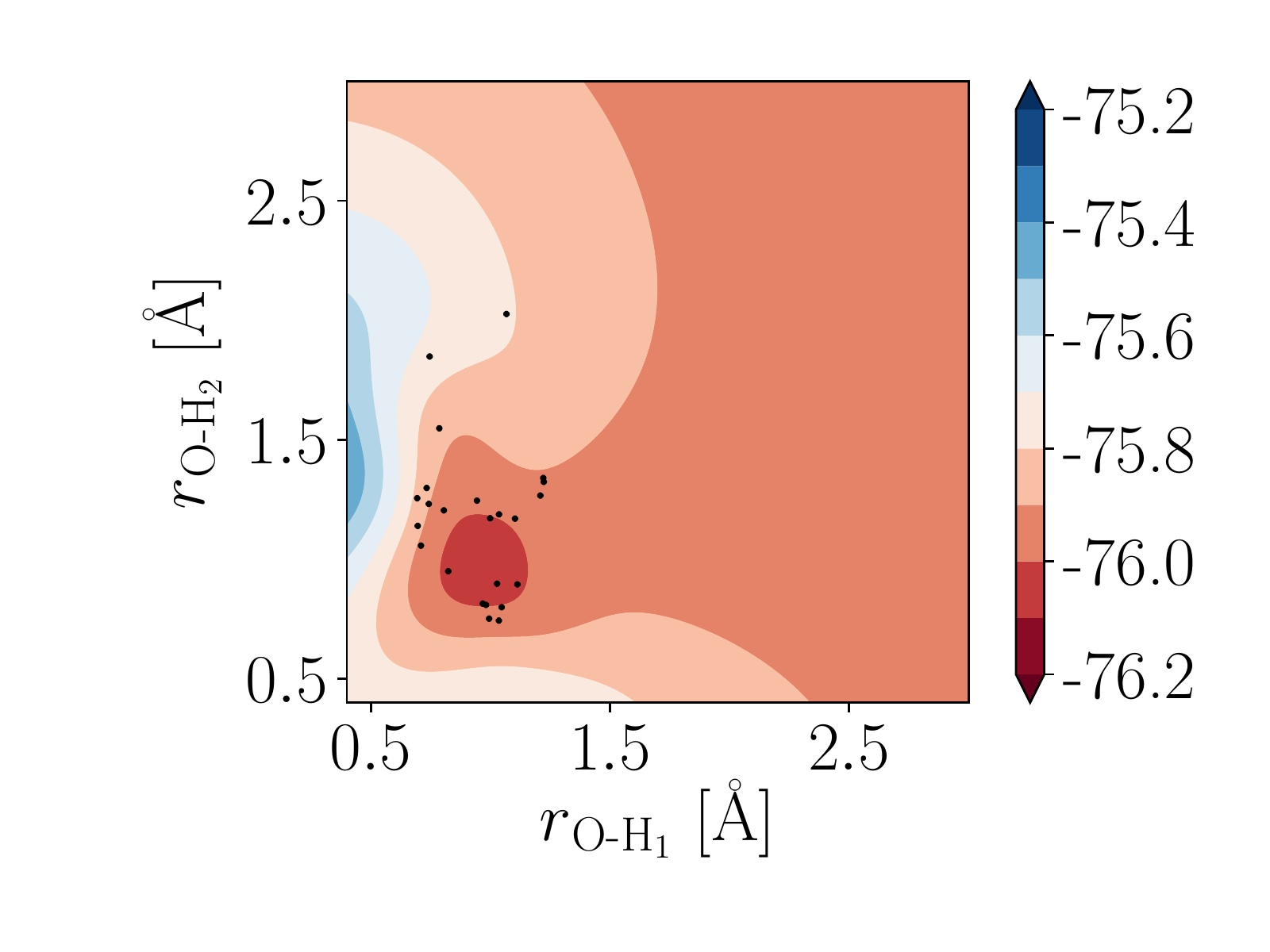}};
	\end{tikzpicture}
	\vspace{0.4cm}
	\caption{Latent functions trained on non-local normal mode coefficients.  The description seems more local than the NM GP whose latent function is shown in figure \ref{fig:nm-pes}. }
	\label{fig:nlnm-pes}
\end{figure}

The NLNM feature space is still struggling with the sparse data and not reproducing an accurate model of the PES.  Both the MAEs,  8.3/38 mHa for the testing sets,  and the latent function of the NLNM GP are very similar to the ID and MID sets.
\par
\begin{figure}[H]
	\centering
	\captionsetup[subfigure]{labelformat=empty}
	\subfloat[NM]{\includegraphics[height=1.6cm, trim=3.7cm 0 3.7cm 0, clip]{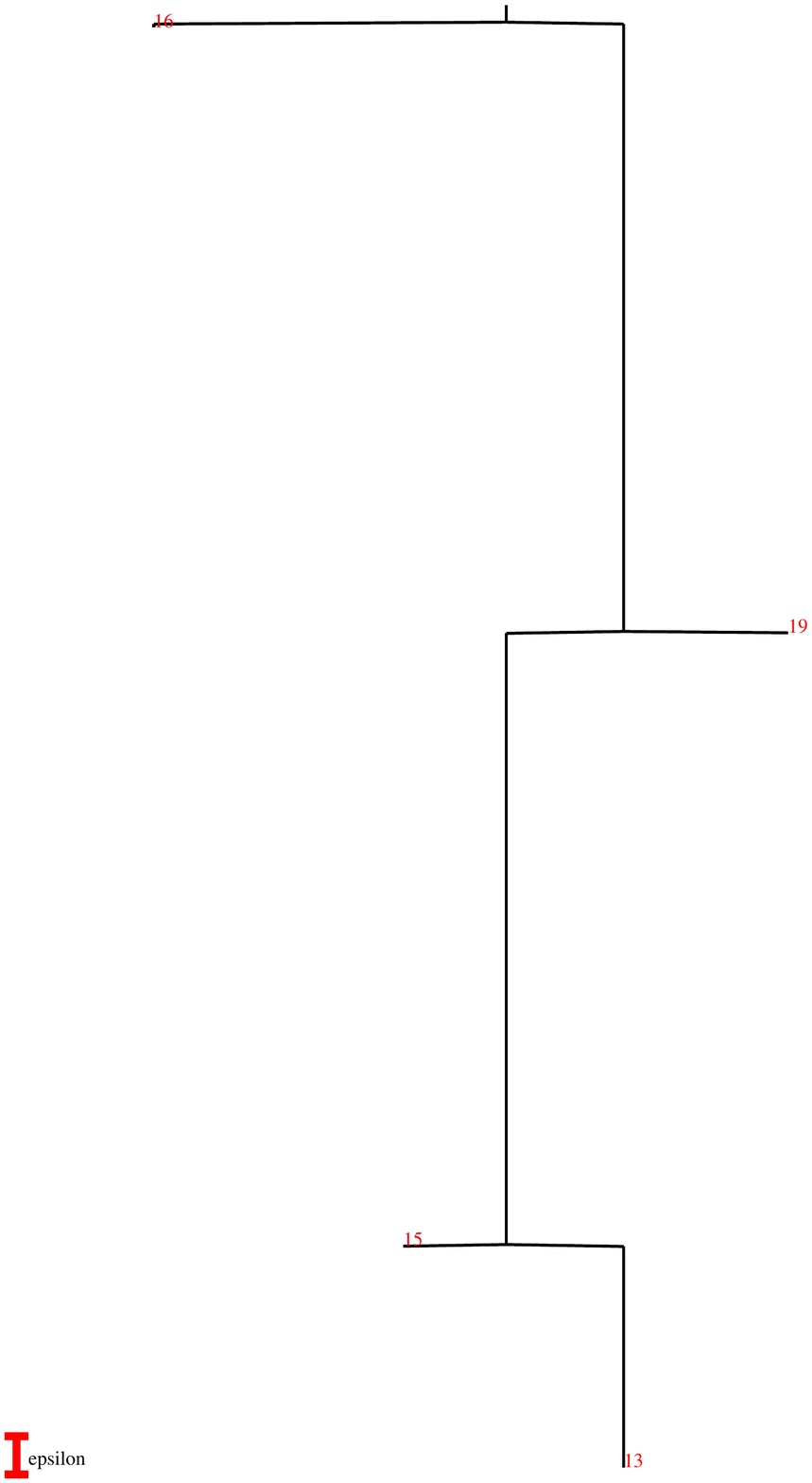}}
	\hspace{2cm}
	\subfloat[MNM]{\includegraphics[height=1.6cm, trim=3.7cm 0 3.7cm 0, clip]{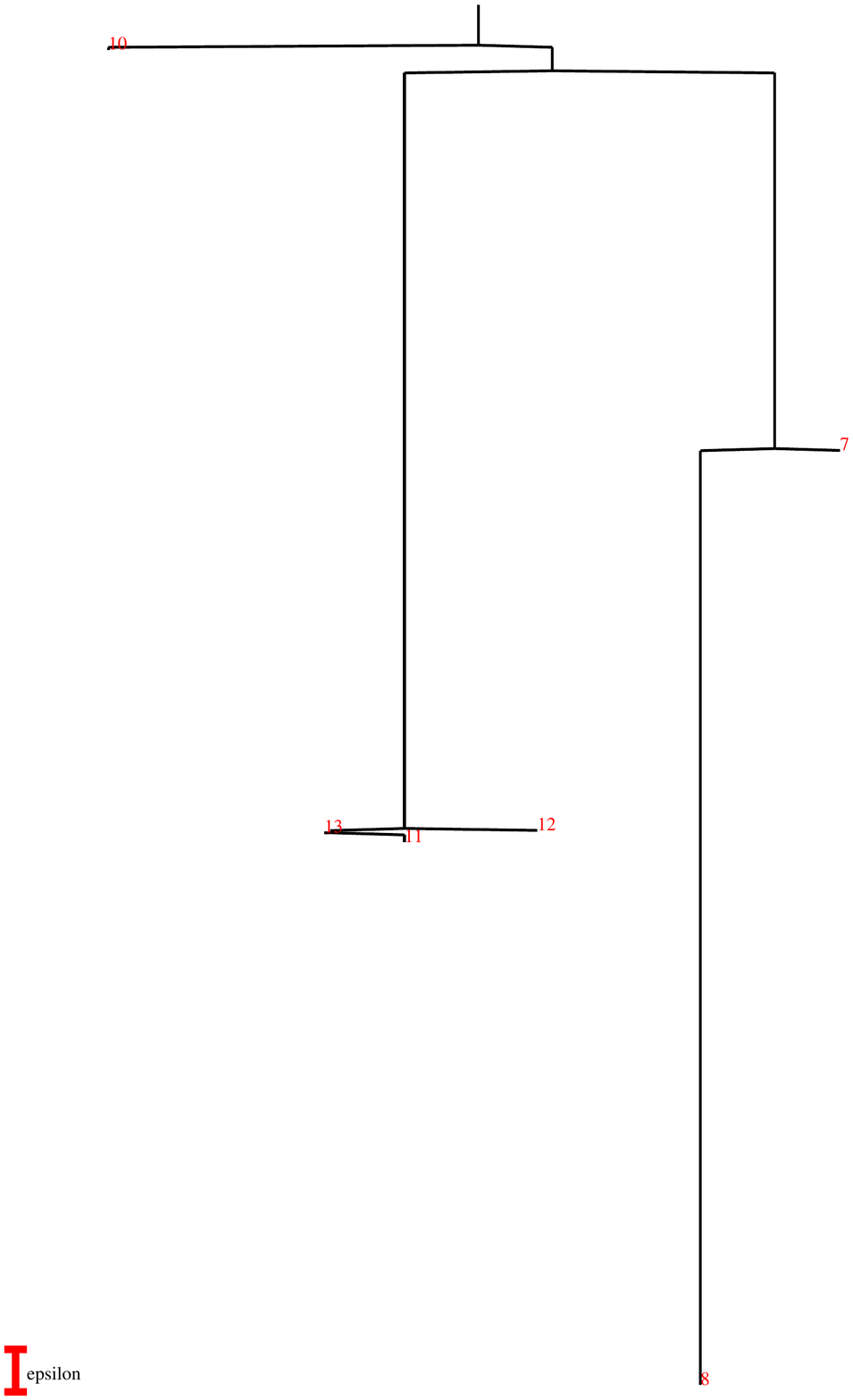}}
	\hspace{2cm}
	\subfloat[NLNM]{\includegraphics[height=1.6cm, trim=3.7cm 0 3.7cm 0, clip]{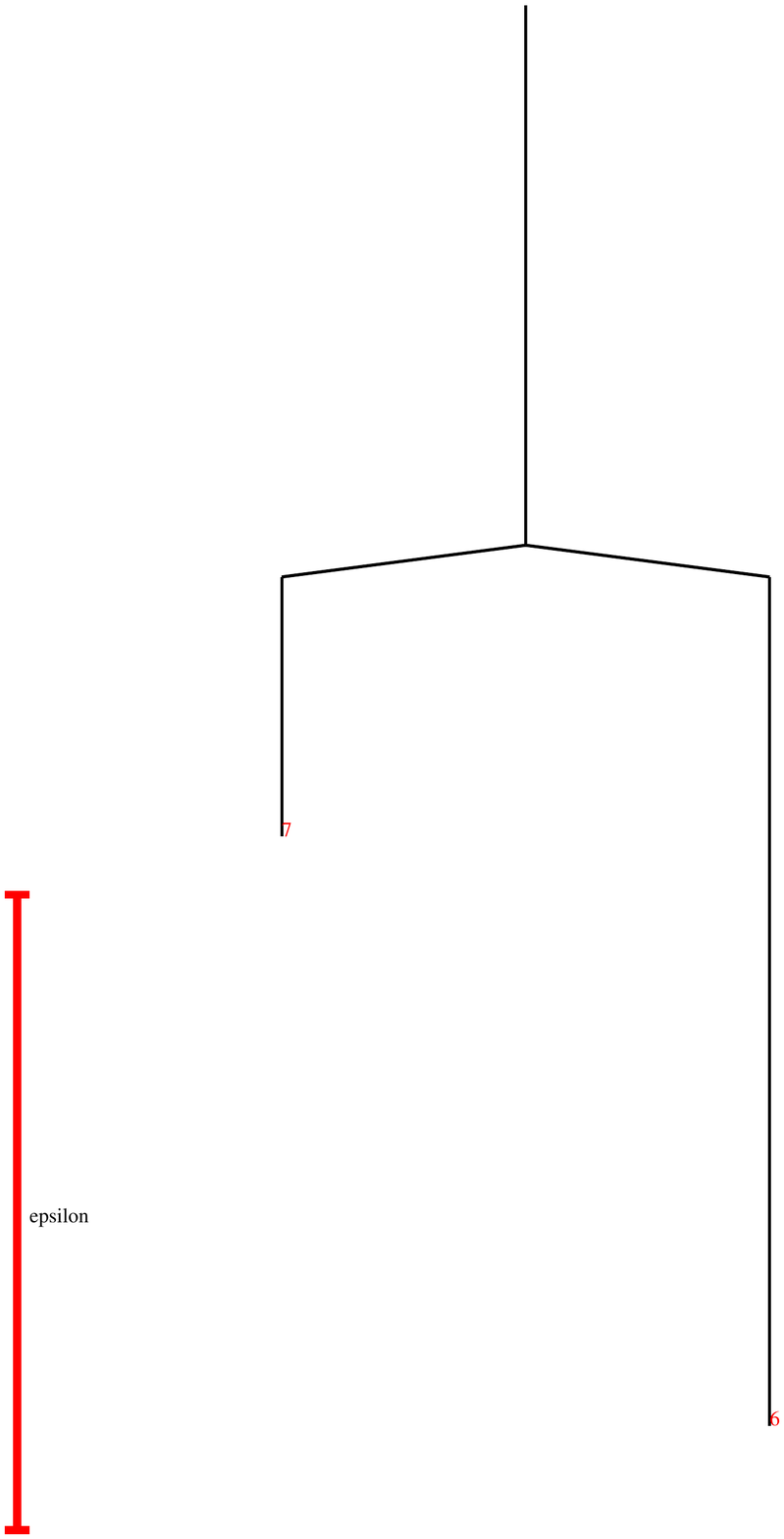}}
	\caption{LML disconnectivity graphs for the original NM feature space,  as well as the Morse transformed and the NLNM feature spaces.}
	\label{fig:nm-lml}
\end{figure}

The LMLs of all three ``flavours'' of NM-based feature space present similarly complex landscapes for NM and NLNM as seen in figure \ref{fig:nm-lml}. On the other hand,  the NLNM landscape is much simpler suggesting that the model is likely to not be stable.  This is surprising since the model found is not accurate and additional data would not improve the most likely model by defining it with better hyperparameters. 

The stability of each LML surface with respect to additional data is not a straightforward matter.  One would expect that LMLs follow the catastrophe theory (where changes to nonlinear systems cause sudden changes to the attracting power of local minima).  With each datum,  local minima merge until a single minimum of the LML is seen.  This does not to be always the case.

\subsection{Symmetrising training data}
We will consider here the MID,  NM and NLNM feature spaces,  which miss the PES symmetry in the dimension.  For a water molecule with two equivalent protons,  new training data can be added by a simple proton permutation which is not equivalent in the feature space but is equivalent in molecular properties\footnote{We will call this process symmetrising the training set.}.  We assess the performance improvement of GP trained on the MID feature space.
\par
\begin{figure}[H]
\vspace{0.4cm}
\centering
	\begin{tikzpicture}[scale=0.5]
	\node[inner sep=0pt,label=above:{\small MAE: 7.2/37.2 mHa}] (graph1) at (0,-2.2) {\includegraphics[width=0.22\textwidth]{midcont_n0}};
	\node[inner sep=0pt,label=above:{\small MAE: 1.6/3.2 mHa}] (graph2) at (8,-2.2) {\includegraphics[width=0.22\textwidth]{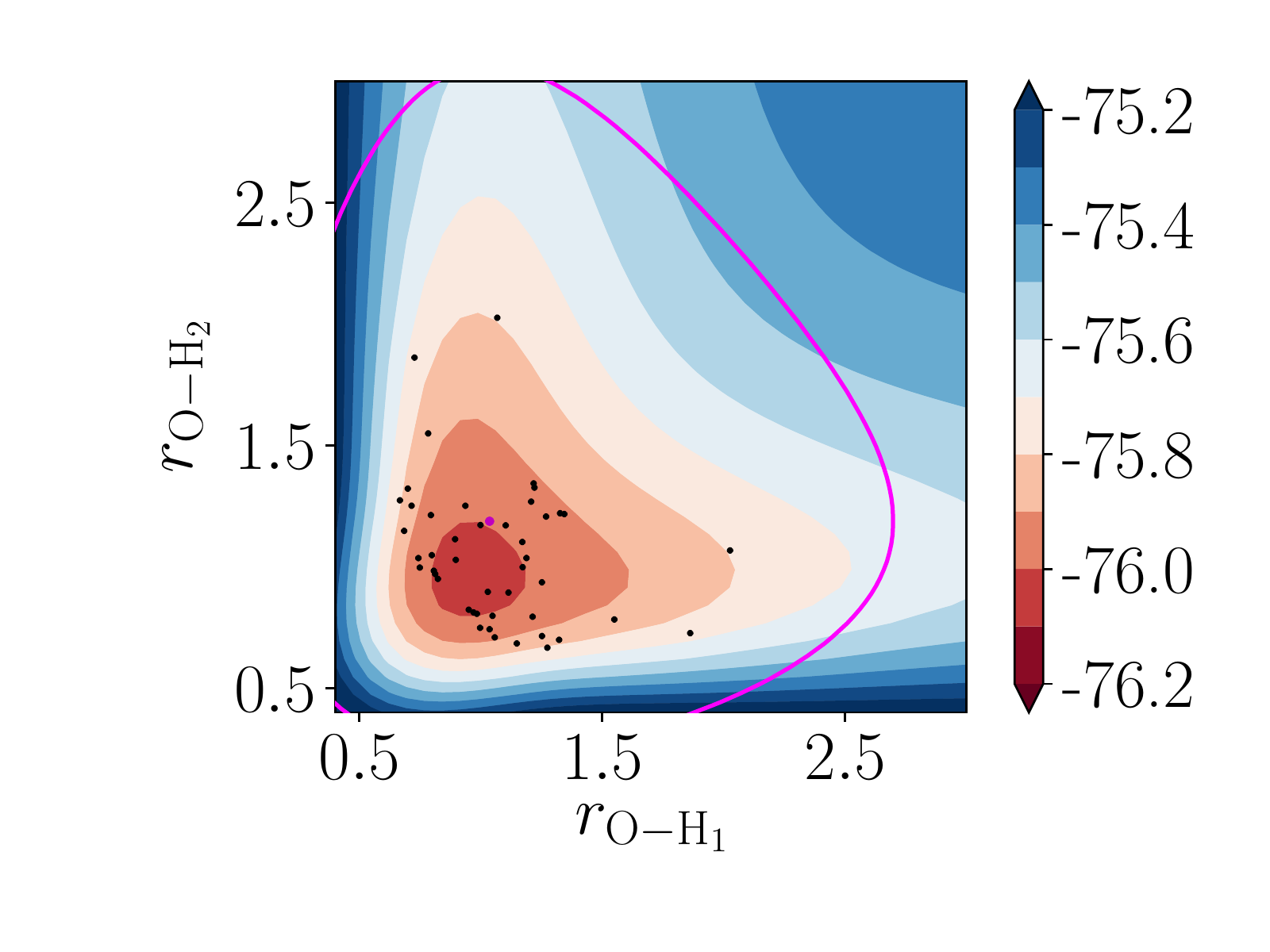}};
	\end{tikzpicture}
	\vspace{0.4cm}
	\caption[Predicted PES for the MID feature spaces with and without the symmetrised training data.]{Projected latent functions along the two O-H stretches (these do not correspond directly to the used feature dimensions) for the original training set and the symmetrised training sets.  The respective MAE on the two Boltzmann-weighted testing sets are of 7.2 / 37.2 mHa and 1.58 / 3.16 mHa respectively which shows a great improvement of the performance of the Gaussian process. Moreover,  the latent function describes a symmetrical PES from the non-symmetrical feature space.  The magenta lines are isovalue contours of the kernel function.}
	\label{fig:latentfunctions-h2o}
\end{figure}

We write the original training set as $\chi = \{ \mathbf{X}_{\mathrm{MID}}, \mathbf{y}\}$ and the symmetrised training set as $\chi = \{ \mathbf{X}_{\mathrm{MID}} \cup \mathbf{X}_{\mathrm{MID}}^{\mathrm{symm}}, \mathbf{y} \cup \mathbf{y}\}$\footnote{Technically this notation is not correct as the sets are ordered.  We use this symbol to mean that the new matrix of $N$\textsubscript{}$\times N$\textsubscript{dimensions} is given by
\begin{equation}
\mathbf{X}_{\mathrm{MID}} \cup \mathbf{X}_{\mathrm{MID}}^{\mathrm{symm}} =
\begin{bmatrix*}
\mathbf{X}_{\mathrm{MID}} \\[0.4cm]
\mathbf{X}_{\mathrm{MID}}^{\mathrm{symm}}
\end{bmatrix*} \qquad \mathrm{and} \qquad \mathbf{y}=
\begin{bmatrix*}
\mathbf{y} \\[0.4cm]
\mathbf{y}
\end{bmatrix*}
\end{equation}
},  which we will call symmMID.  The GP trained with the symmMID set shows,  in figure \ref{fig:latentfunctions-h2o},  the symmetry appearing in the latent function. 

The much longer length scale of the optimised GP kernel (as discussed in figure \ref{fig:latentfunctions-h2o}) improves the model substantially.  One could assume,  that it is simply the case of increasing the training set size that allows the GP trained with symmMID to improve on the MID learning.  We thus assess the effect of the size of the training set on the GP by the symmetrised data points sequentially.

\subsubsection{Fully symmetrised training data}
Before considering the progression,  we will shortly discuss the GP that produces the symmetrical PES seen in figure \ref{fig:latentfunctions-h2o}(b).  Although it might seem obvious at first that a dataset containing all data points related by symmetry creates symmetry in the resulting function,  it is not obvious that a GP without explicit symmetry will optimise to models that produce symmetrical latent functions.  However,  in terms of the LML surface the effect of symmetrising the training set allows one to rewrite,  due to the kernel function properties,  the covariance of the training set to itself is as a block matrix of the form
\begin{equation}
\begin{aligned}
&\mathrm{K}(\mathbf{X} \cup \mathbf{X}^{\mathrm{symm}}, \mathbf{X} \cup \mathbf{X}^{\mathrm{symm}}) =\\
&\begin{bmatrix*}[r]
\mathrm{K}(\mathbf{X},  \mathbf{X})                                 & \mathrm{K}(\mathbf{X},  \mathbf{X}^{\mathrm{symm}})\\
\mathrm{K}(\mathbf{X},  \mathbf{X}^{\mathrm{symm}}) & \mathrm{K}(\mathbf{X}^{\mathrm{symm}},  \mathbf{X}^{\mathrm{symm}})
\end{bmatrix*}
\end{aligned}
\end{equation}
For general centro-symmetric molecules AB$_n$,  the consequence on the LML is that the gradient along the 3 length scales of the MID feature space  is of the form $\partial_{\boldsymbol{\rho}} \mathrm{LML} = [g_0, g_0, g_1]$.  The LML is also convex along the $\rho_0=\rho_1$ direction ensuring that all minima have equal $\rho_0$ and $\rho_1$ hyperparameters.  In addition,  with the fact that,  along the two equivalent feature dimensions,  each sample in the MID feature space $[\tilde{\mathrm{X}}_0, \tilde{\mathrm{X}}_1]$ also has a sample with $[\tilde{\mathrm{X}}_1, \tilde{\mathrm{X}}_0]$,  this implies that GP latent functions are indeed symmetrical as we see in figure \ref{fig:latentfunctions-h2o}.

\subsubsection{Partially symmetrised training data}
\begin{figure}[H]
	\centering
	\subfloat[MAEs]{\includegraphics[width=0.2\textwidth]{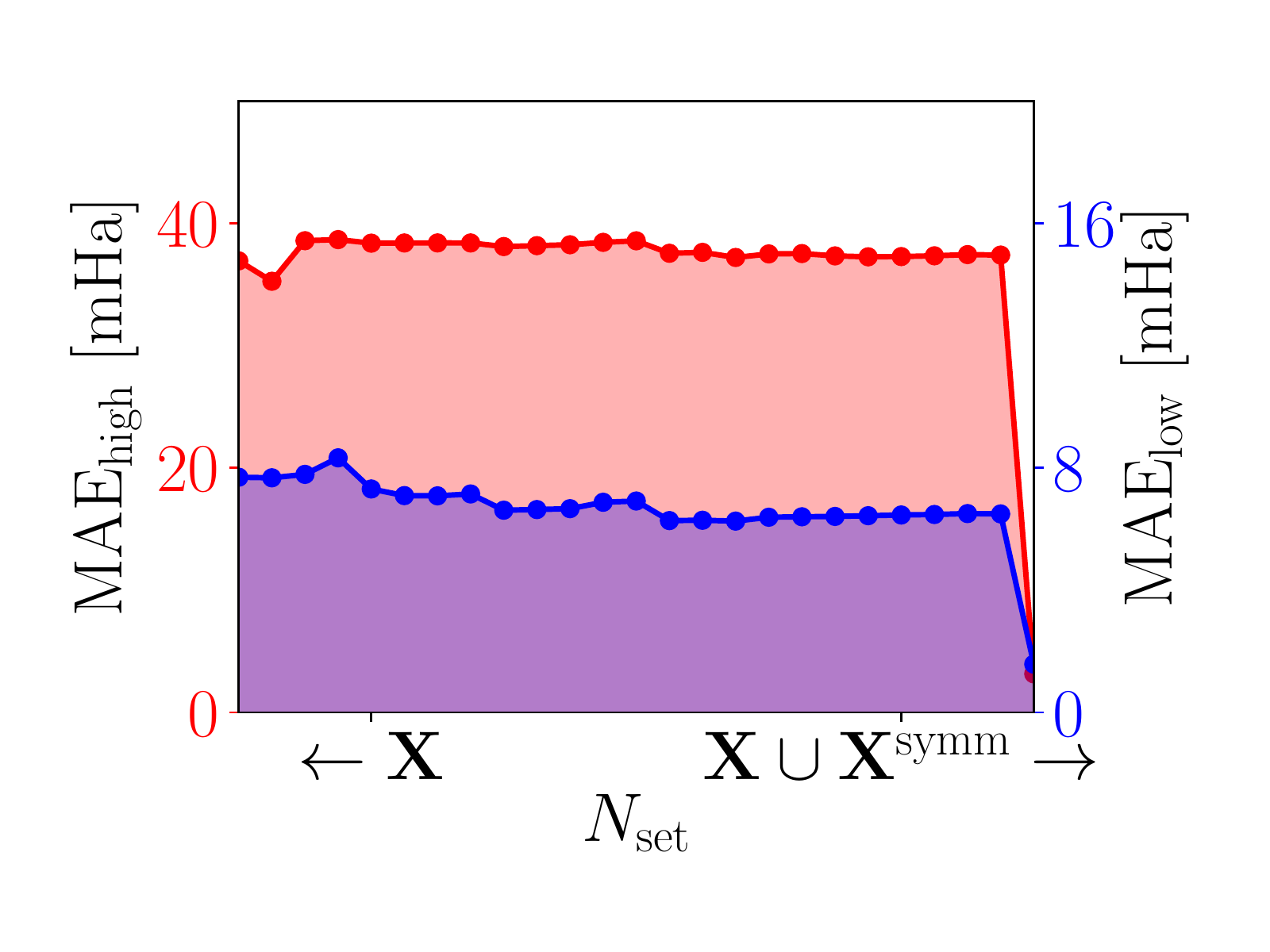}}
	\subfloat[Optimised $\rho_0/\rho_1$]{\includegraphics[width=0.2\textwidth]{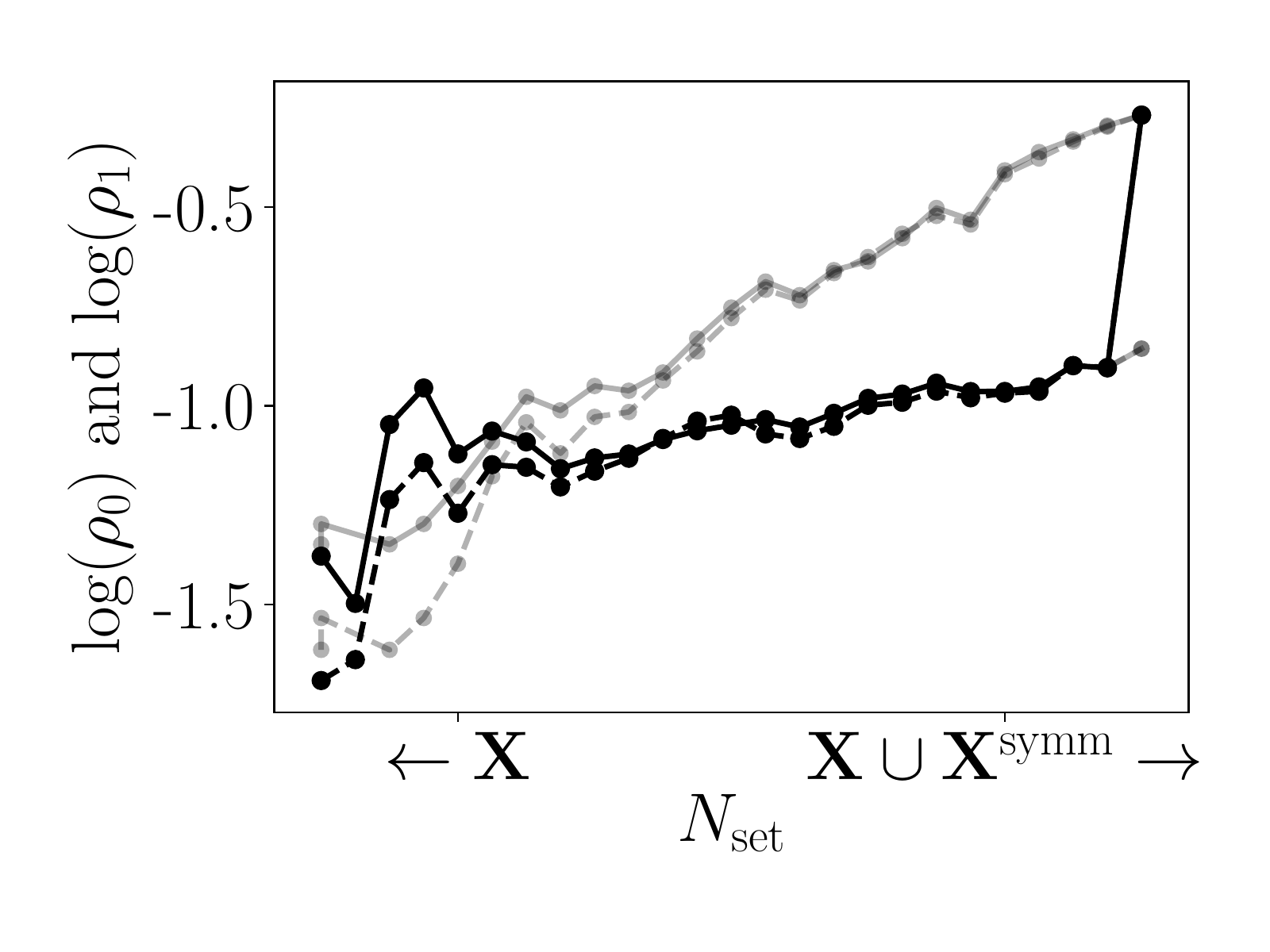}}
	\hfill
	\caption{Evolution of the performance and the hyperparameters of Gaussian process model as data points are added to the training set and the MID set tends to the symmMID set.  One can see that the improvement on the MAE is mostly slow (barely seen on the scale of the sharp drop at symmMID) but then sharply drops from > 5 mHa to below chemical accuracy for the low energy Boltzmann testing set (shown in blue) and from > 35 mHa to < 4 mHa for the high energy Boltzmann testing set (shown in red).  The overall drop in MAE for each set is represented by the double-headed arrows next to their respective axes.  In the second panel,  one can see that the sharp drop in MAEs are accompanied and explained by a sharp increase in the length scale of the best model selected by the Gaussian process.}
	\label{fig:symmetry-slow}
\end{figure}

\begin{figure}[H]
	\centering
	\captionsetup[subfigure]{labelformat=empty}
	\subfloat[24]{\includegraphics[width=0.05\textwidth, trim=3.7cm 0 3.7cm 0, clip]{symm_dg_n0}}
	\hfill
	\subfloat[25]{\includegraphics[width=0.05\textwidth, trim=3.7cm 0 3.7cm 0, clip]{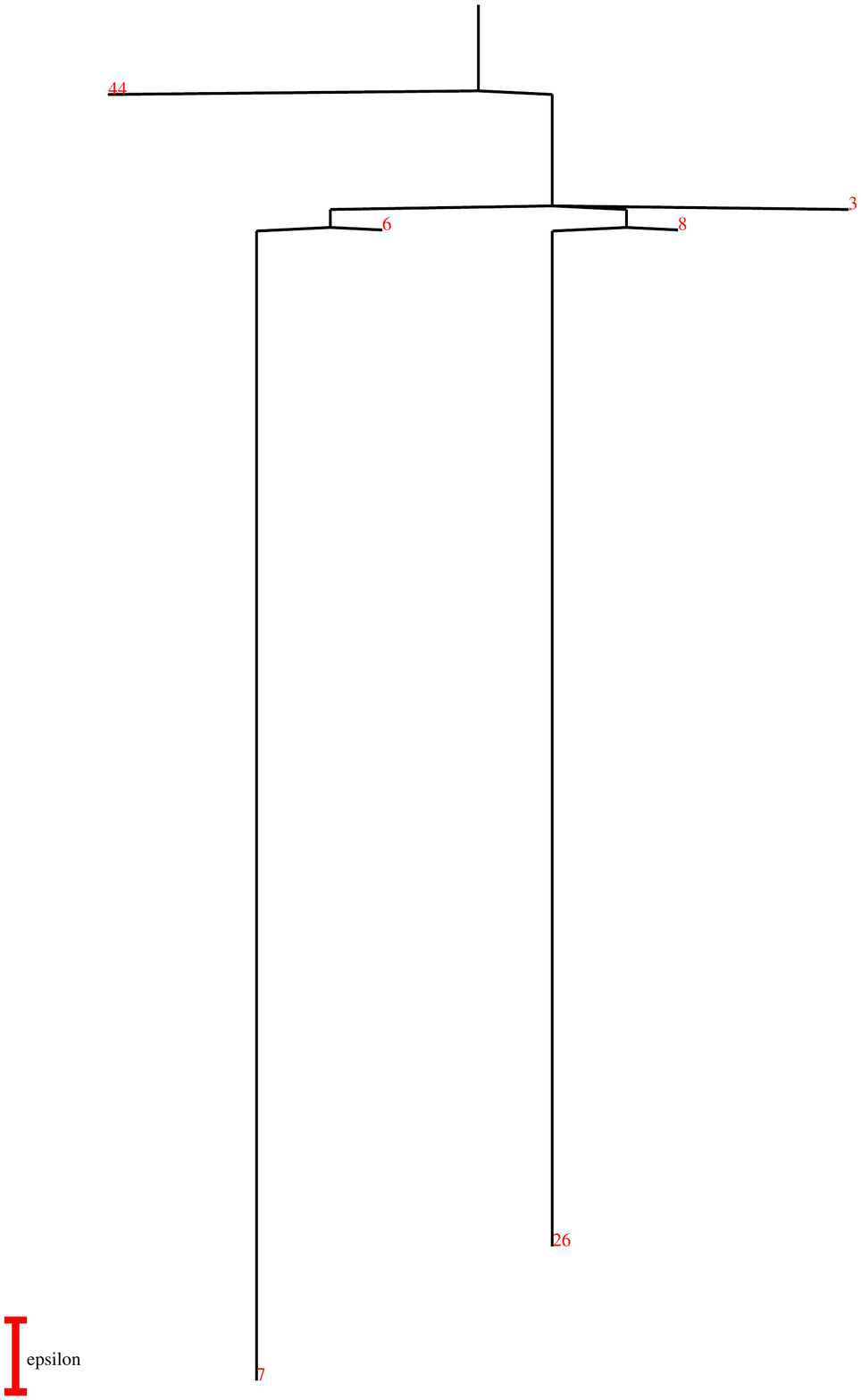}}
	\hfill
	\subfloat[26]{\includegraphics[width=0.05\textwidth, trim=3.7cm 0 3.7cm 0, clip]{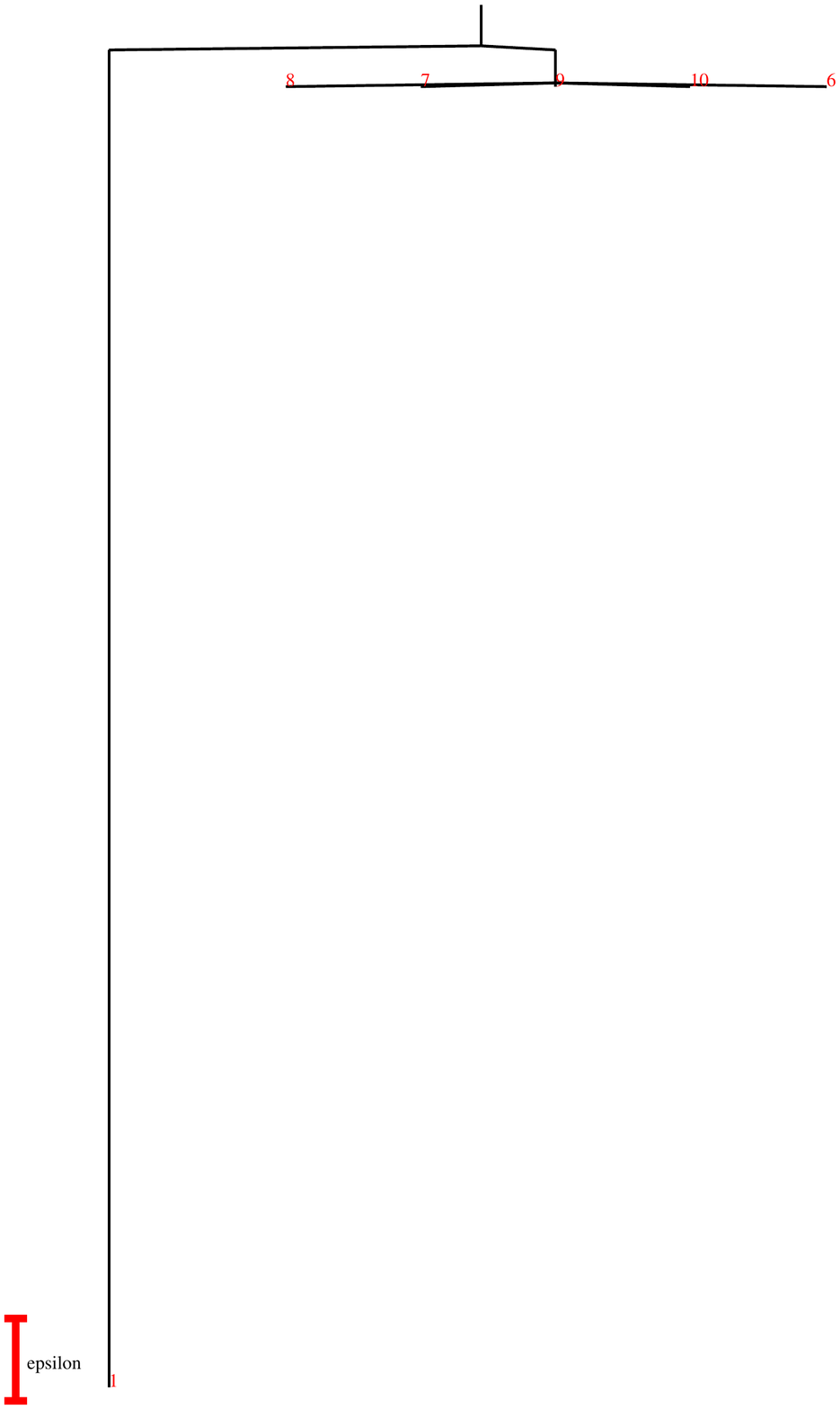}}
	\hfill
	\subfloat[27]{\includegraphics[width=0.05\textwidth, trim=3.7cm 0 3.7cm 0, clip]{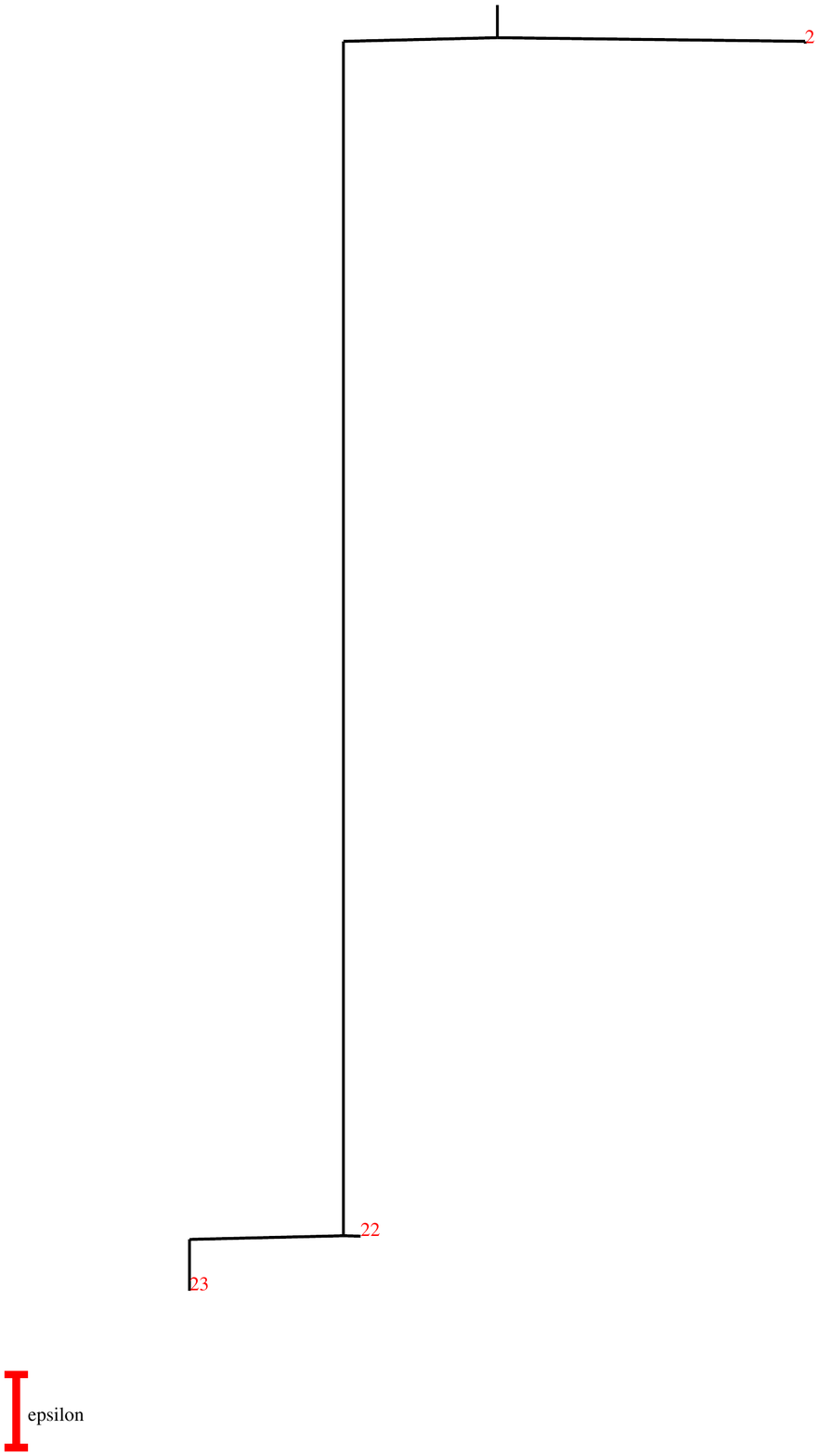}}
	\hfill
	\subfloat[28]{\includegraphics[width=0.05\textwidth, trim=3.7cm 0 3.7cm 0, clip]{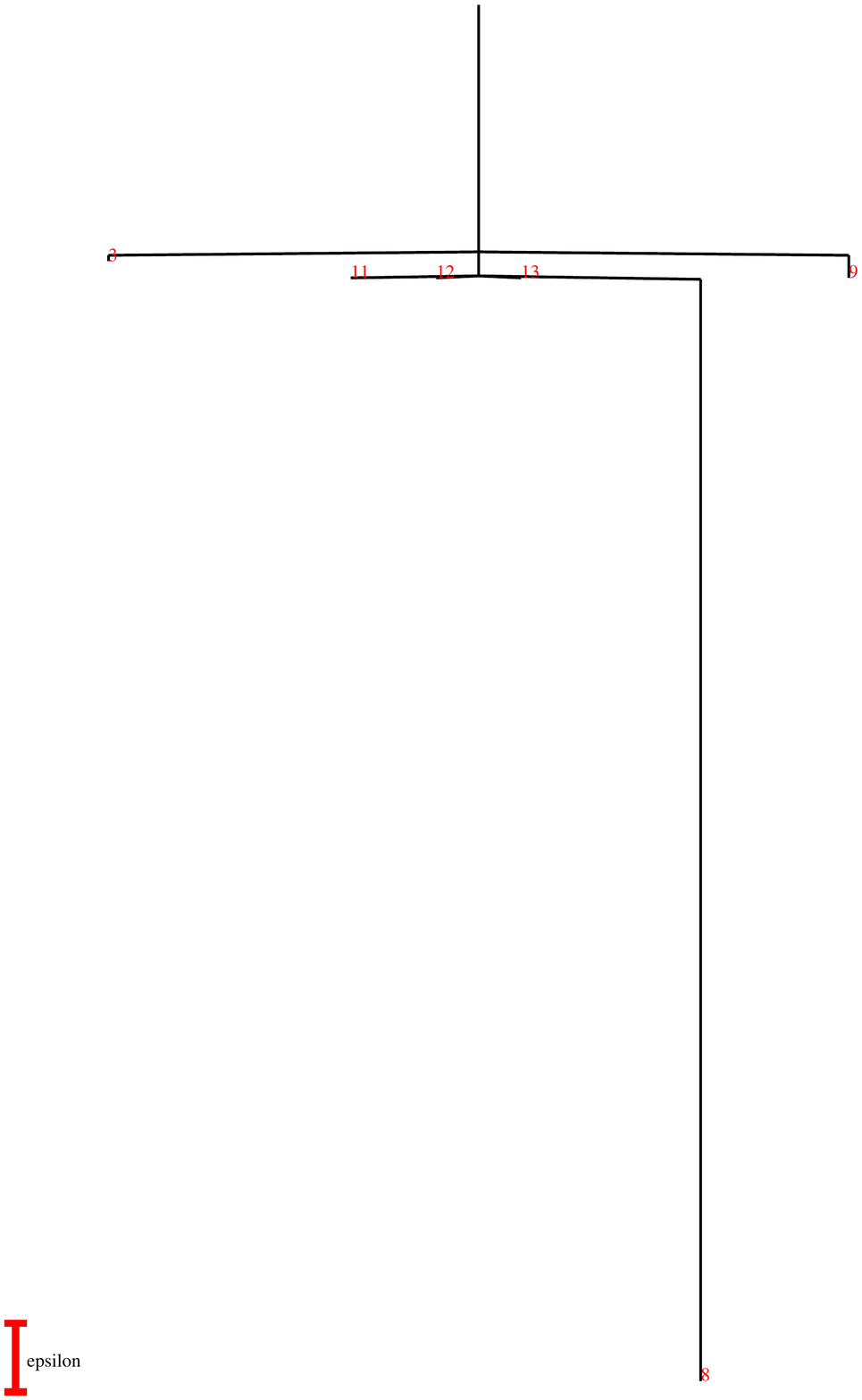}}
	\hfill
	\subfloat[29]{\includegraphics[width=0.05\textwidth, trim=3.7cm 0 3.7cm 0, clip]{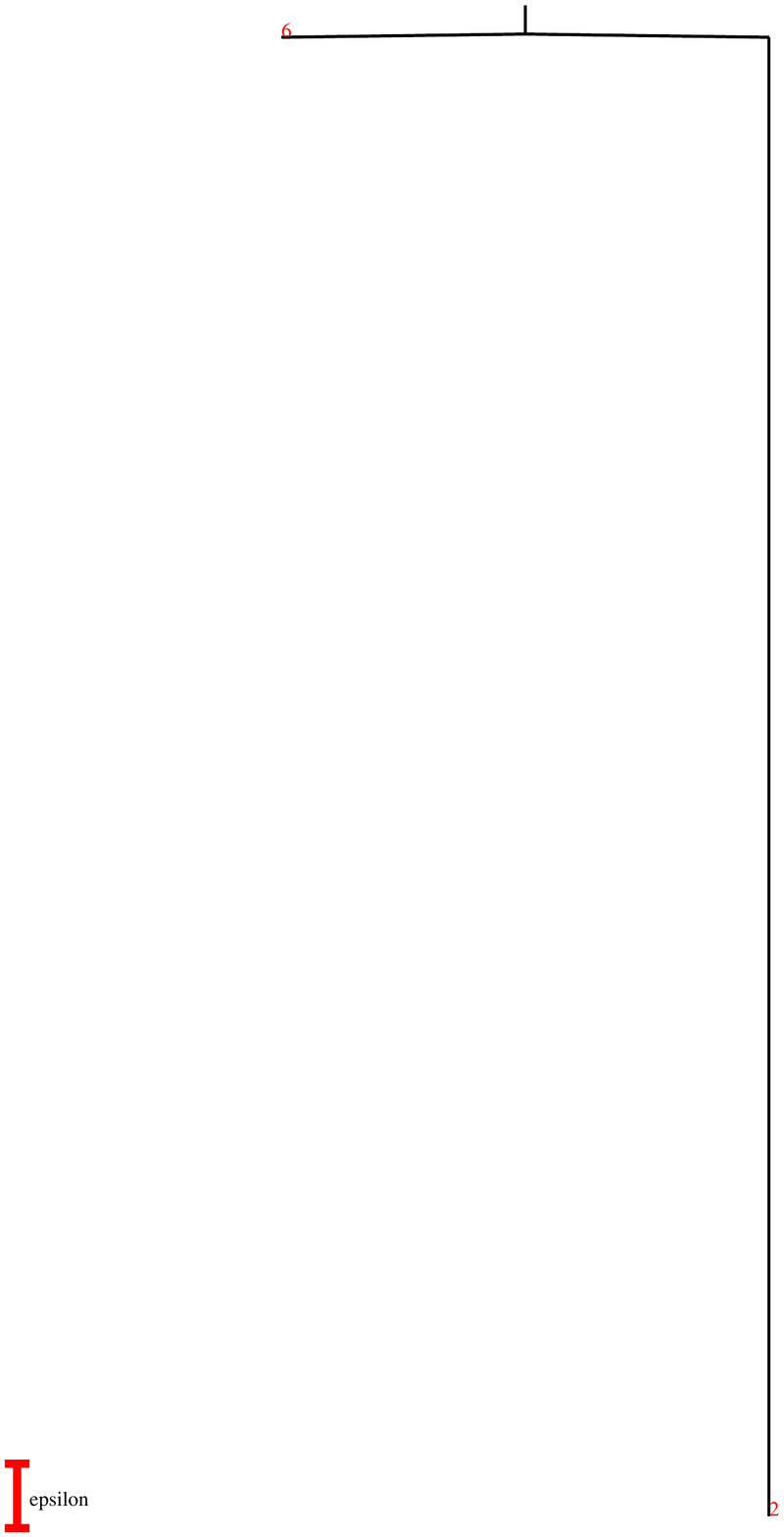}}
	\hfill
	\subfloat[30]{\includegraphics[width=0.05\textwidth, trim=3.7cm 0 3.7cm 0, clip]{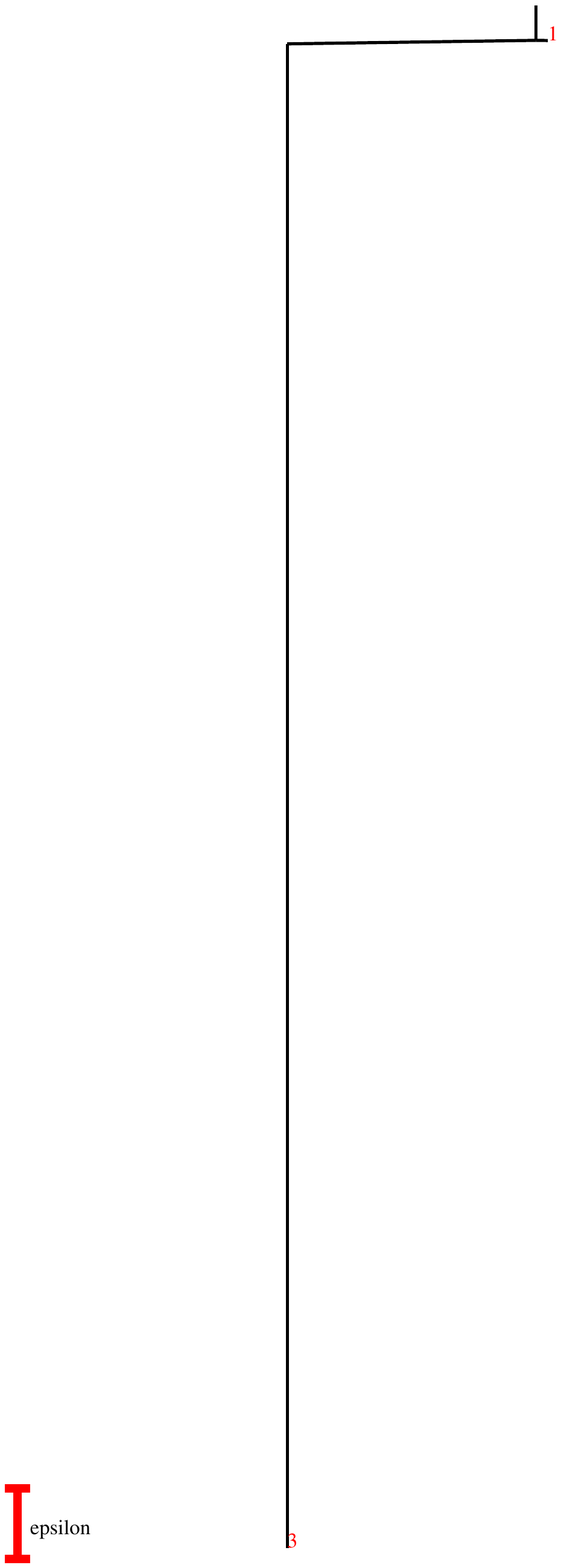}}
	\hfill
	\subfloat[32]{\includegraphics[width=0.05\textwidth, trim=3.7cm 0 3.7cm 0, clip]{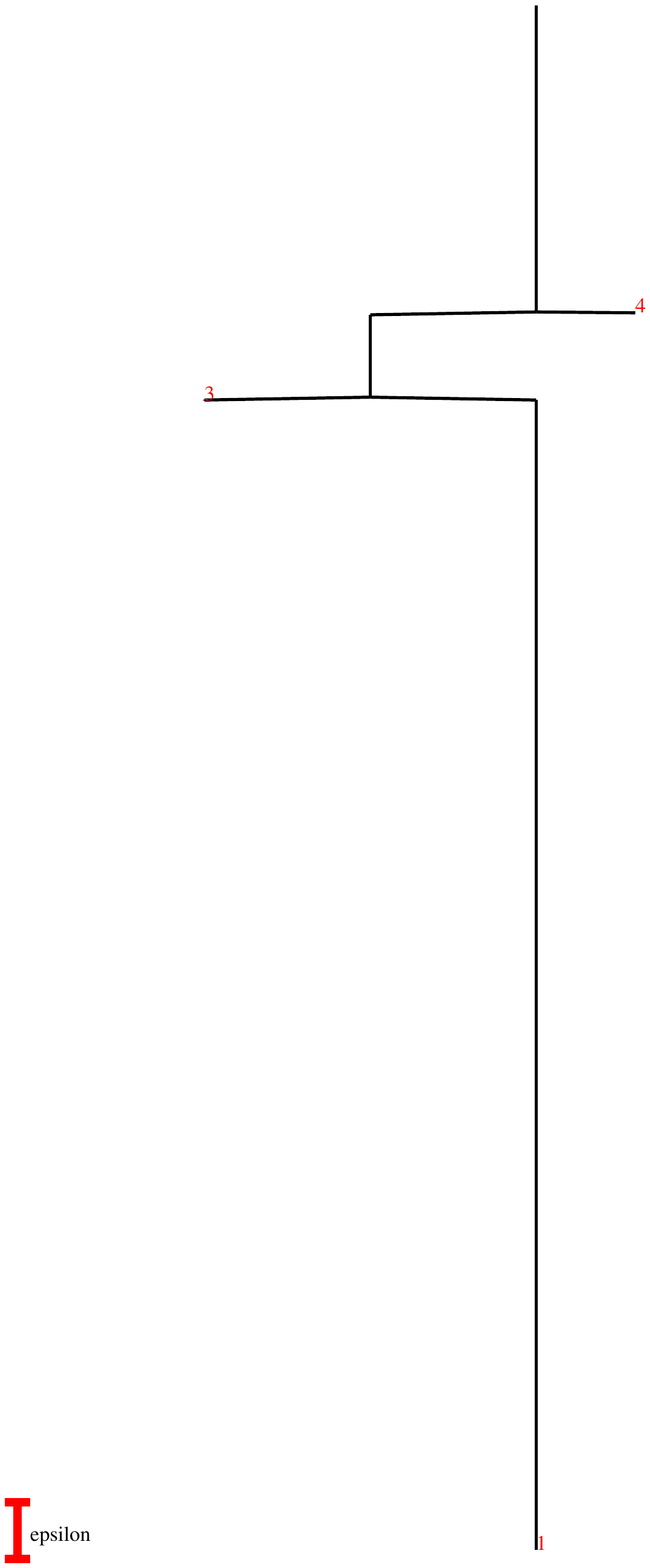}}
	\hfill \par
	\subfloat[33]{\includegraphics[width=0.05\textwidth, trim=3.7cm 0 3.7cm 0, clip]{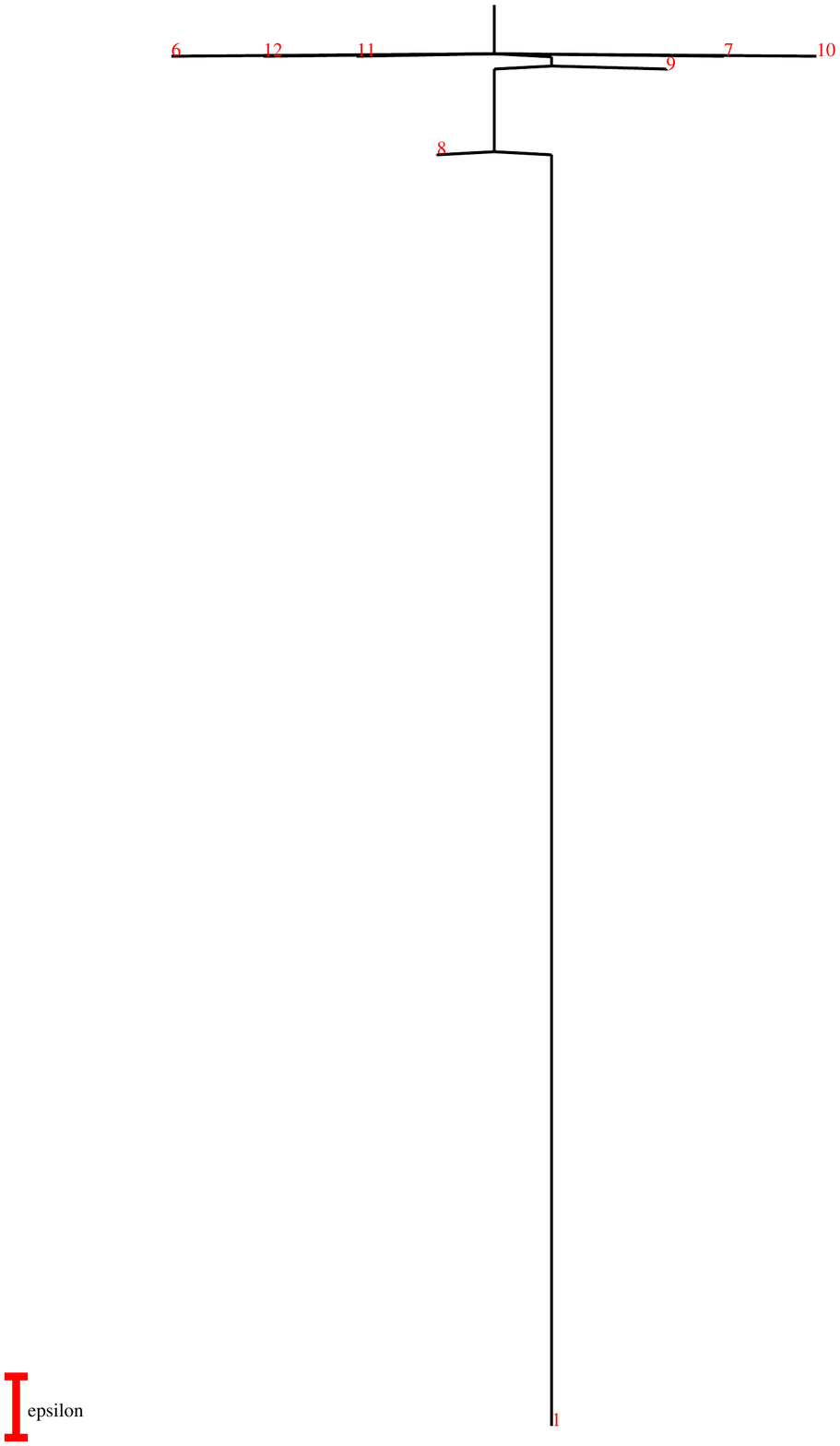}}
	\hfill
	\subfloat[35]{\includegraphics[width=0.05\textwidth, trim=3.7cm 0 3.7cm 0, clip]{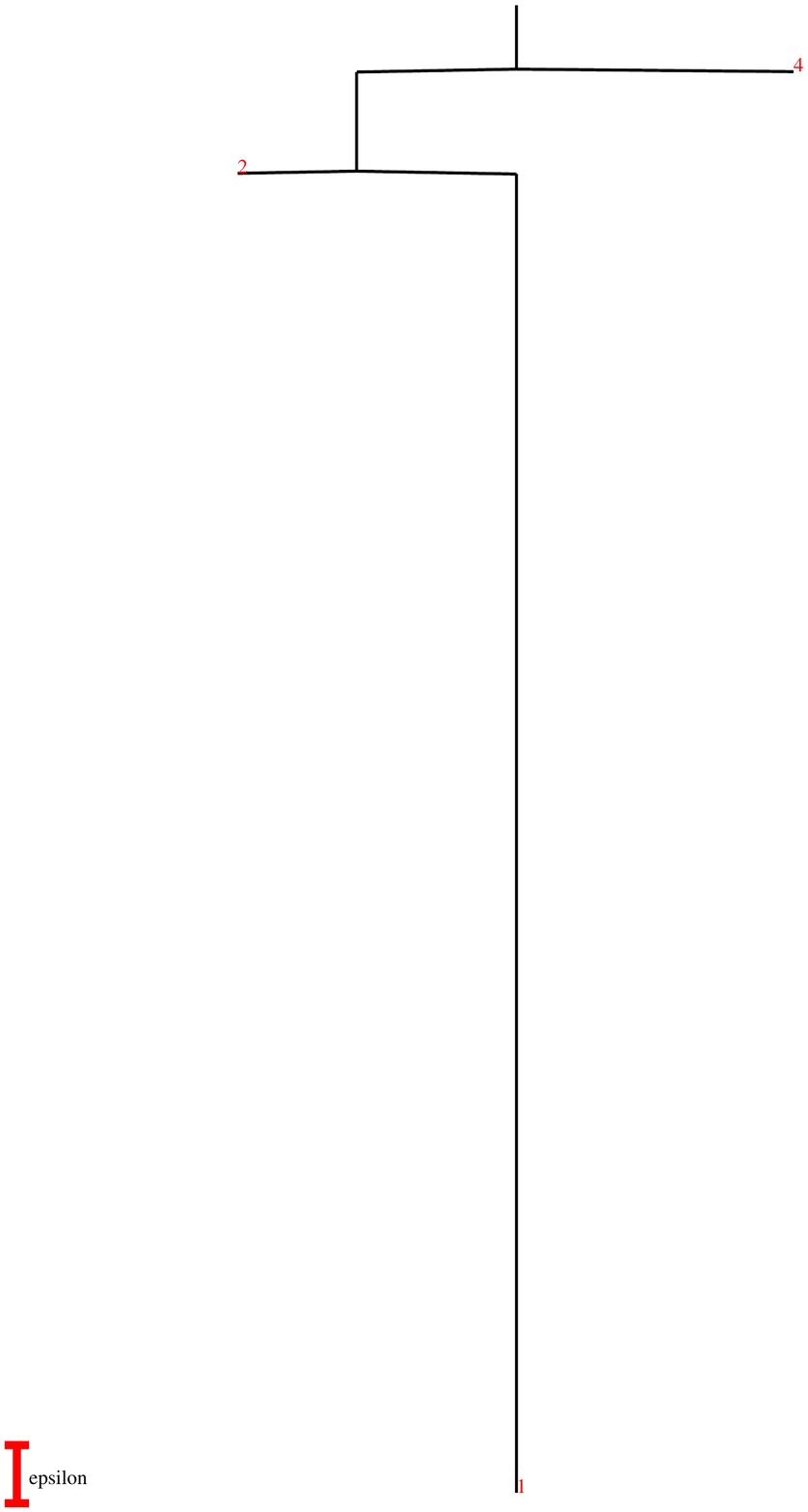}}
	\hfill
	\subfloat[36]{\includegraphics[width=0.05\textwidth, trim=3.7cm 0 3.7cm 0, clip]{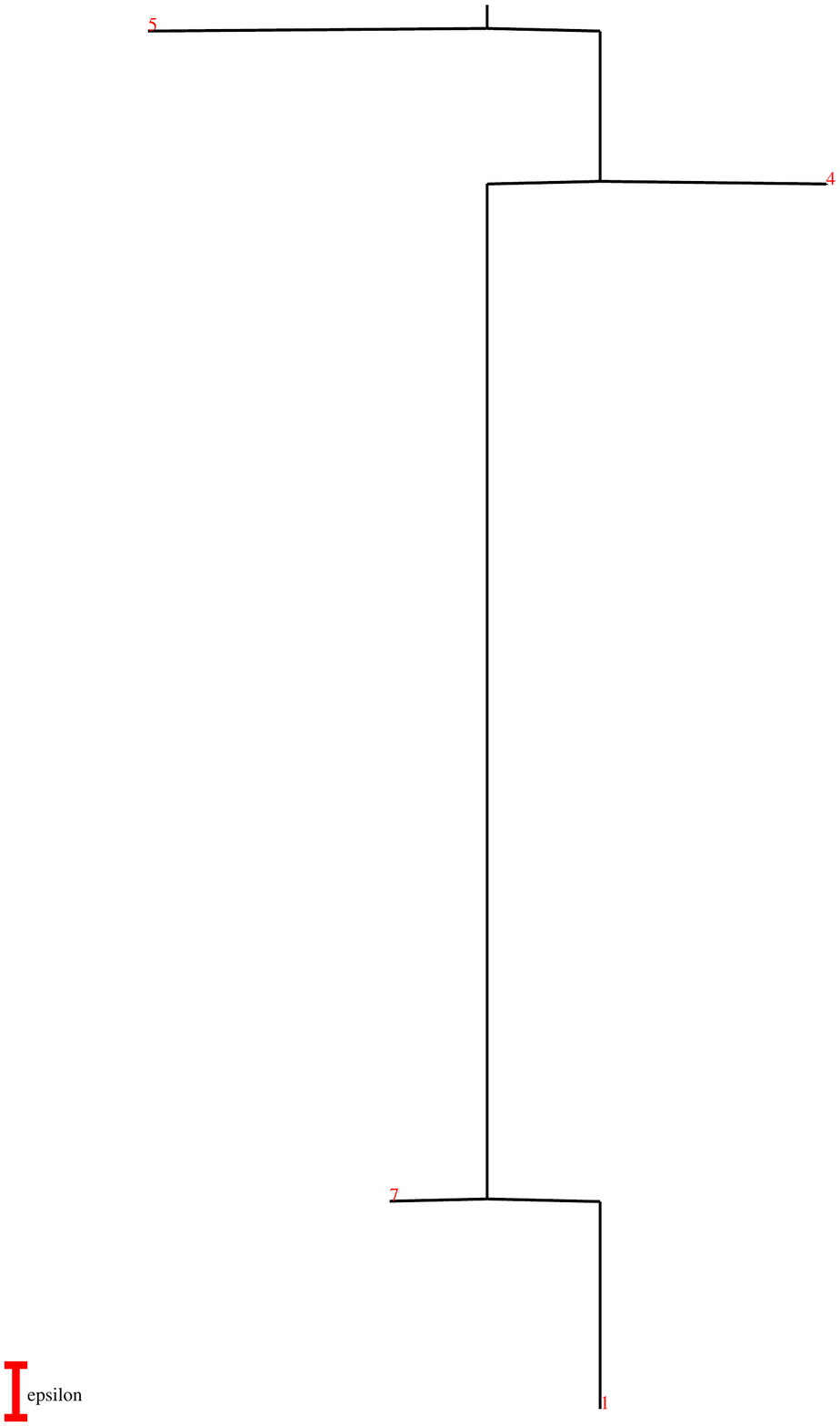}}
	\hfill
	\subfloat[37]{\includegraphics[width=0.05\textwidth, trim=3.7cm 0 3.7cm 0, clip]{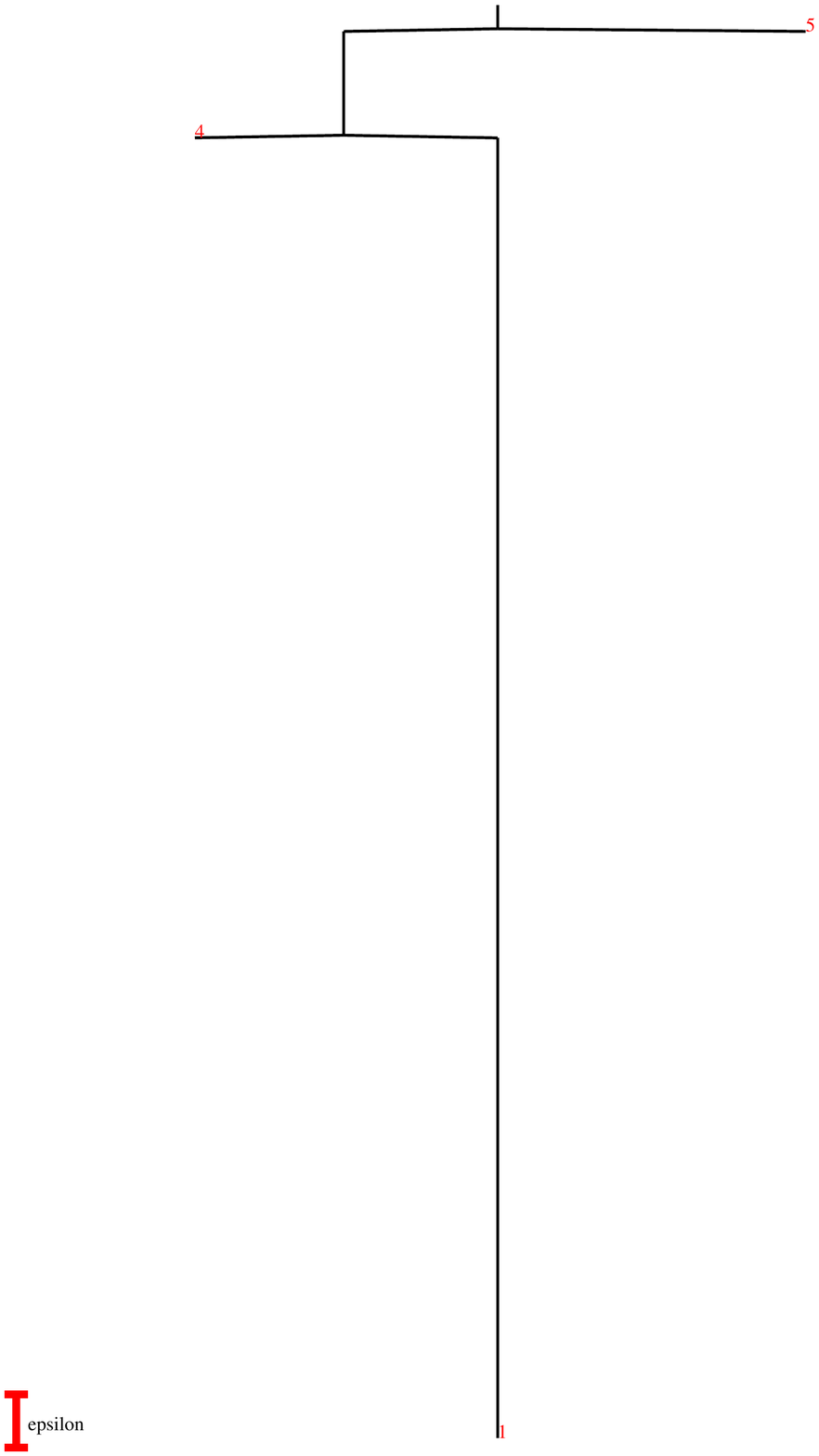}}
	\hfill
	\subfloat[38]{\includegraphics[width=0.05\textwidth, trim=3.7cm 0 3.7cm 0, clip]{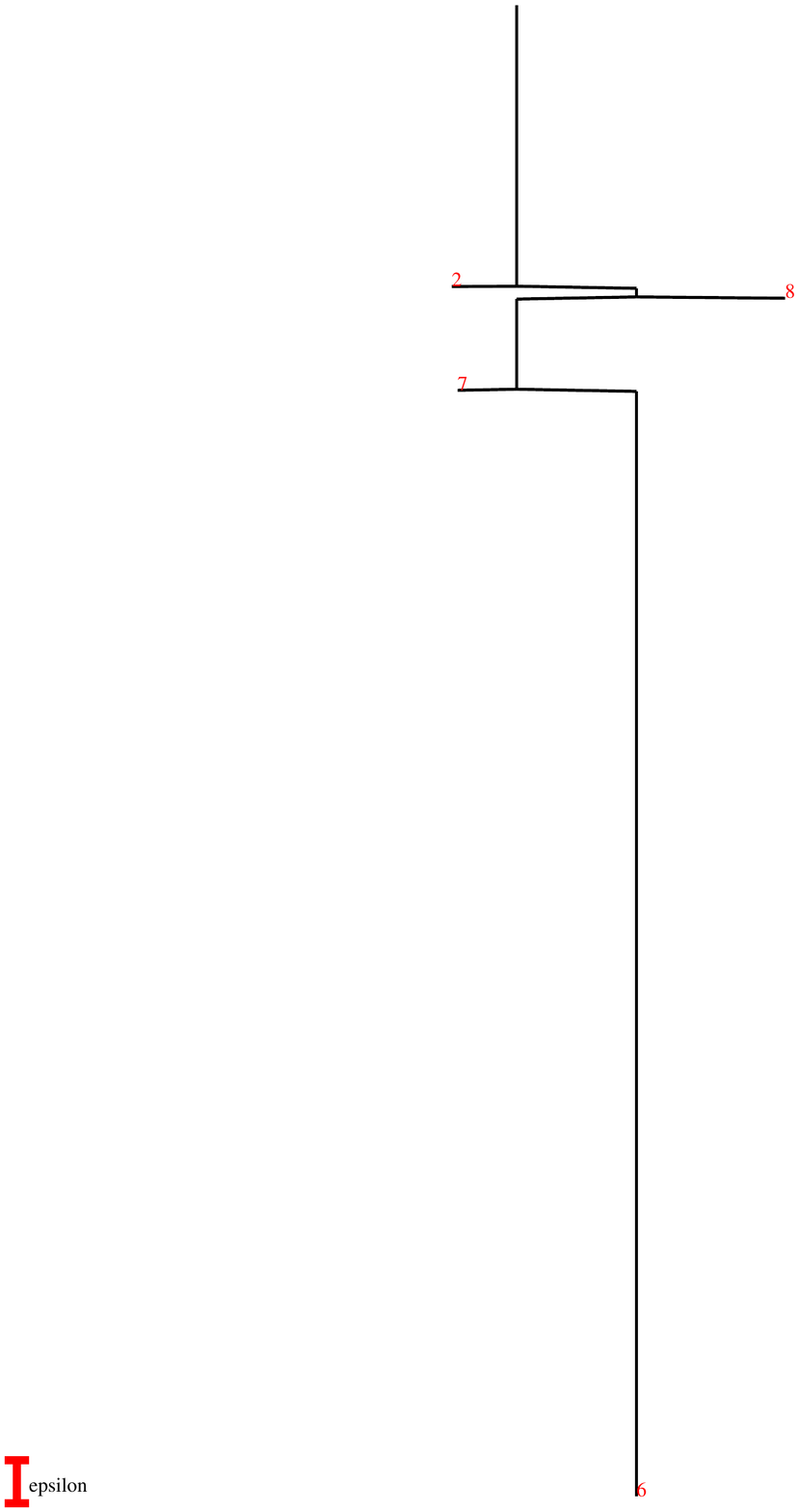}}
	\hfill
	\subfloat[39]{\includegraphics[width=0.05\textwidth, trim=3.7cm 0 3.7cm 0, clip]{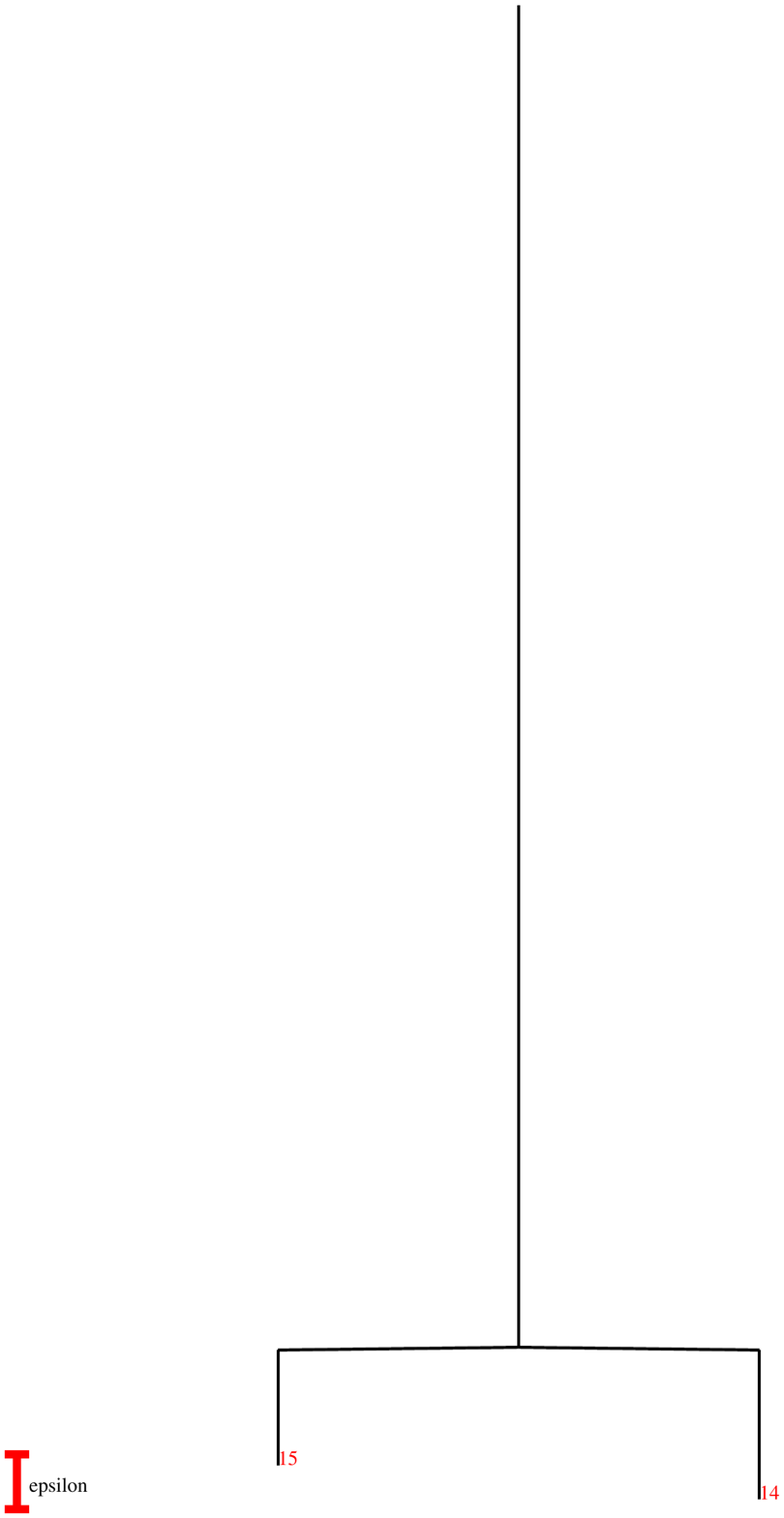}}
	\hfill
	\subfloat[40]{\includegraphics[width=0.05\textwidth, trim=3.7cm 0 3.7cm 0, clip]{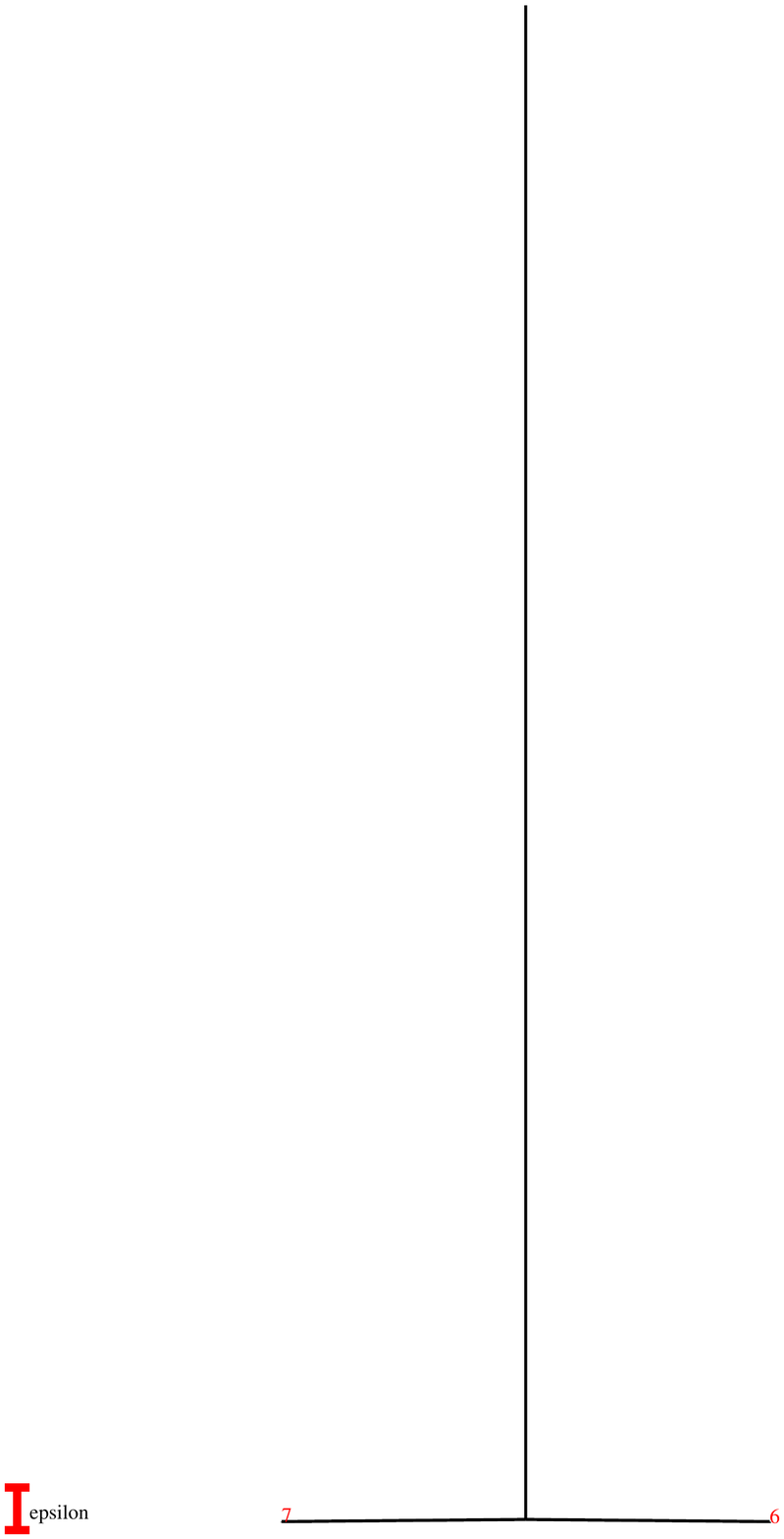}}
	\hfill
	\subfloat[41]{\includegraphics[width=0.05\textwidth, trim=3.7cm 0 3.7cm 0, clip]{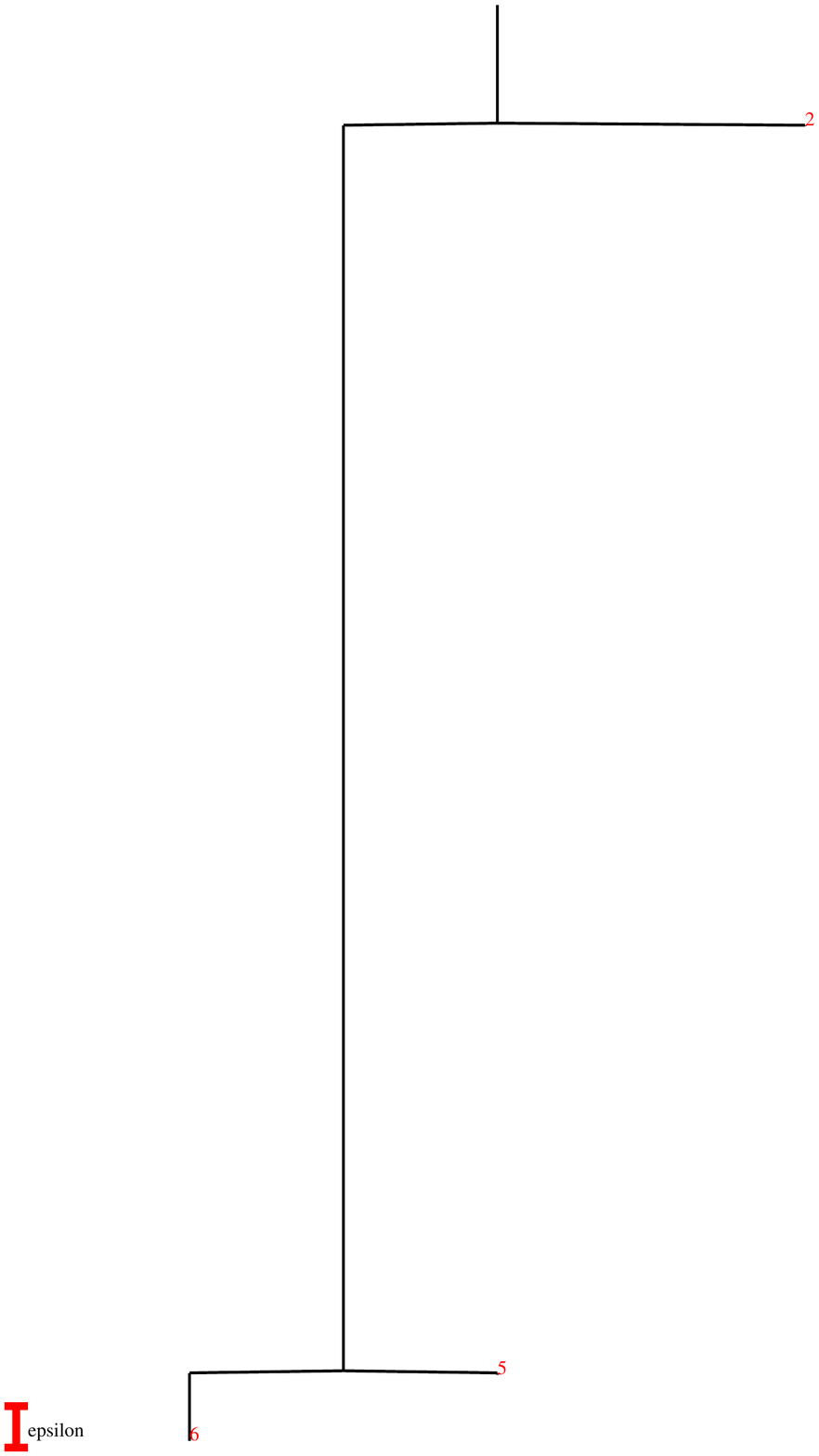}}
	\hfill
	
	\subfloat[42]{\includegraphics[width=0.05\textwidth, trim=3.7cm 0 3.7cm 0, clip]{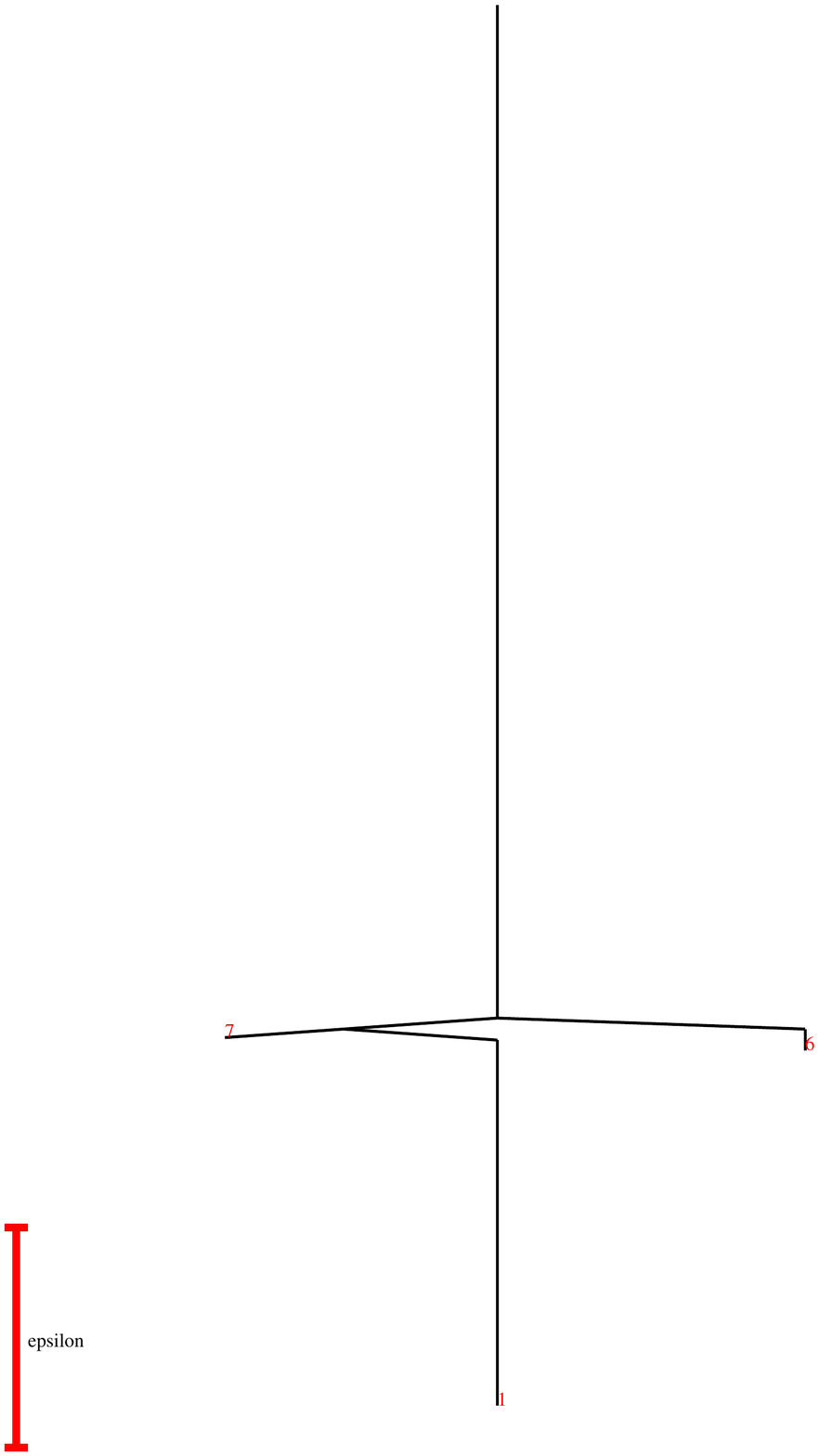}}
	\hfill
	\subfloat[43]{\includegraphics[width=0.05\textwidth, trim=3.7cm 0 3.7cm 0, clip]{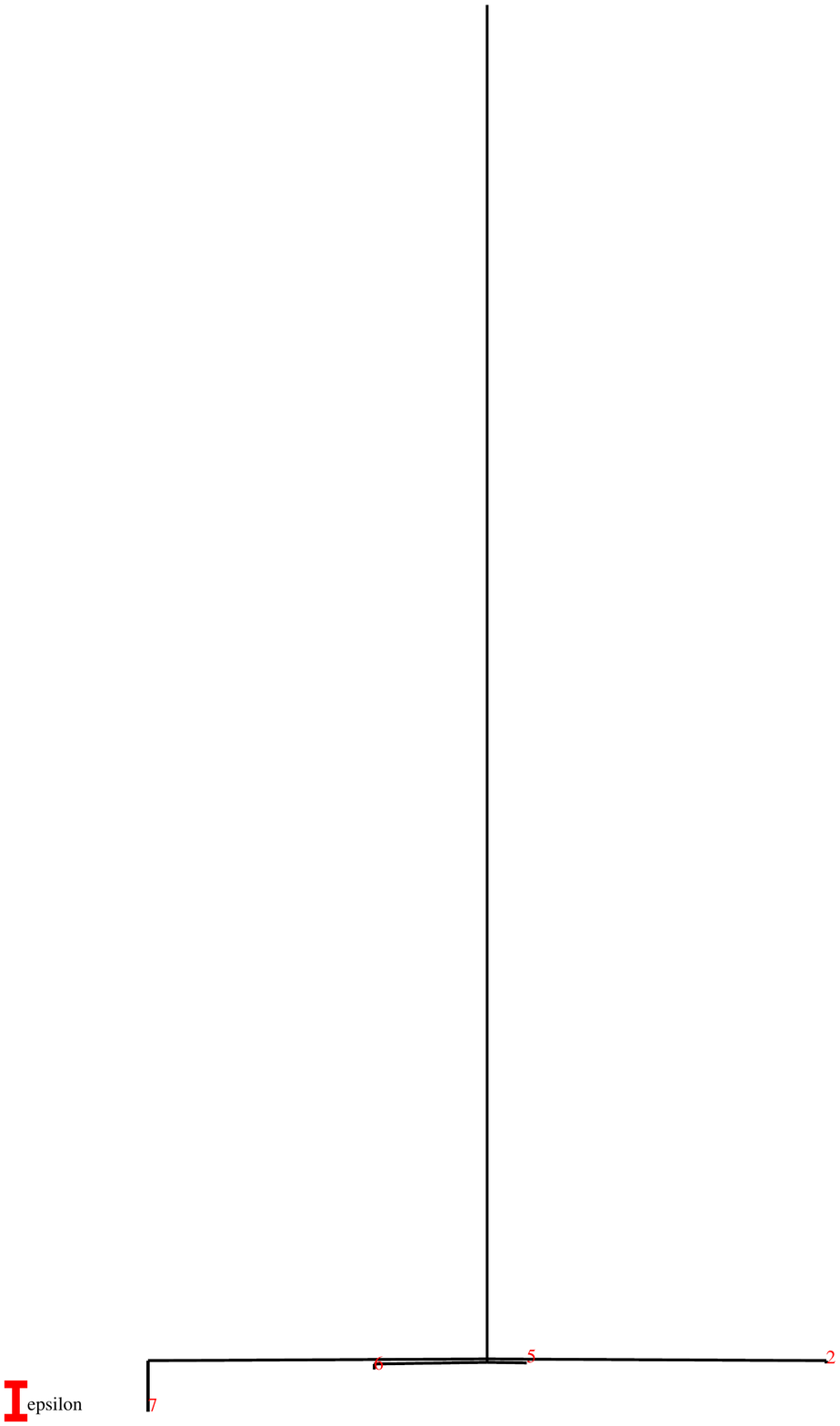}}
	\hfill
	\subfloat[44]{\includegraphics[width=0.05\textwidth, trim=3.7cm 0 3.7cm 0, clip]{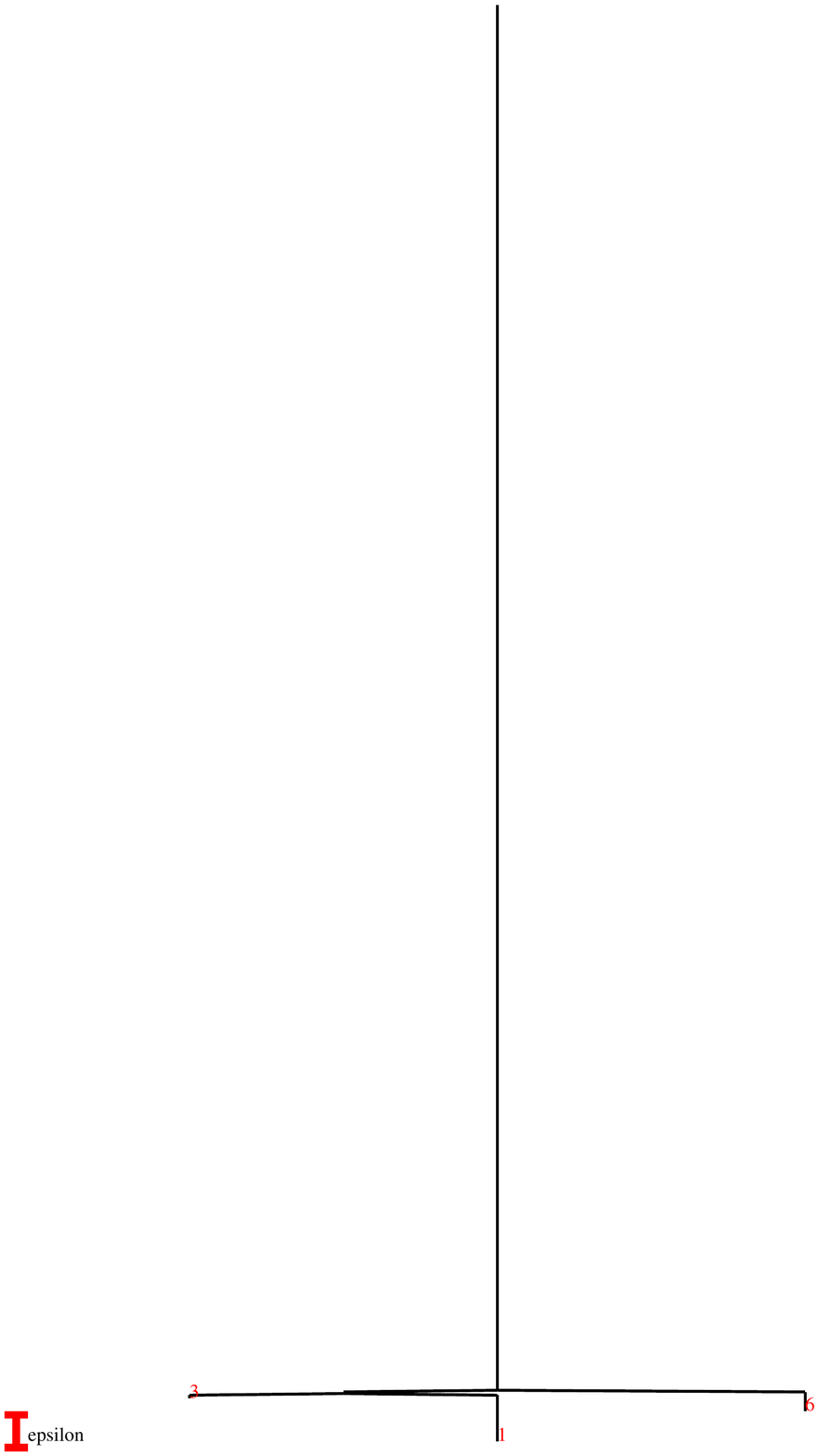}}
	\hfill
	\subfloat[45]{\includegraphics[width=0.05\textwidth, trim=3.7cm 0 3.7cm 0, clip]{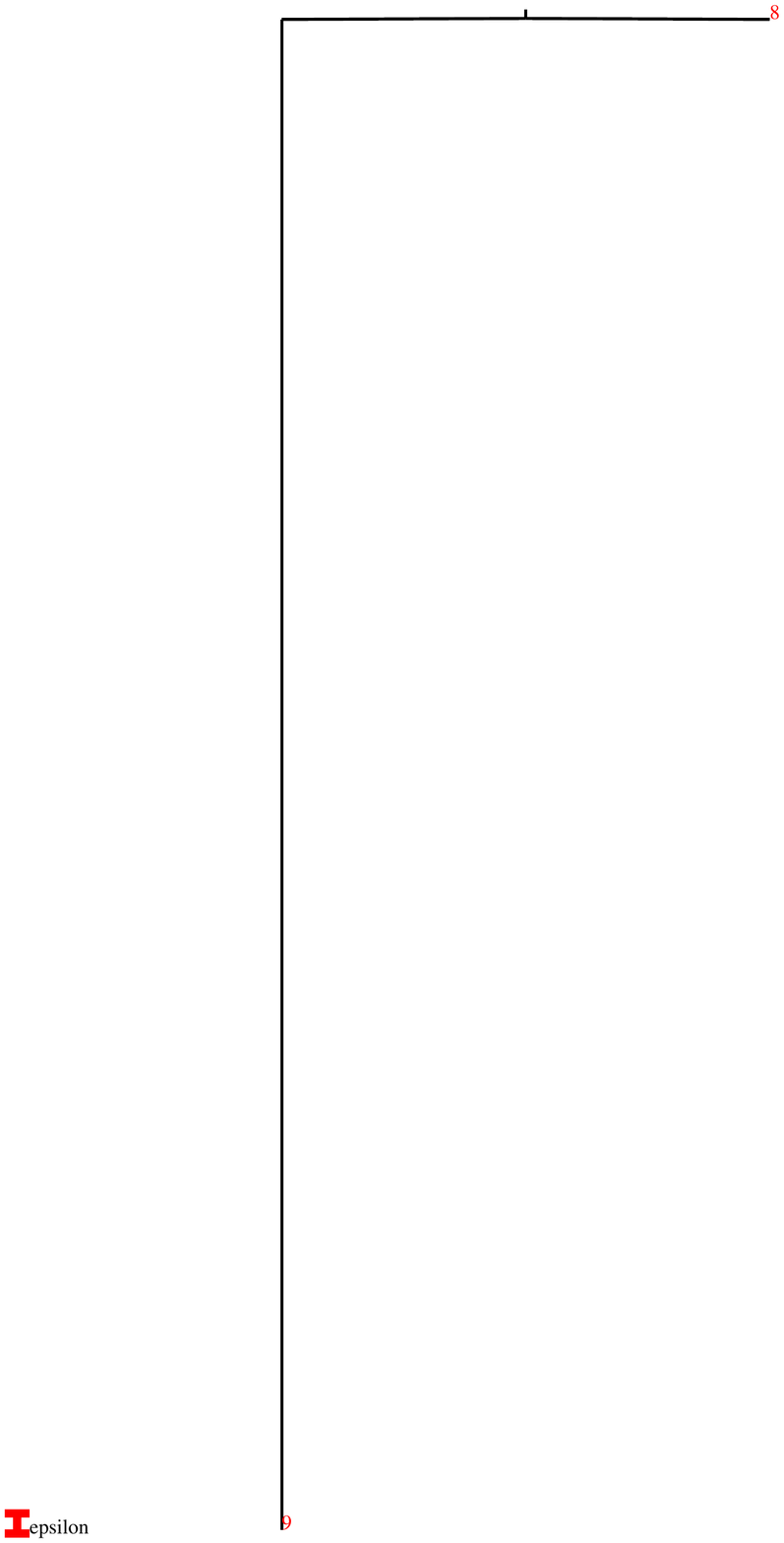}}
	\hfill
	\subfloat[46]{\includegraphics[width=0.05\textwidth, trim=3.7cm 0 3.7cm 0, clip]{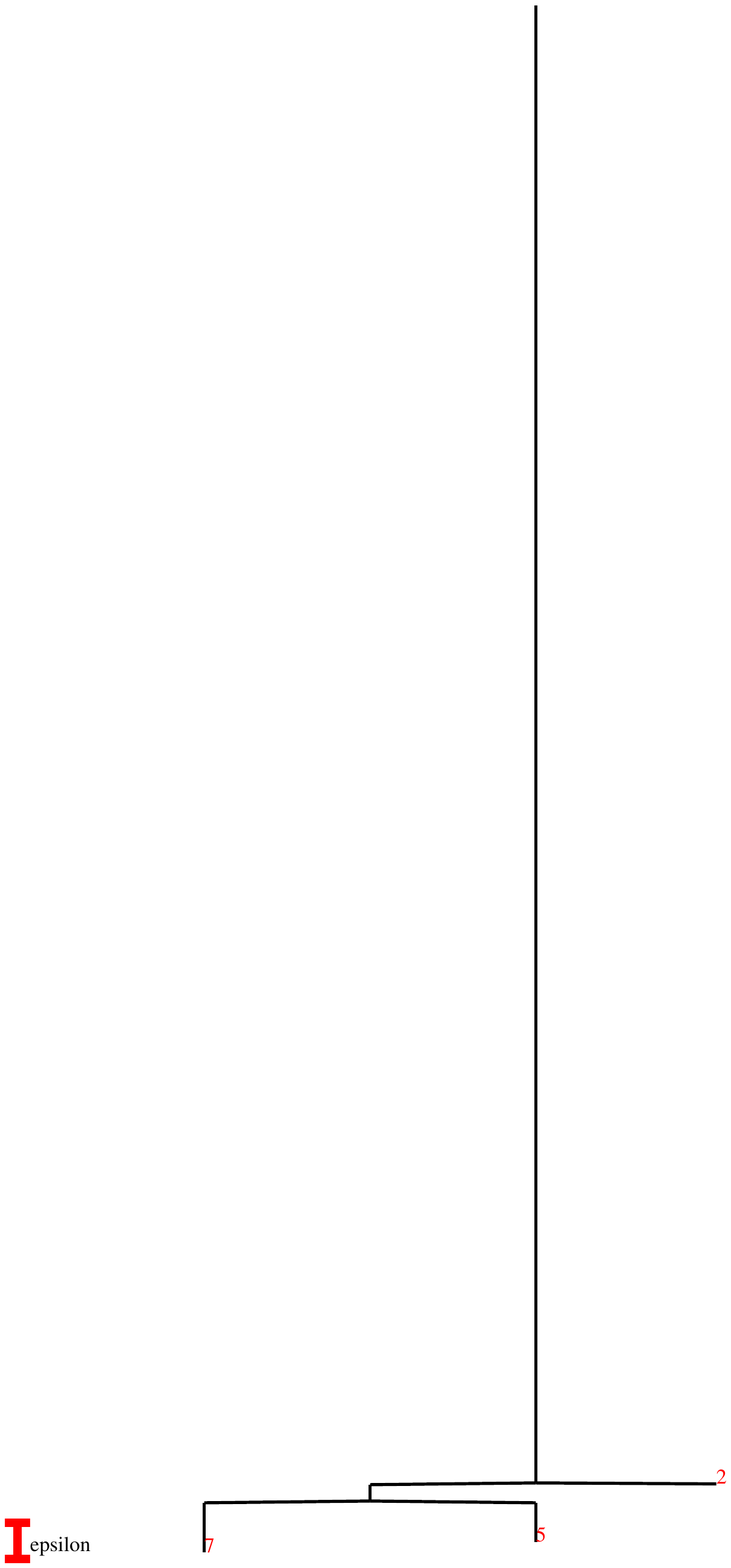}}
	\hfill
	\subfloat[47]{\includegraphics[width=0.05\textwidth, trim=3.7cm 0 3.7cm 0, clip]{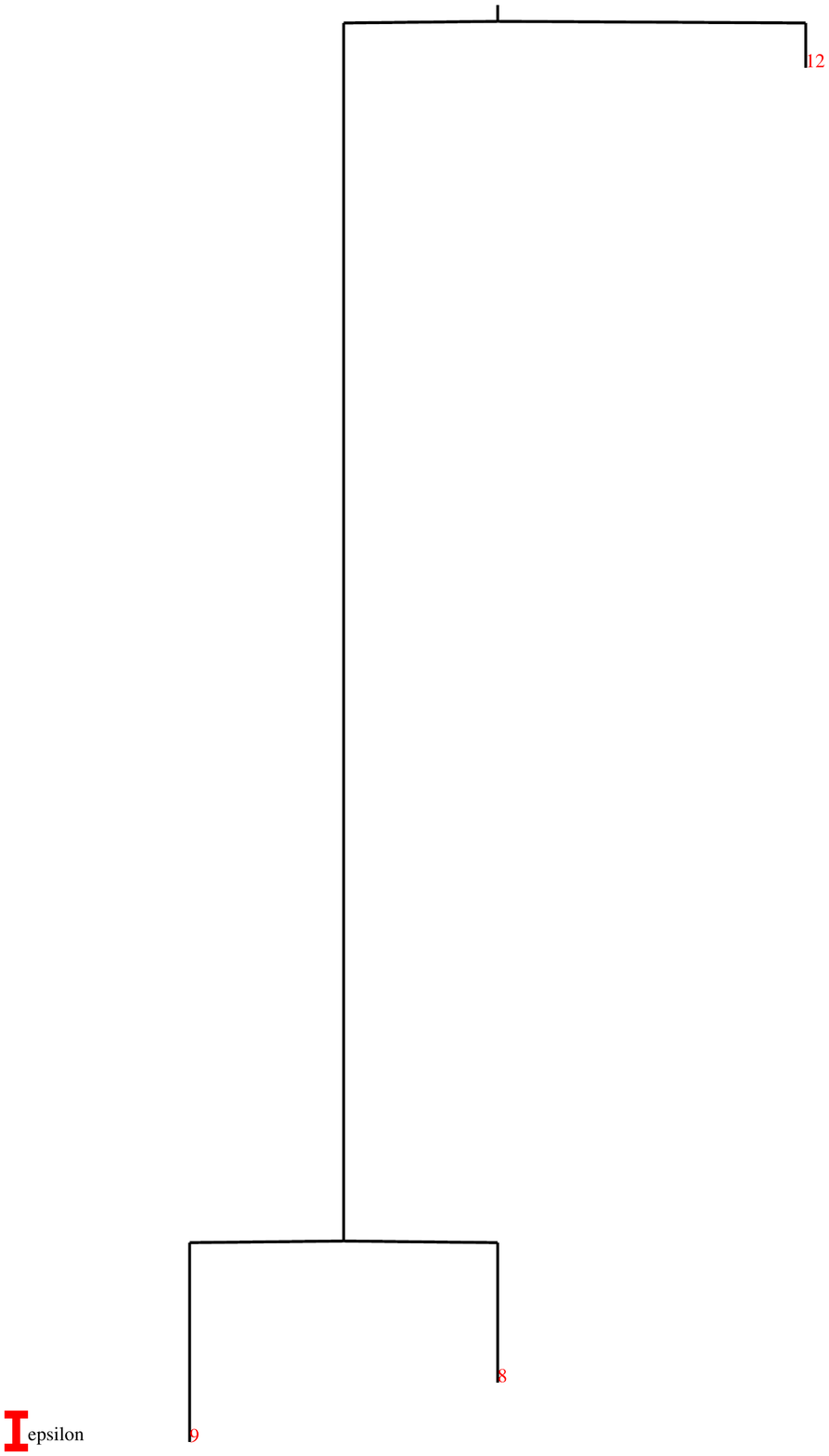}}
	\hfill
	\subfloat[48]{\includegraphics[width=0.05\textwidth, trim=3.7cm 0 3.7cm 0, clip]{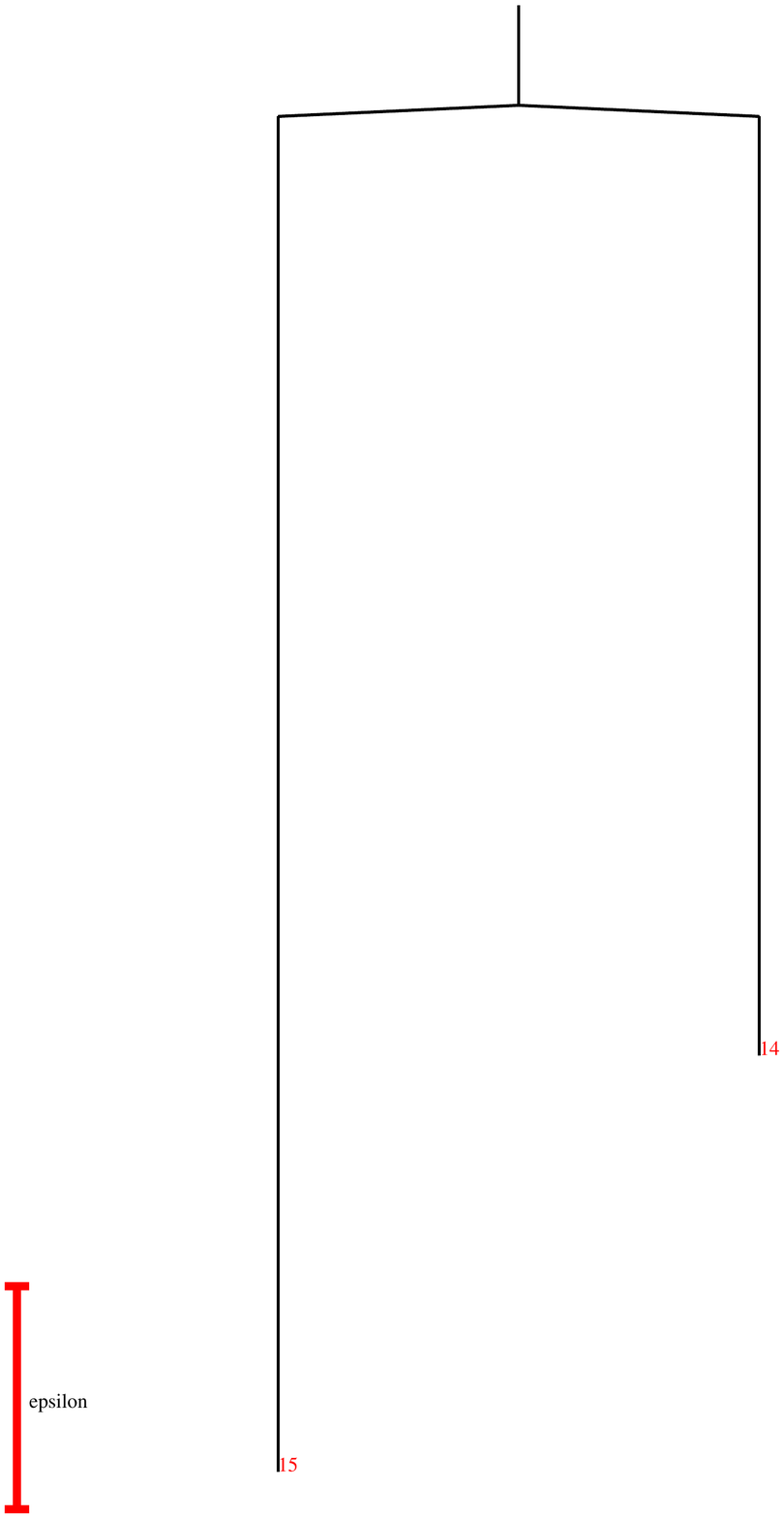}}
	\hfill
	\caption[Disconnectivity graphs of the LML surfaces for GP trained with Mat\'{e}rn ($\nu = 2.5$) kernels. ]{Disconnecitivity graphs of the LML surfaces for GP trained with Mat\'{e}rn ($\nu = 2.5$) kernels.  The red bars represent the same LML interval to give a sense of the gaps between different minima and the different TSs and $N$ is the total number of points in the training data.}
	\label{fig:ds-graphs-mat}
\end{figure}

Along the changes in the LML landscape shown in figure \ref{fig:ds-graphs-mat},  there is not a correspondence of lowest minima on the LML to model with lowest MAE.  For this particular training data,  one has two minima that are similar in terms of error: the one with the lowest error is only lower on the LML landscape when the set is fully symmetrised while the rest of the partial training sets partial training set always select the slightly less performant model.  One can follow,  as data is added to the training set,  each minima and its corresponding hyperparameters.  It results that longer length scales of the model lower the MAE and as shown on figure \ref{fig:symmetry-slow} only the last training set picks the longer length scale model.
\par
It can also be said that the sharp change in length scales and MAEs shown in figure \ref{fig:symmetry-slow} is also a telling sign of multiple competing minima on the LML as we expect additional data to smoothly change the LML of the GP and not produce sharp switches of global minimum.

It is rather unexpected to see a sharp change in performance and even more so a sharp change in hyperparameters as we expect additional data to smoothly change the log-marginal likelihood of the Gaussian process.  However,  upon closer investigation,  one can follow the best model selected in the symmMID set when removing data points.  It is then not a different minimum on the log-marginal likelihood but just a sharp change in the scoring of those.  This ``better'' model starts in very similar length scales to the best model for the MID set in figure \ref{fig:symmetry-slow}(b) and then proceeds to smoothly evolve towards the symmMID best model (this can be seen in the shaded models in figure \ref{fig:symmetry-slow}). Moreover,  one can also follow the smooth change in the best model for sub-symmetrical training sets to the final symmMID and find the equivalent model which,  as it is lower scoring on the log-marginal likelihood,  is not selected. 

\begin{figure}[H]
	\centering
	\captionsetup[subfigure]{labelformat=empty}
	\subfloat[$33$]{\includegraphics[width=0.05\textwidth,trim = 3.7cm -5cm 3.7cm 0cm,clip]{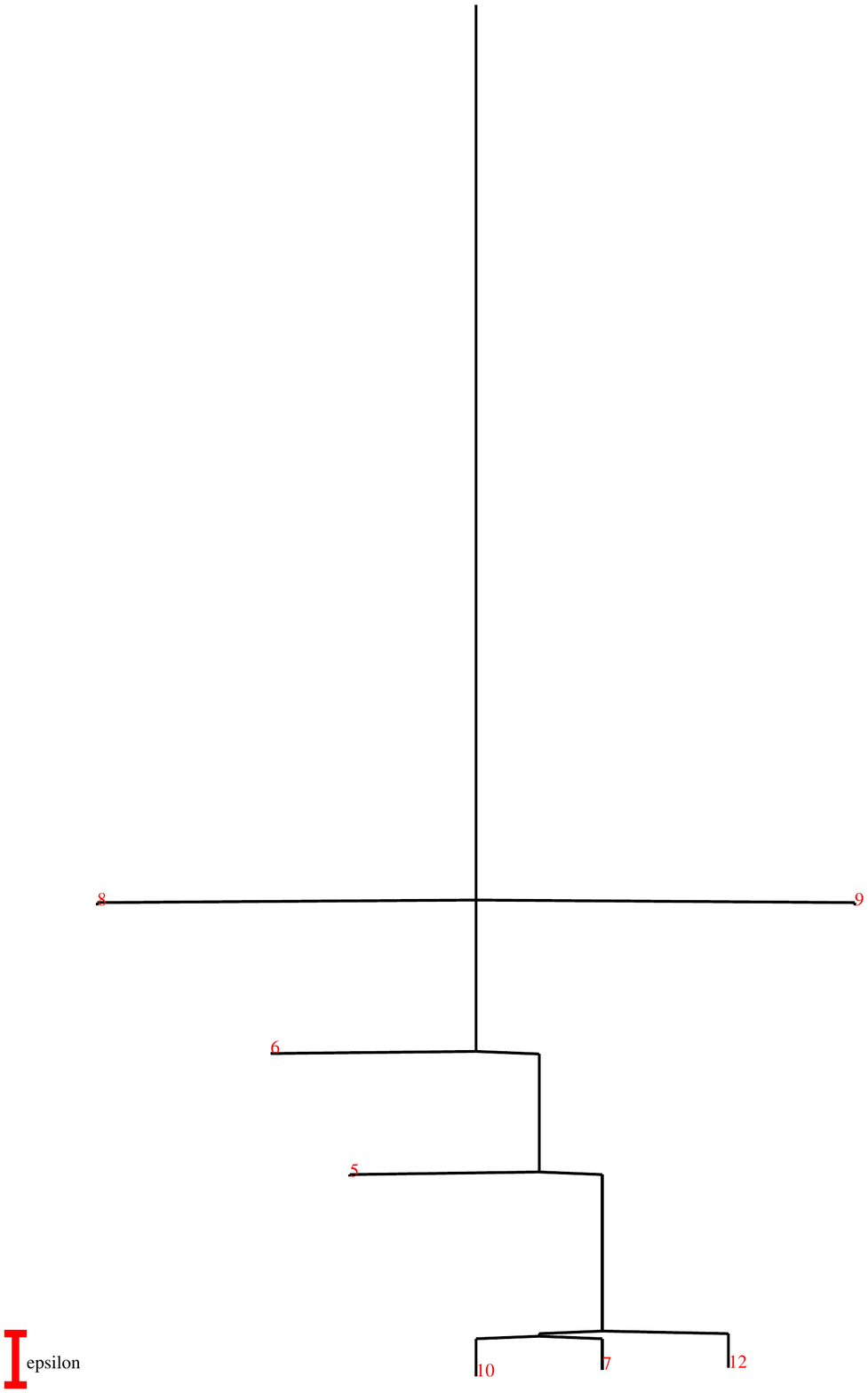}}
	\hfill
	\subfloat[$34$]{\includegraphics[width=0.05\textwidth,trim = 3.7cm -5cm 3.7cm 0cm,clip]{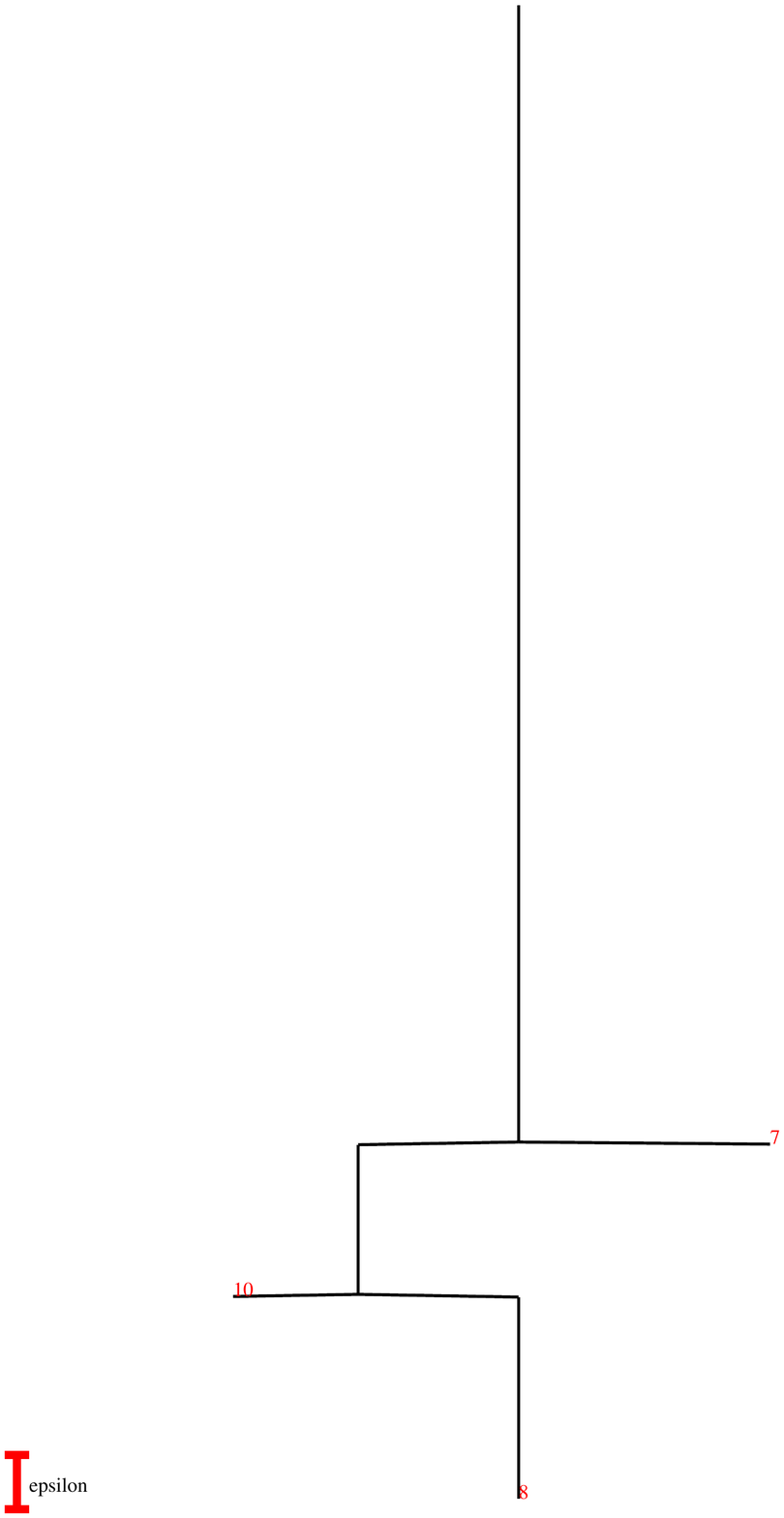}}
	\hfill
	\subfloat[$35$]{\includegraphics[width=0.05\textwidth,trim = 3.7cm -5cm 3.7cm 0cm,clip]{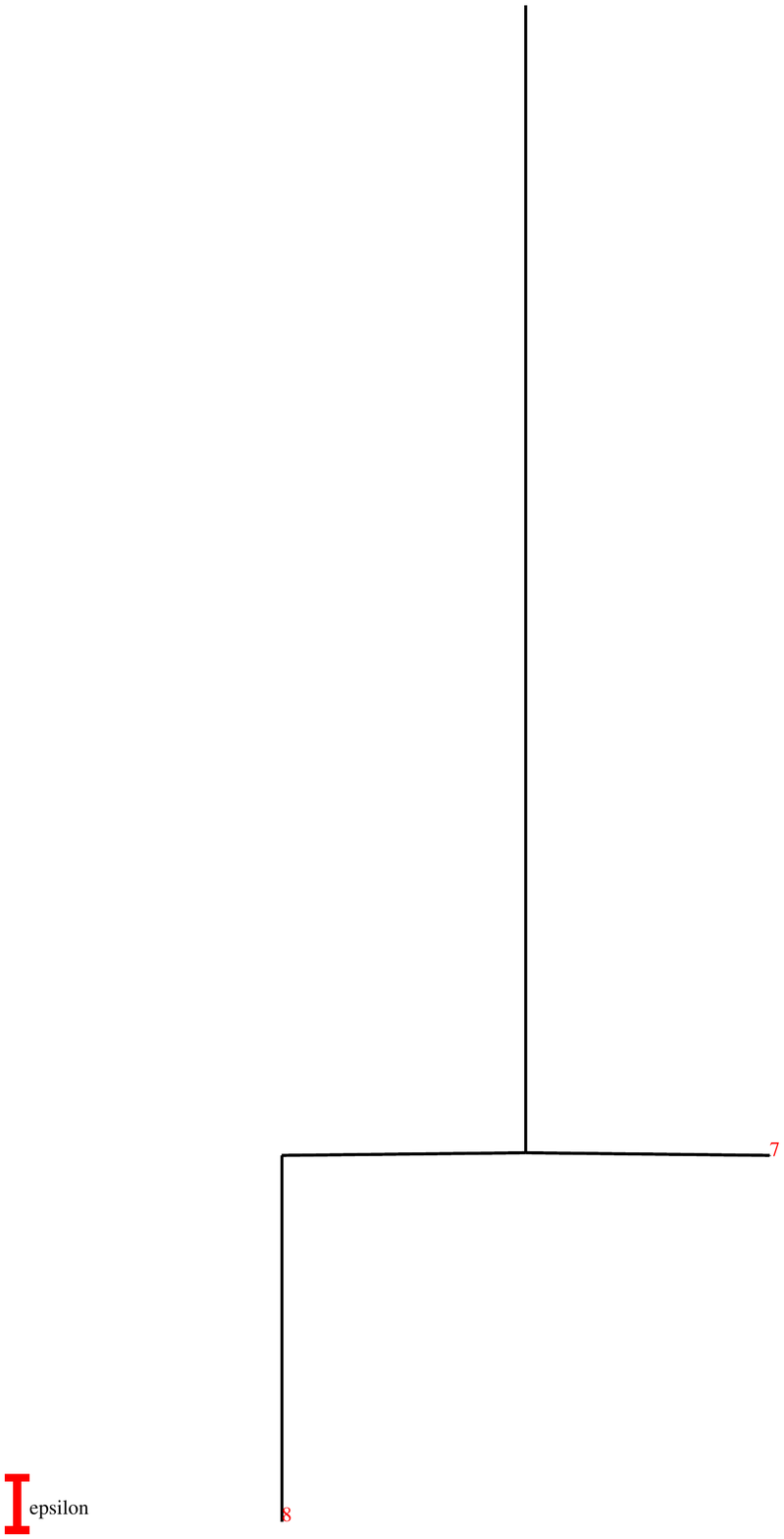}}
	\hfill
	\subfloat[$36$]{\includegraphics[width=0.05\textwidth,trim = 3.7cm -5cm 3.7cm 0cm,clip]{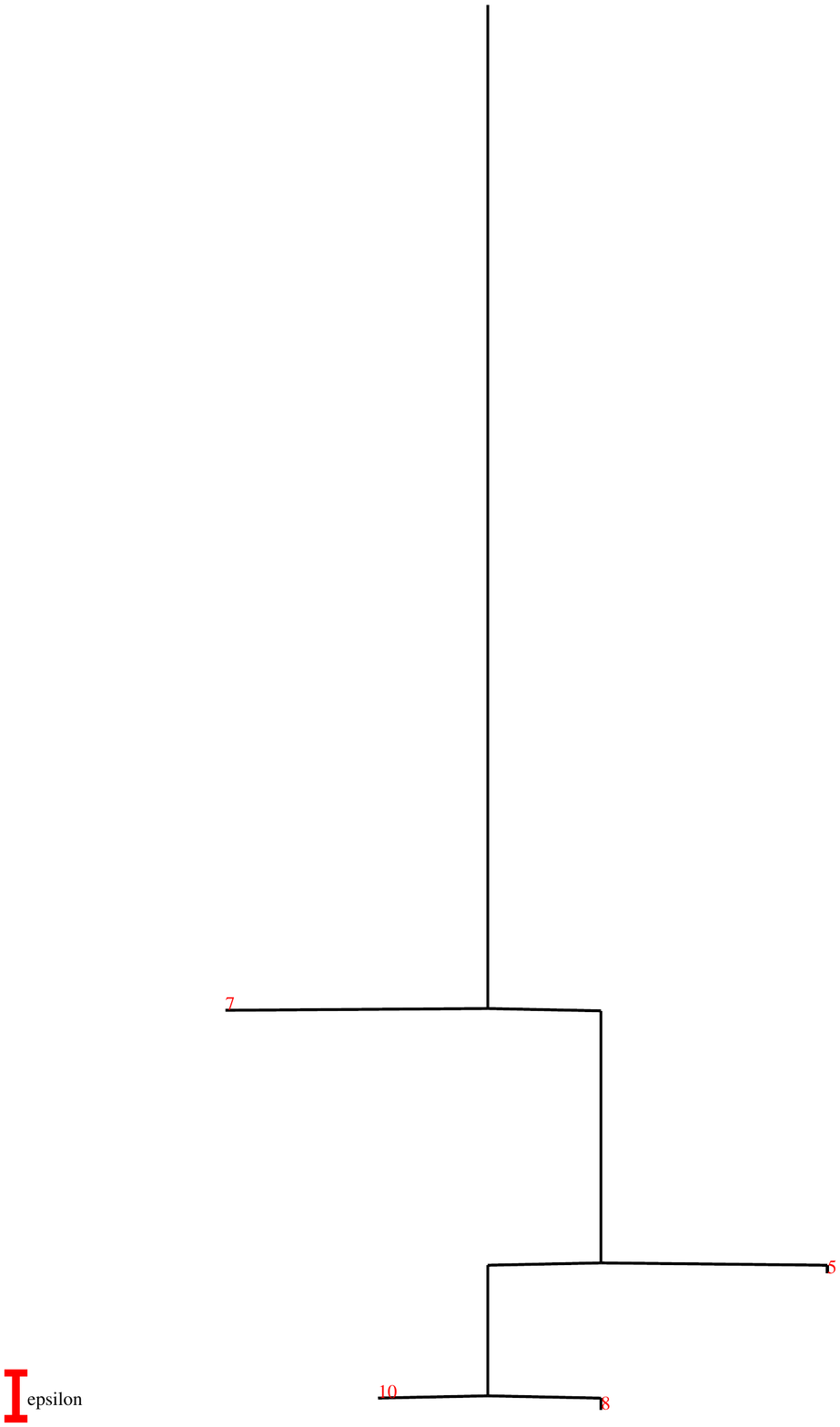}}
	\hfill
	\subfloat[$37$]{\includegraphics[width=0.05\textwidth,trim = 3.7cm -5cm 3.7cm 0cm,clip]{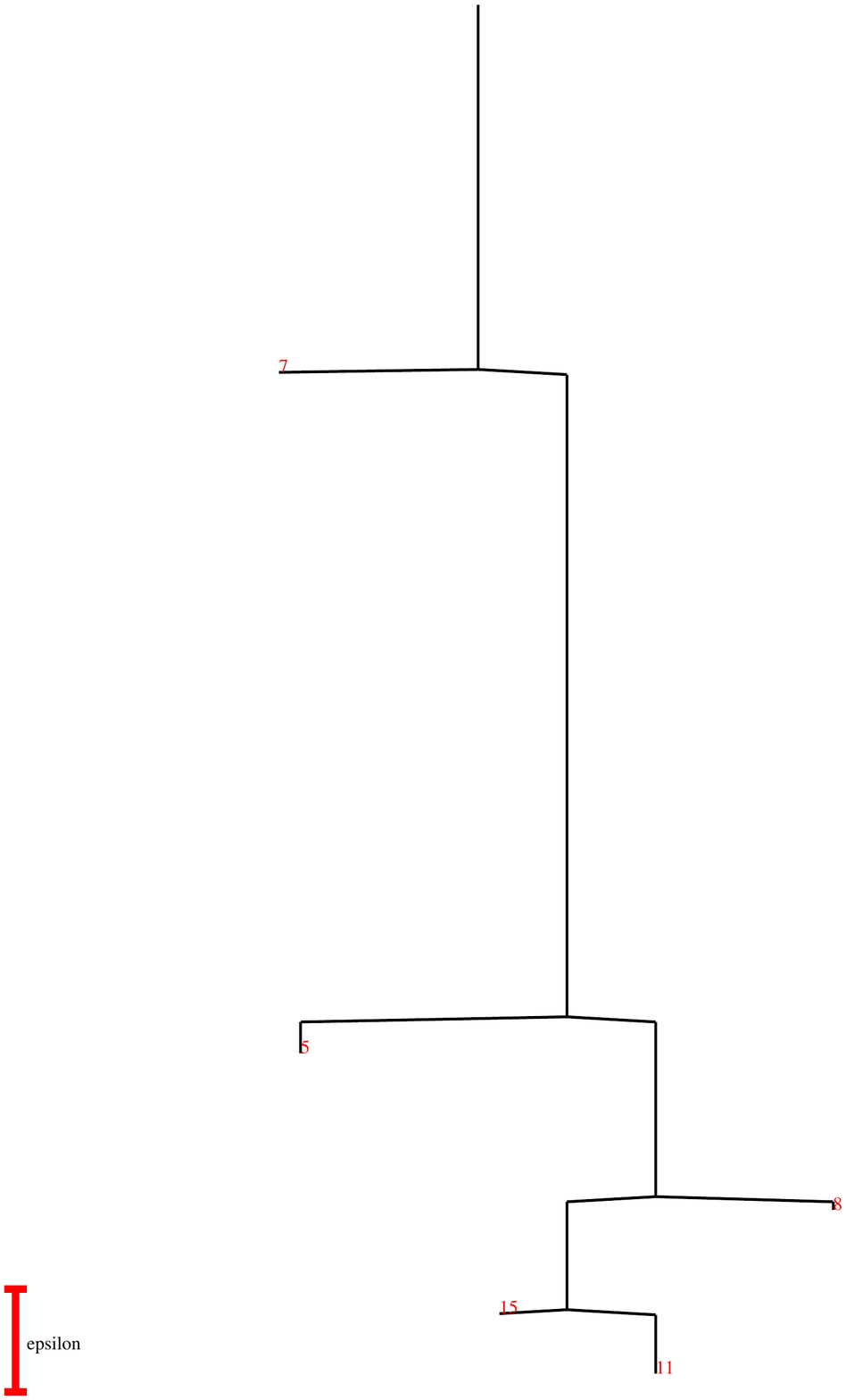}}
	\hfill
	\subfloat[$38$]{\includegraphics[width=0.05\textwidth,trim = 3.7cm -5cm 3.7cm 0cm,clip]{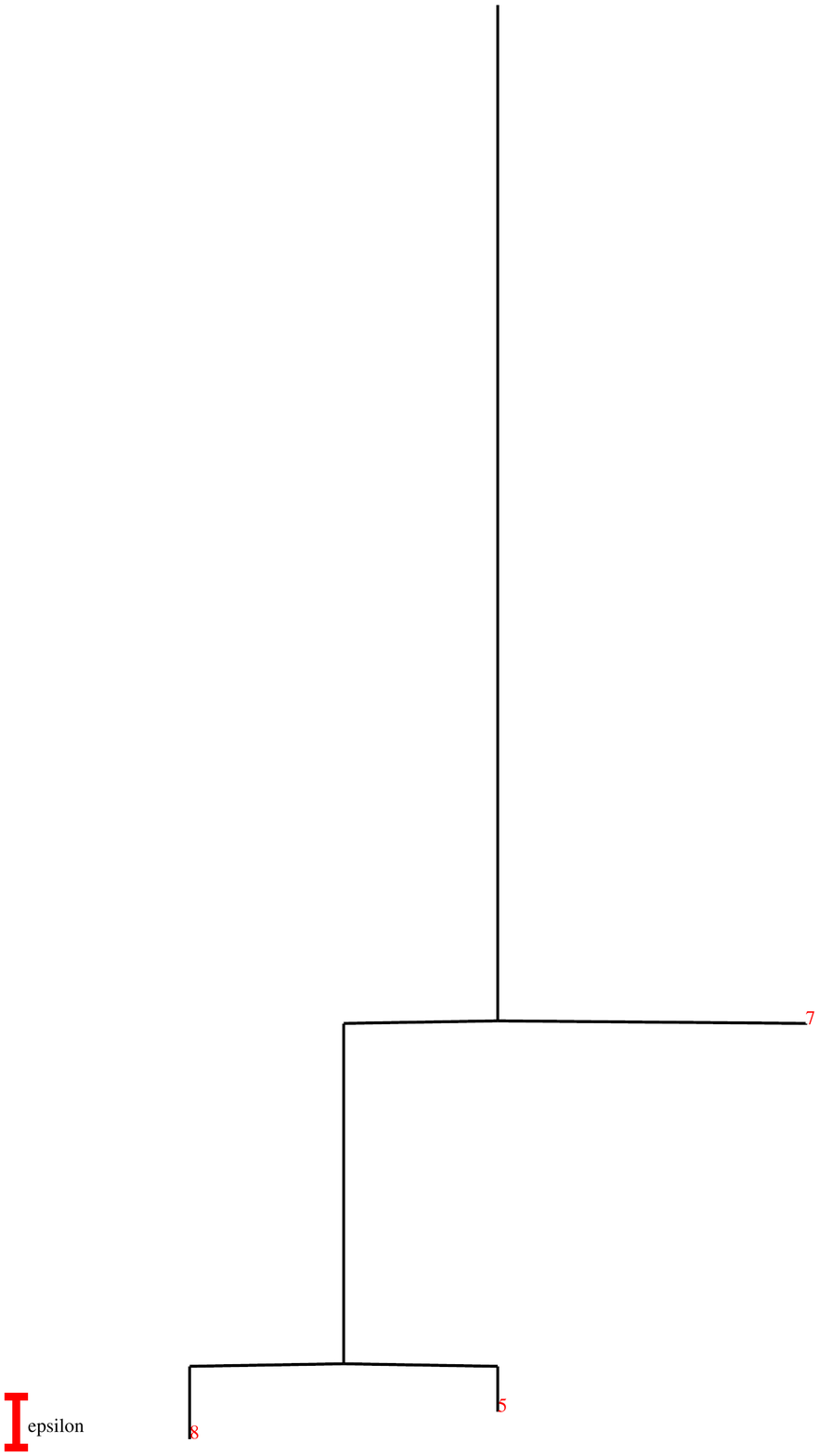}}
	\hfill
	\subfloat[$39$]{\includegraphics[width=0.05\textwidth,trim = 3.7cm -5cm 3.7cm 0cm,clip]{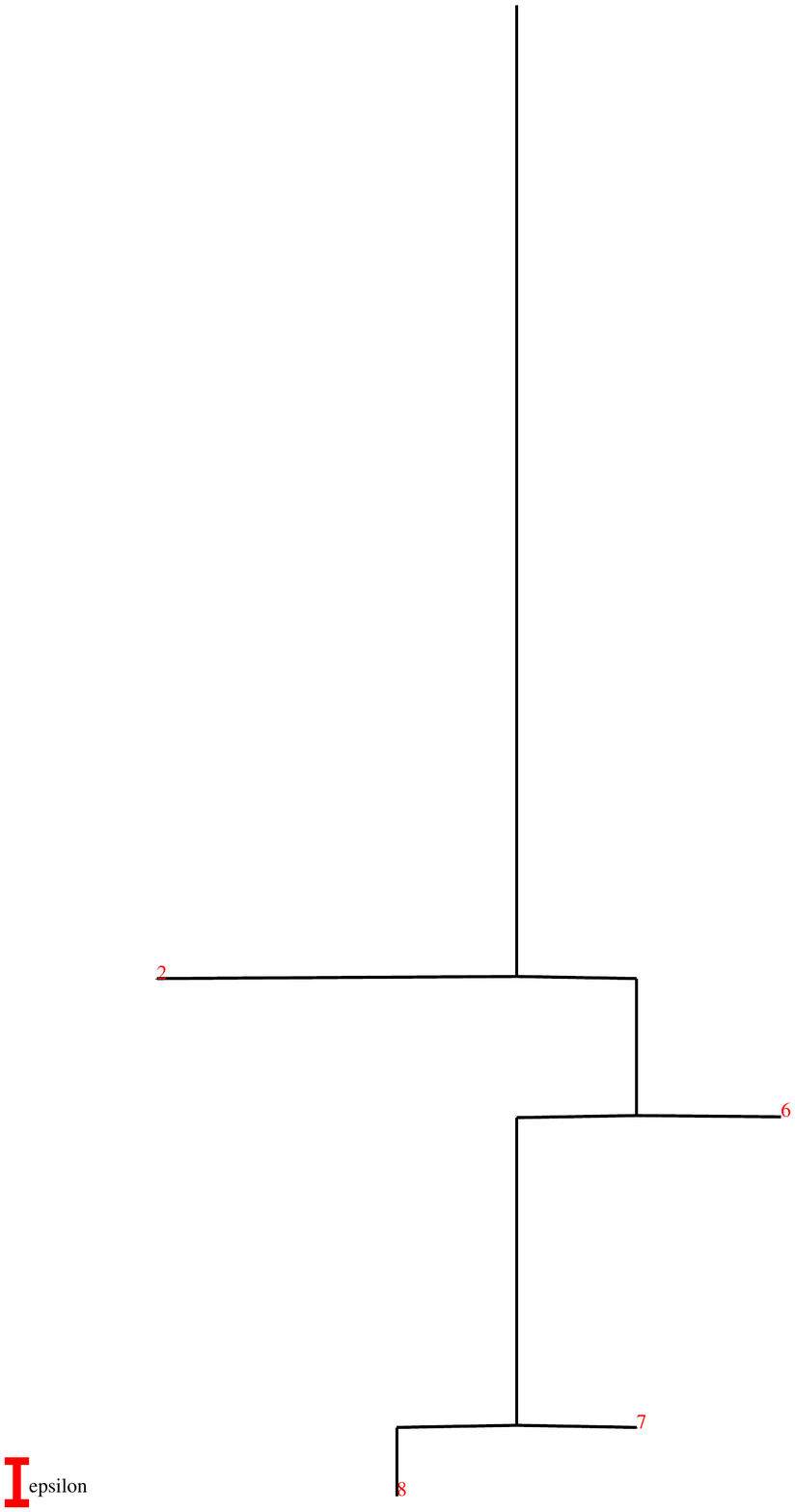}}
	\hfill
	\subfloat[$40$]{\includegraphics[width=0.05\textwidth,trim = 3.7cm -5cm 3.7cm 0cm,clip]{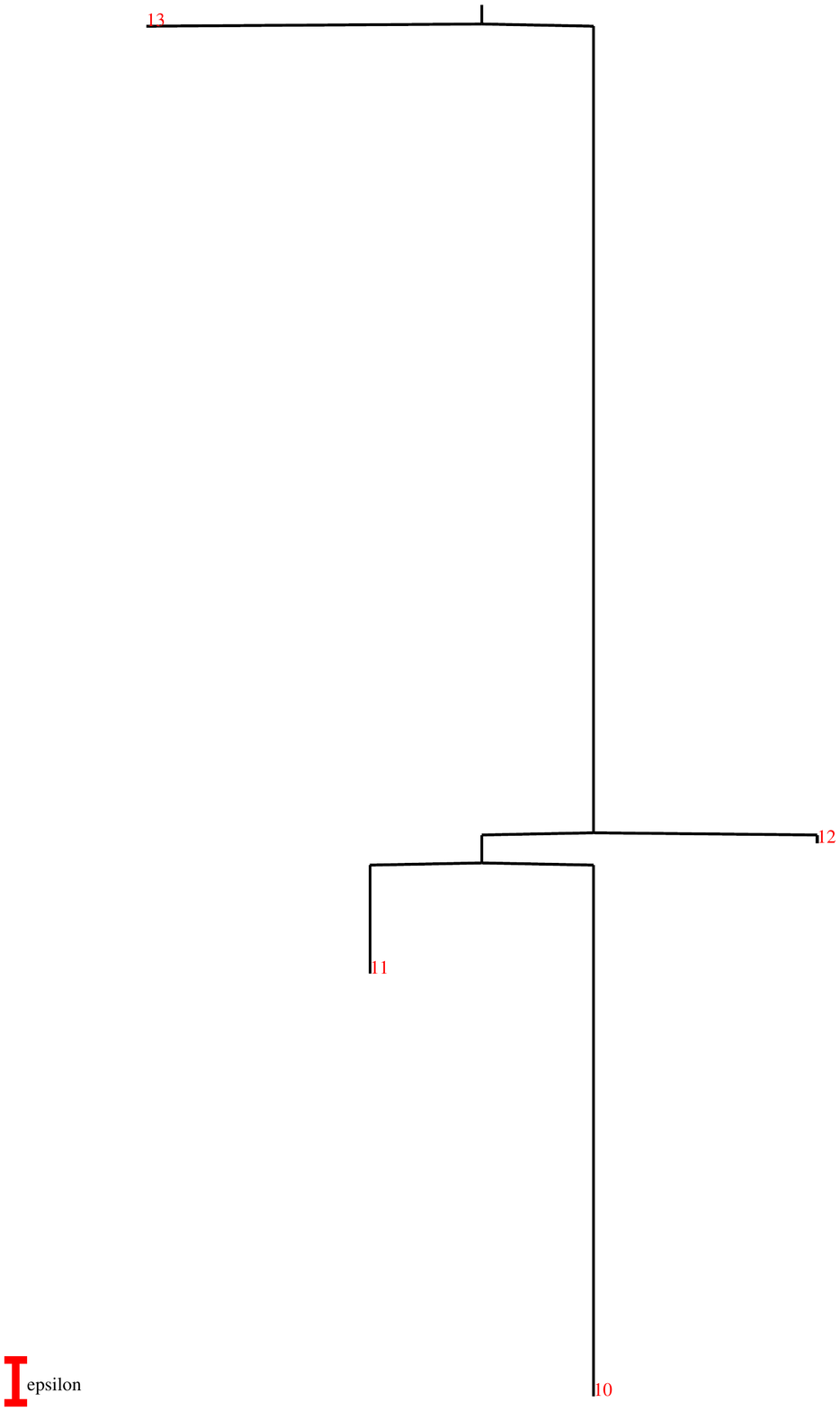}}
	\hfill \par
	\subfloat[$41$]{\includegraphics[width=0.05\textwidth,trim = 3.7cm -5cm 3.7cm 0cm,clip]{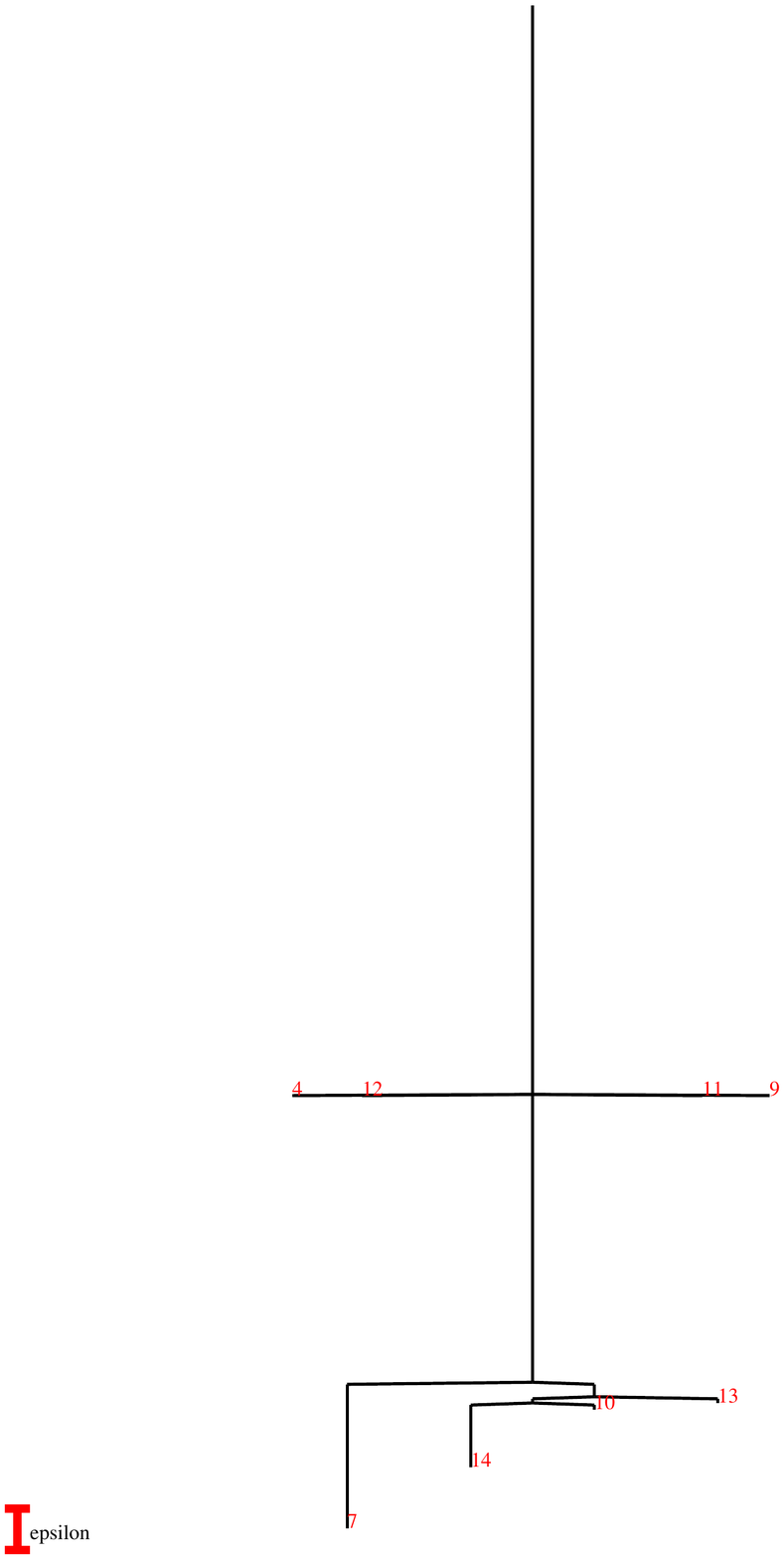}}
	\hfill
	\subfloat[$42$]{\includegraphics[width=0.05\textwidth,trim = 3.7cm -5cm 3.7cm 0cm,clip]{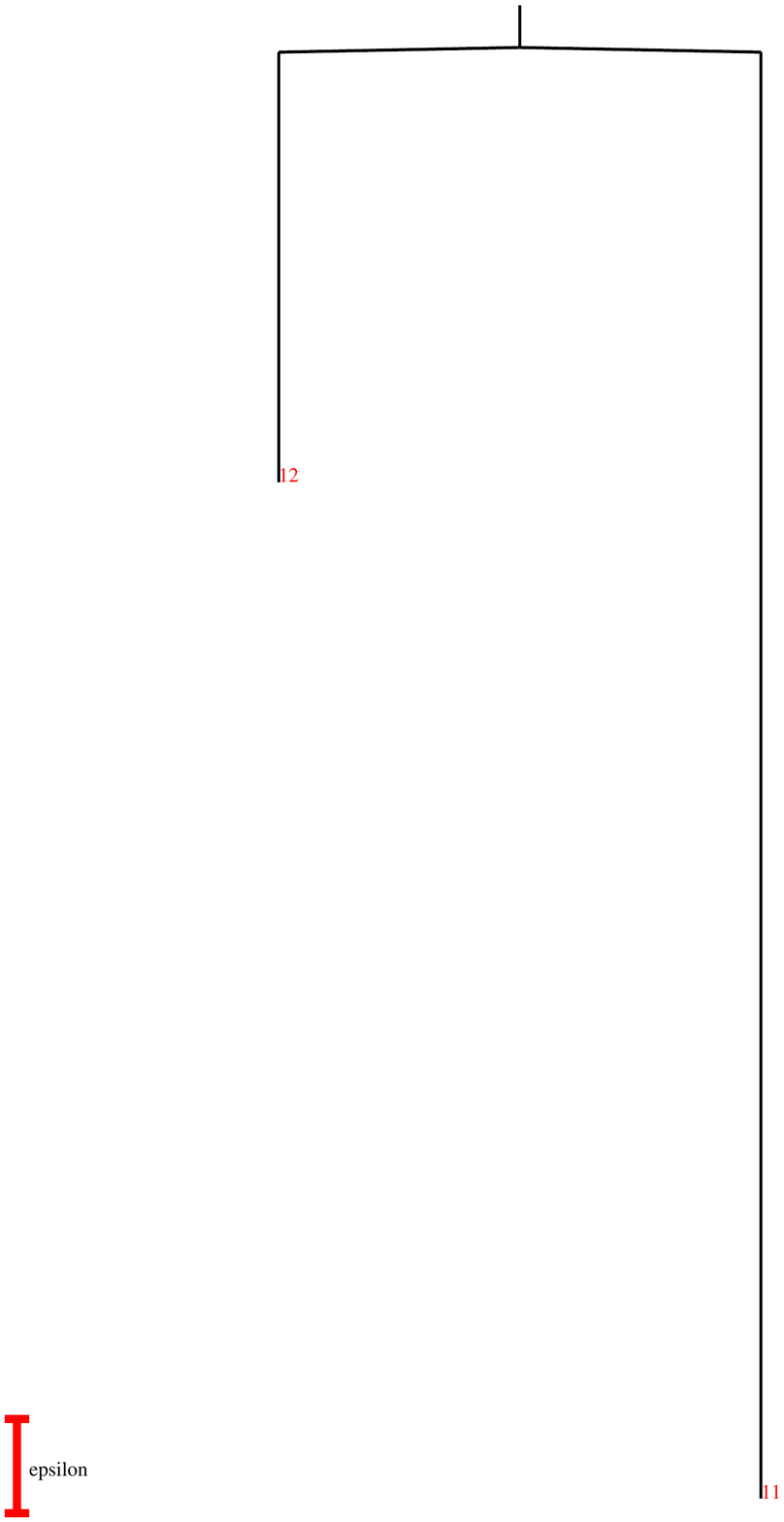}}
	\hfill
	\subfloat[$44$]{\includegraphics[width=0.05\textwidth,trim = 3.7cm -5cm 3.7cm 0cm,clip]{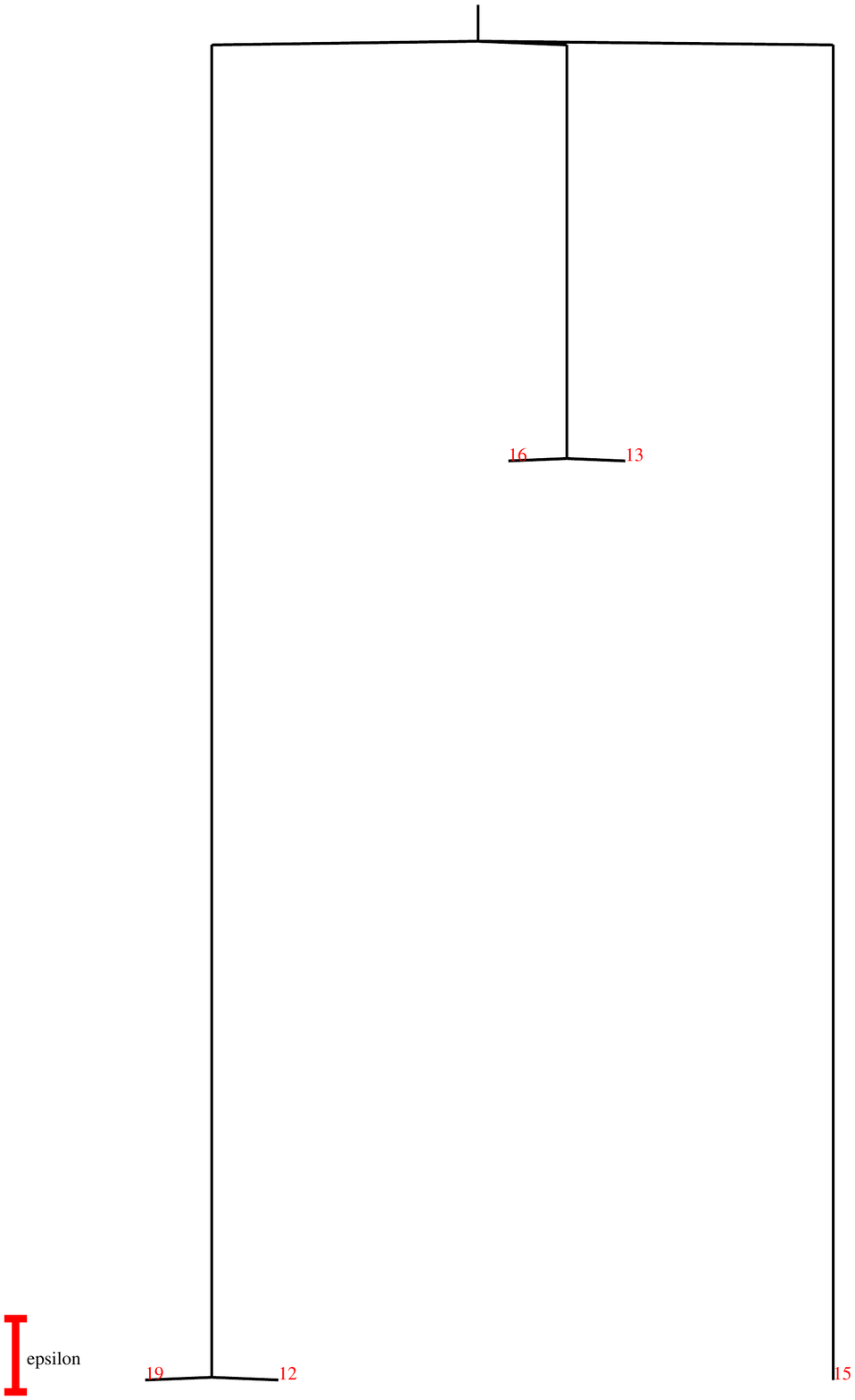}}
	\hfill
	\subfloat[$45$]{\includegraphics[width=0.05\textwidth,trim = 3.7cm -5cm 3.7cm 0cm,clip]{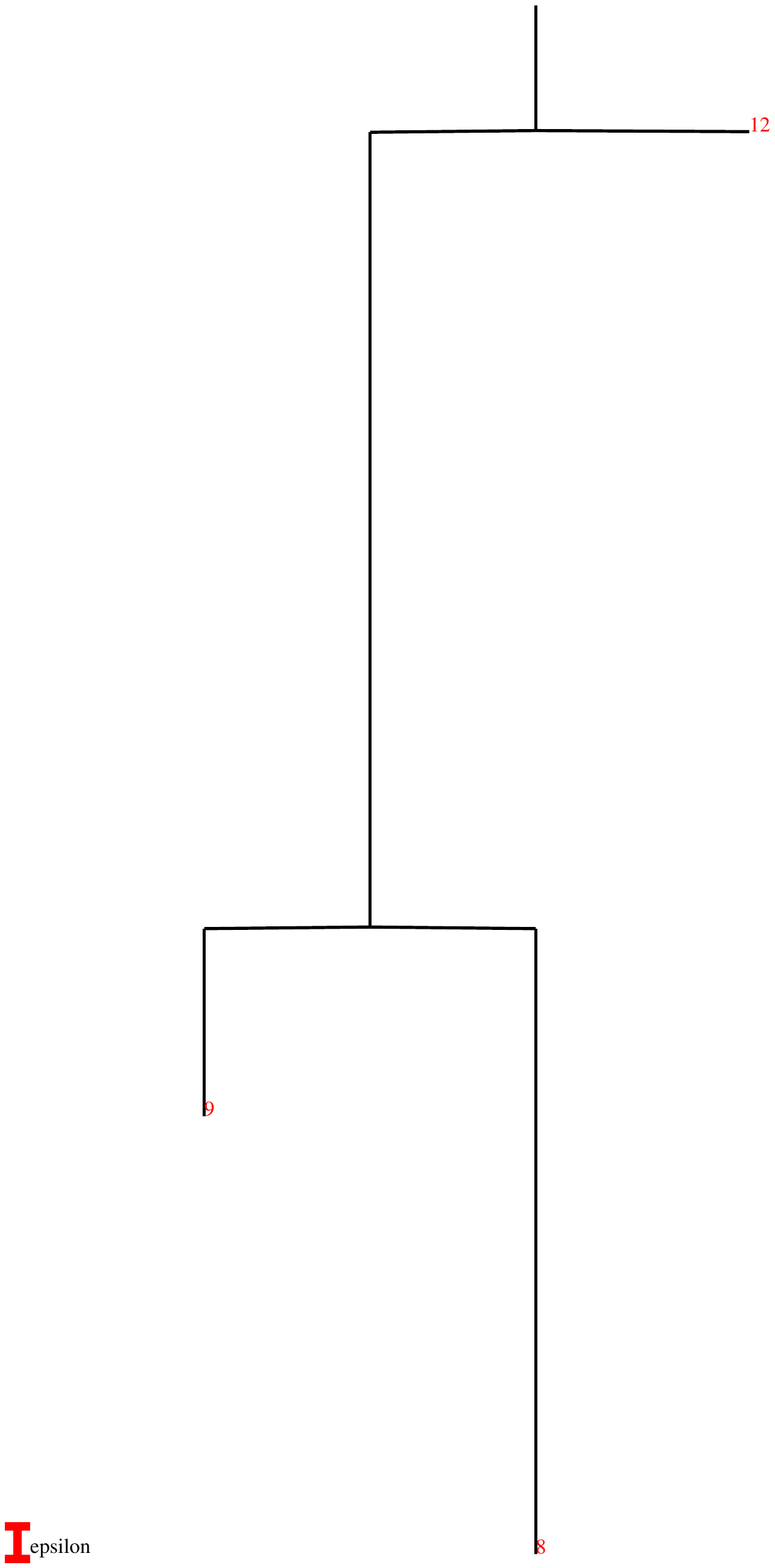}}
	\hfill
	\subfloat[$46$]{\includegraphics[width=0.05\textwidth,trim = 3.7cm -5cm 3.7cm 0cm,clip]{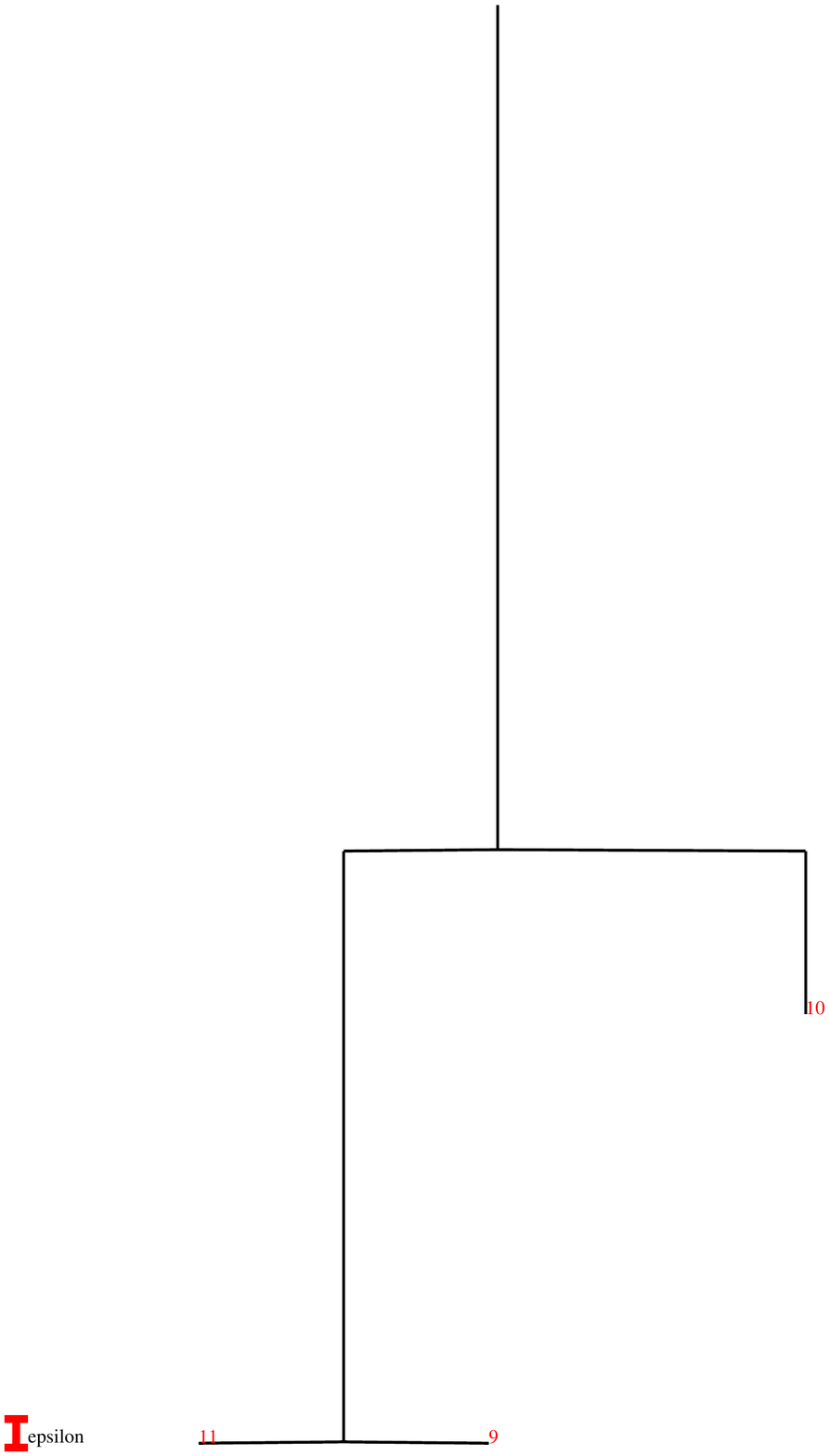}}
	\hfill
	\subfloat[$47$]{\includegraphics[width=0.05\textwidth,trim = 3.7cm -5cm 3.7cm 0cm,clip]{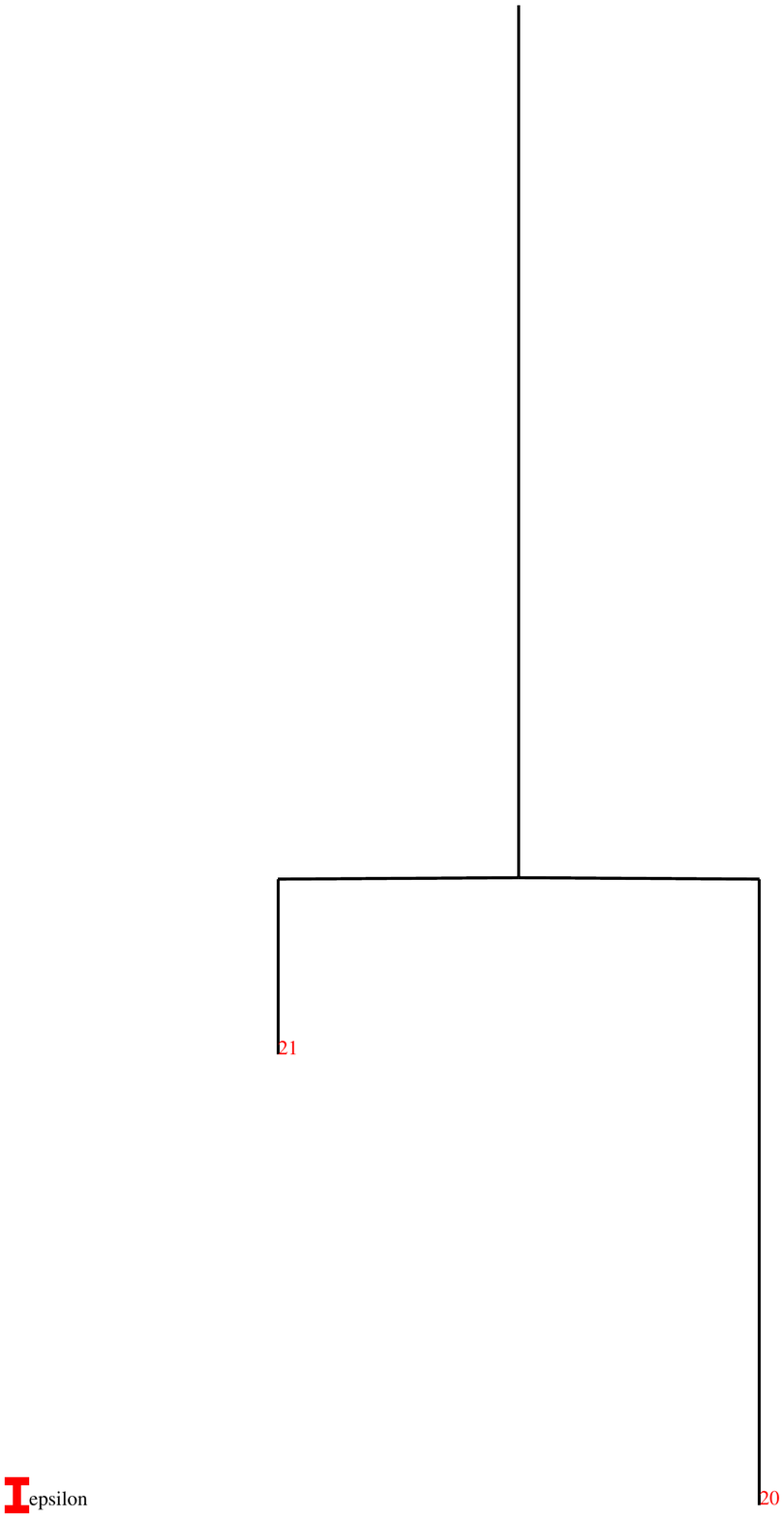}}
	\hfill
	\subfloat[$48$]{\includegraphics[width=0.05\textwidth,trim = 3.7cm -5cm 3.7cm 0cm,clip]{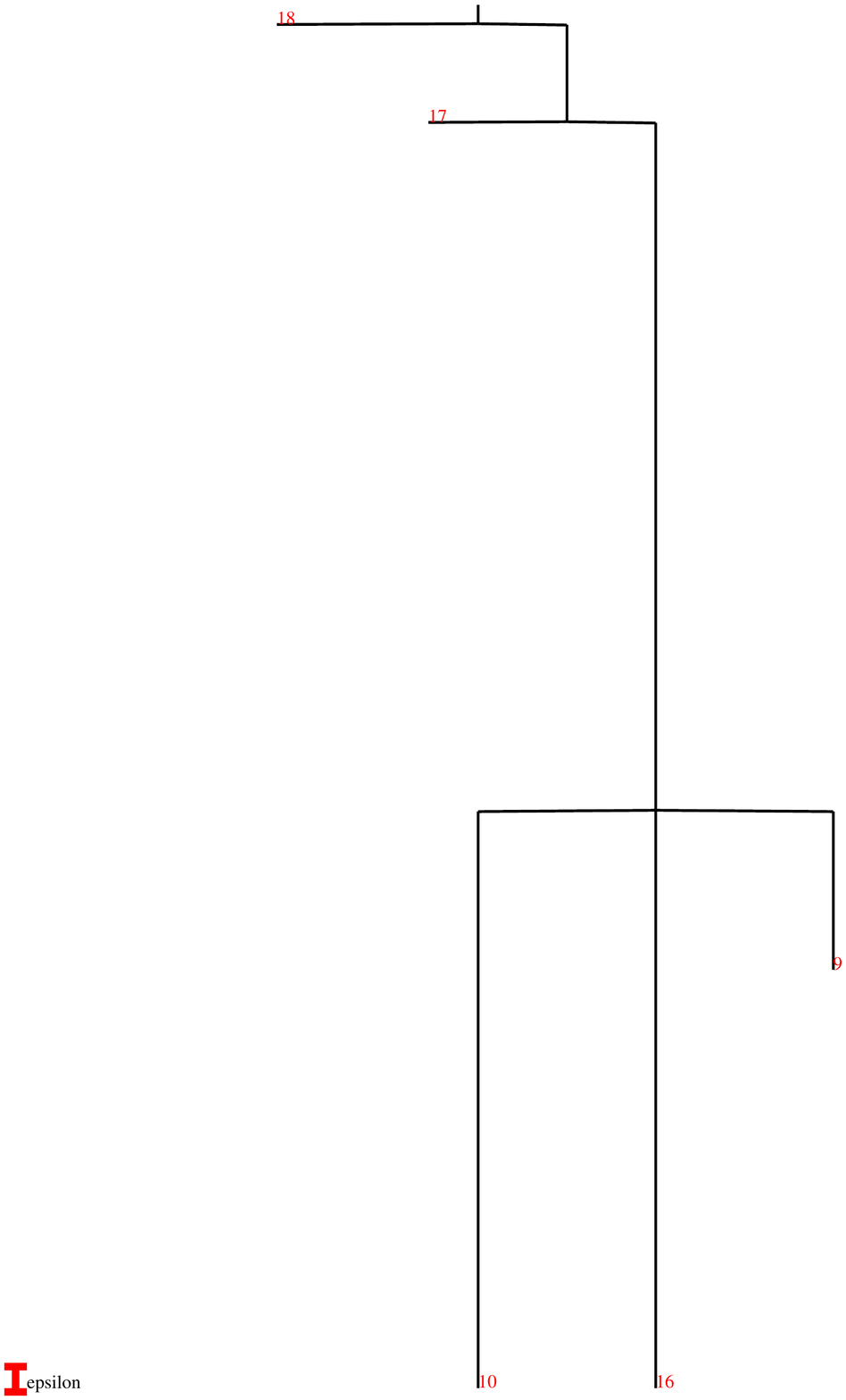}}
	\hfill
	\caption[Disconnectivity graphs of the LML surfaces for GP trained with RBF kernels. ]{Disconnecitivity graphs of the LML surfaces for GP trained with RBF kernels where $N$ is the total number of points in the training data..}
	\label{fig:ds-graphs-rbf}
\end{figure}

The same is seen for other kernel,  such as the RBF seen in figure \ref{fig:ds-graphs-rbf} where the LML landscape is also unstable with changing in the training data.  For the latter,  one essentially observes a similar,  and expected,  trend to the one in figure \ref{fig:symmetry-slow} with length hyperparameters getting larger w.r.t.  number of samples in the training set.  However,  given the much shorter length scales seen with the RBF kernel,  the MAEs are more susceptible to the data than the actual model which is rather over fitted.

\begin{figure}[H]
	\centering
	\captionsetup[subfigure]{labelformat=empty}
	\subfloat[$24$]{\includegraphics[width=0.05\textwidth,trim = 3.7cm -5cm 3.7cm 0cm,clip]{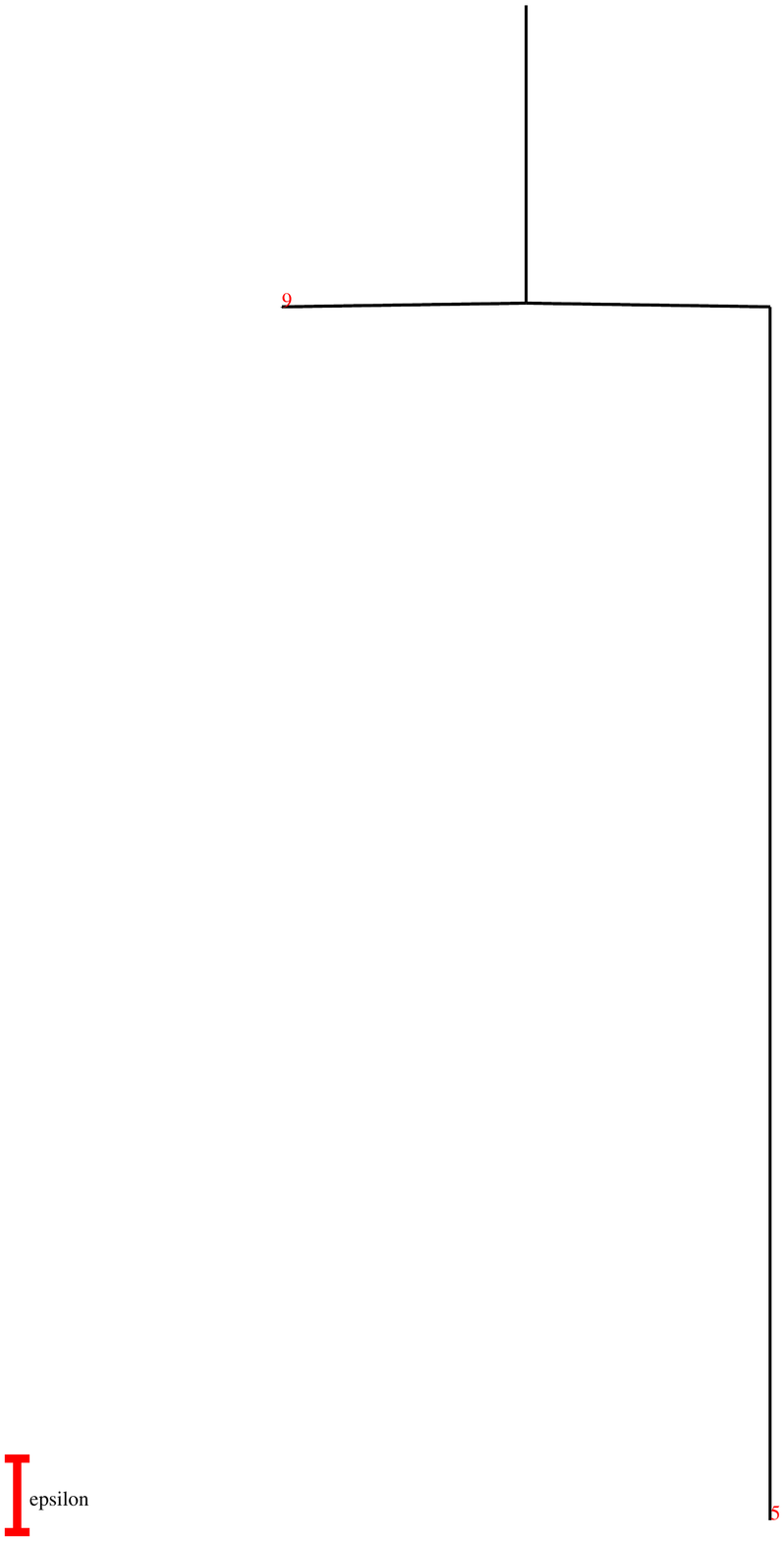}}
	\hfill
	\subfloat[$25$]{\includegraphics[width=0.05\textwidth,trim = 3.7cm -5cm 3.7cm 0cm,clip]{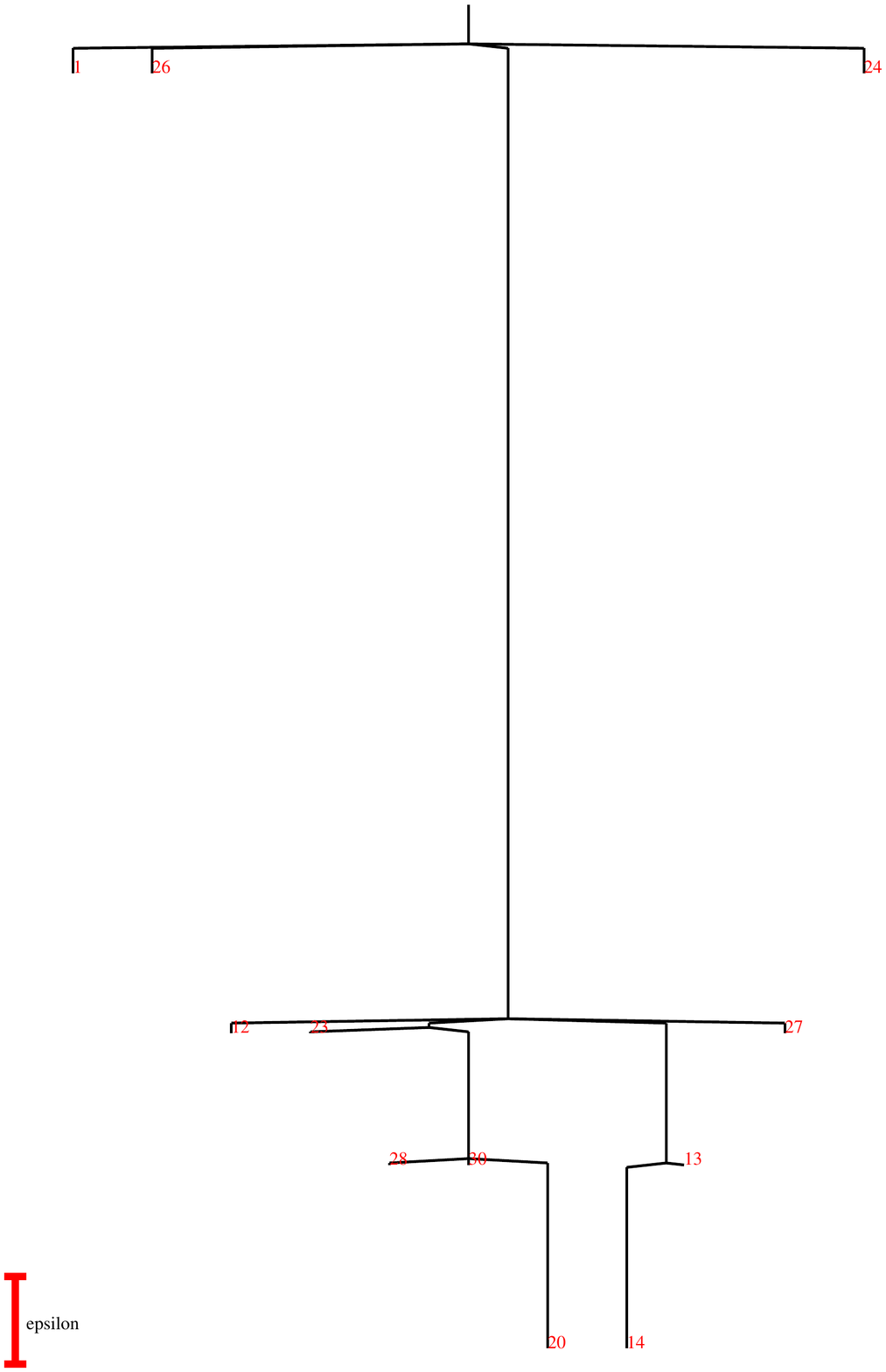}}
	\hfill
	\subfloat[$26$]{\includegraphics[width=0.05\textwidth,trim = 3.7cm -5cm 3.7cm 0cm,clip]{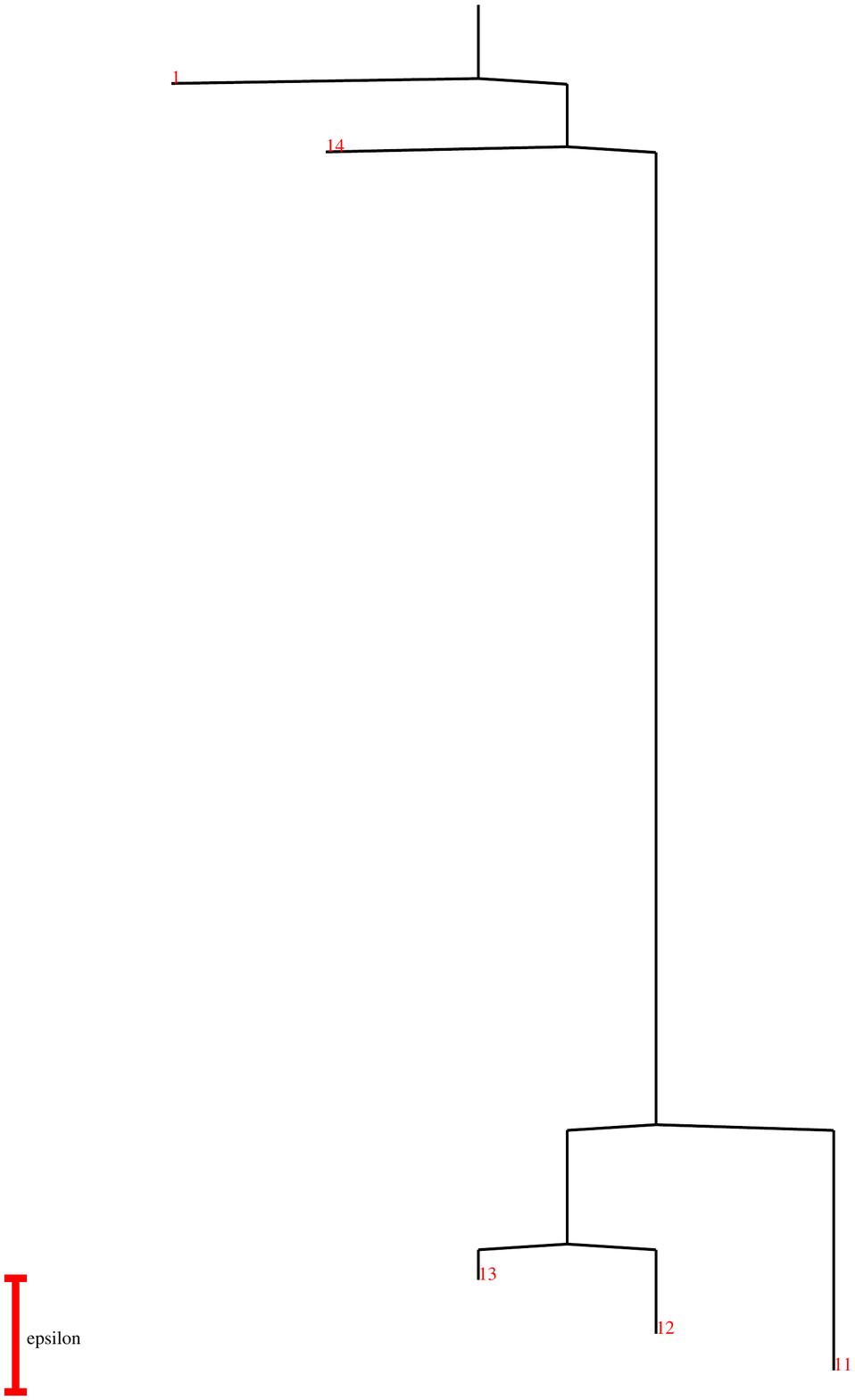}}
	\hfill
	\subfloat[$27$]{\includegraphics[width=0.05\textwidth,trim = 3.7cm -5cm 3.7cm 0cm,clip]{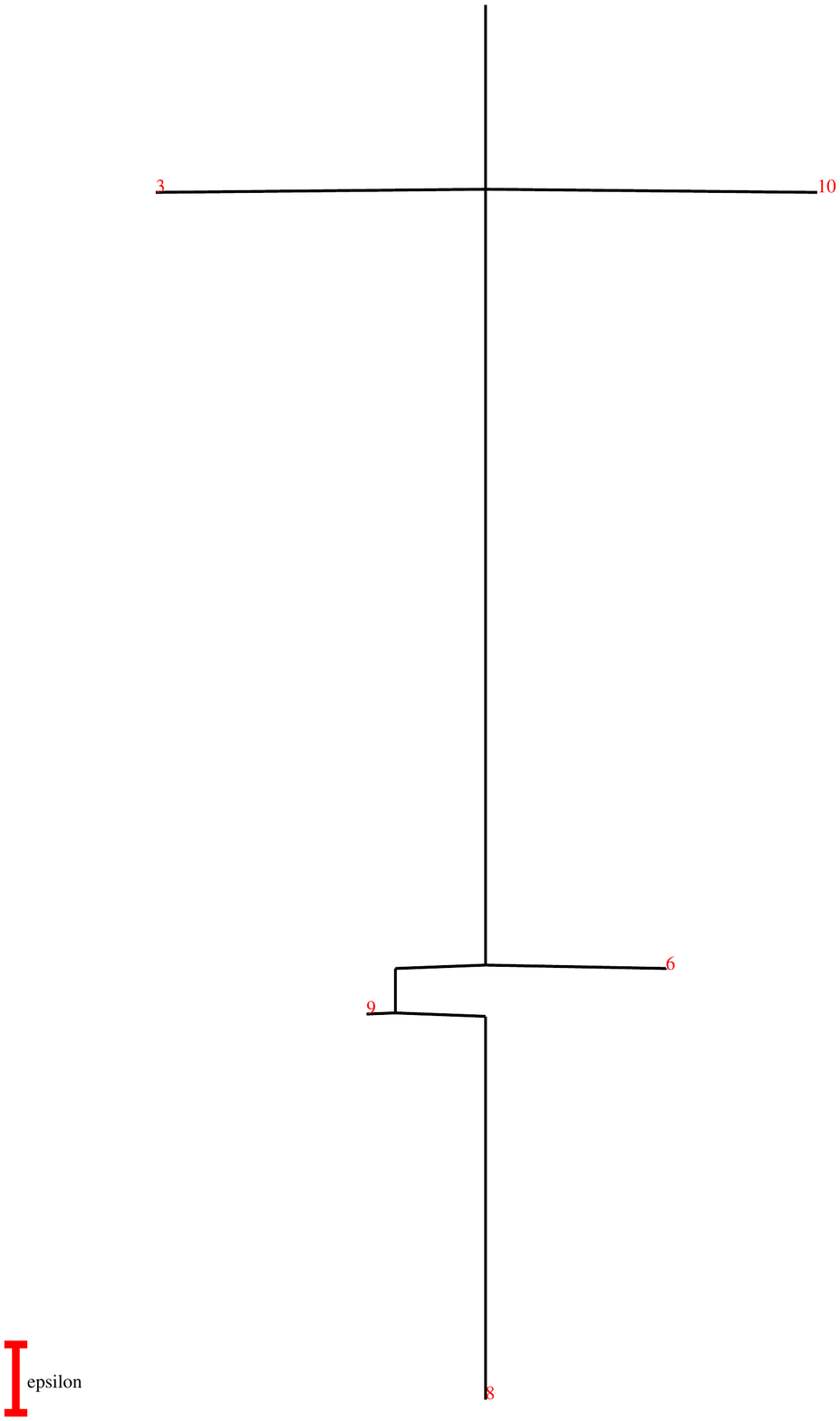}}
	\hfill
	\subfloat[$28$]{\includegraphics[width=0.05\textwidth,trim = 3.7cm -5cm 3.7cm 0cm,clip]{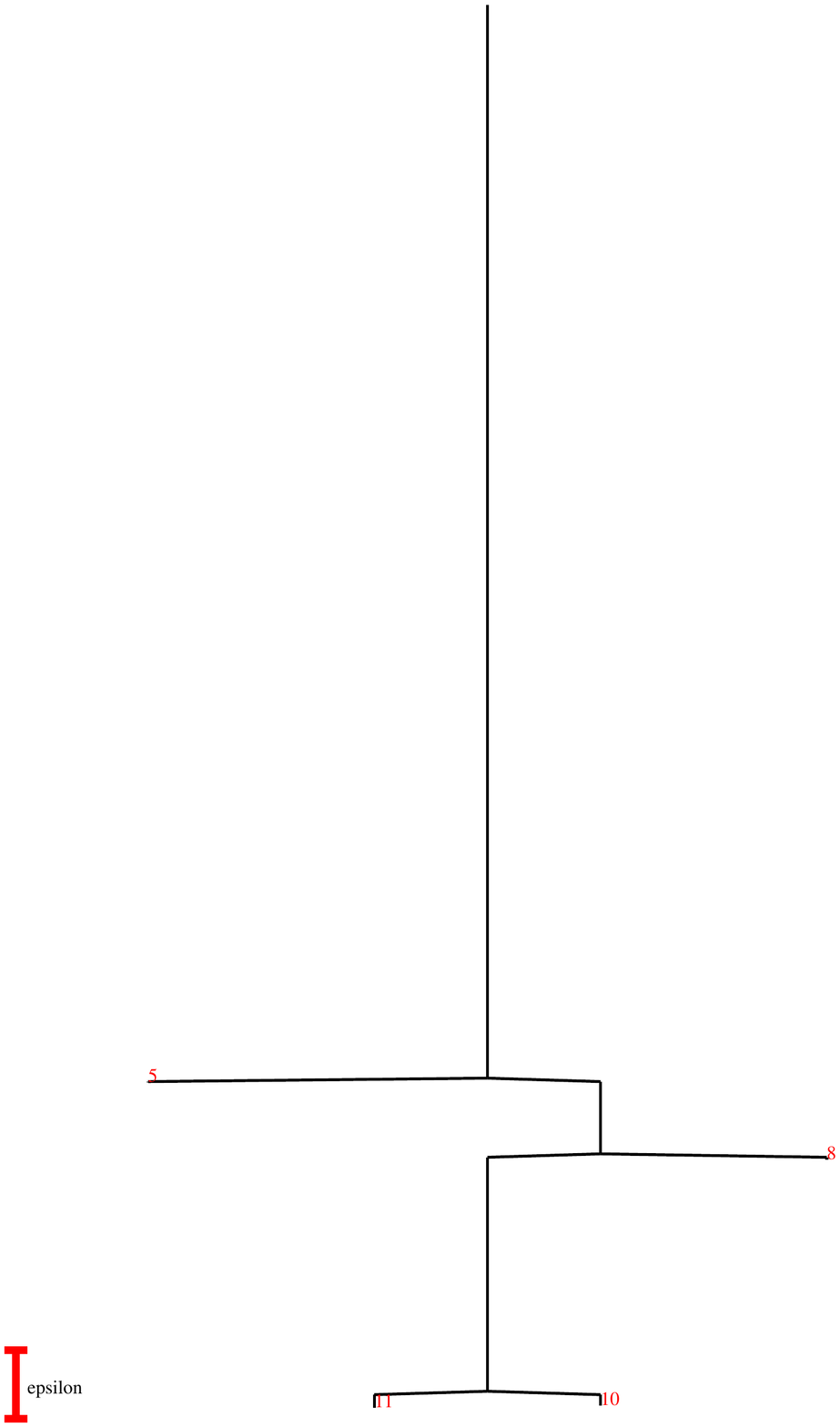}}
	\hfill
	\subfloat[$29$]{\includegraphics[width=0.05\textwidth,trim = 3.7cm -5cm 3.7cm 0cm,clip]{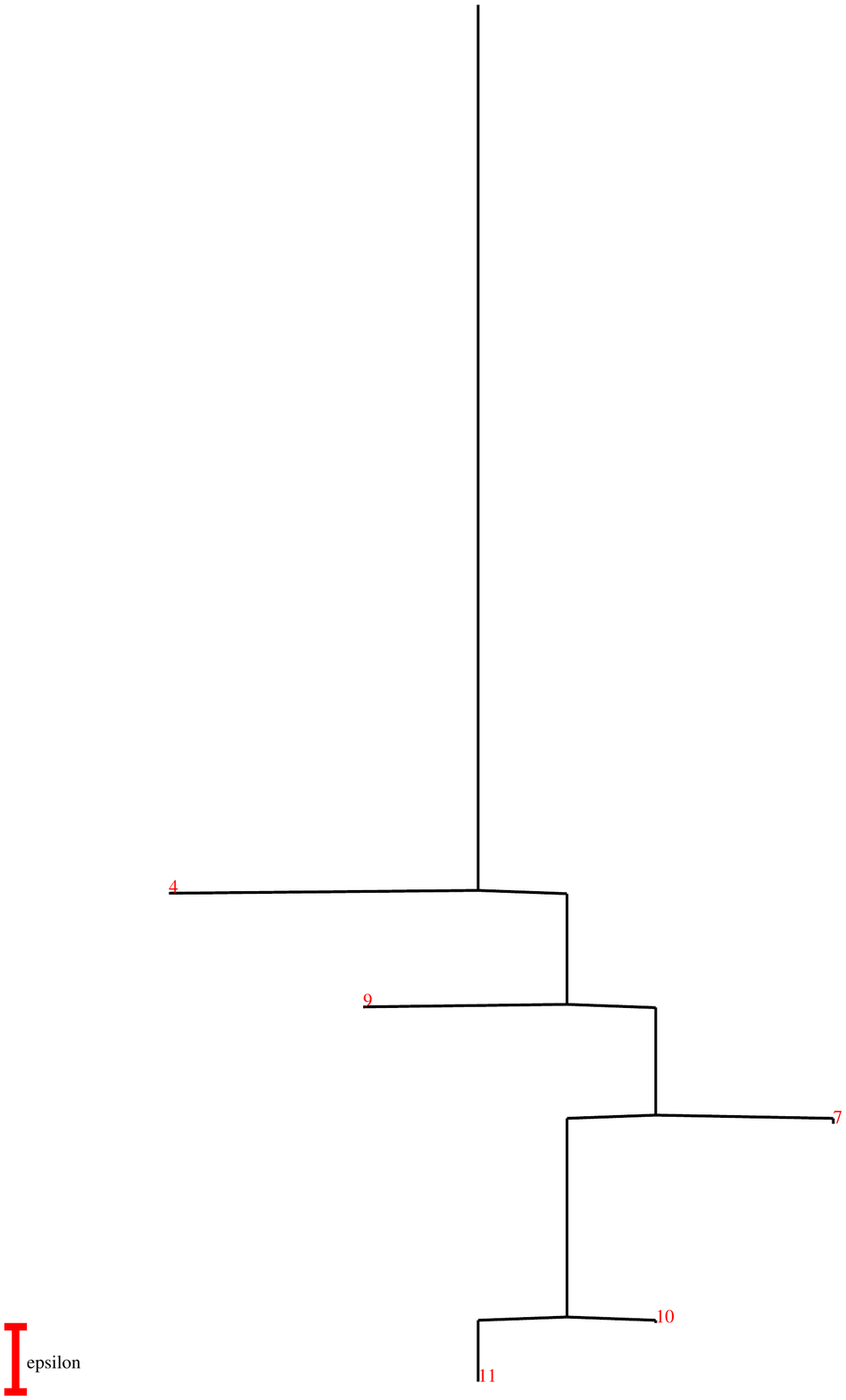}}
	\hfill
	\subfloat[$30$]{\includegraphics[width=0.05\textwidth,trim = 3.7cm -5cm 3.7cm 0cm,clip]{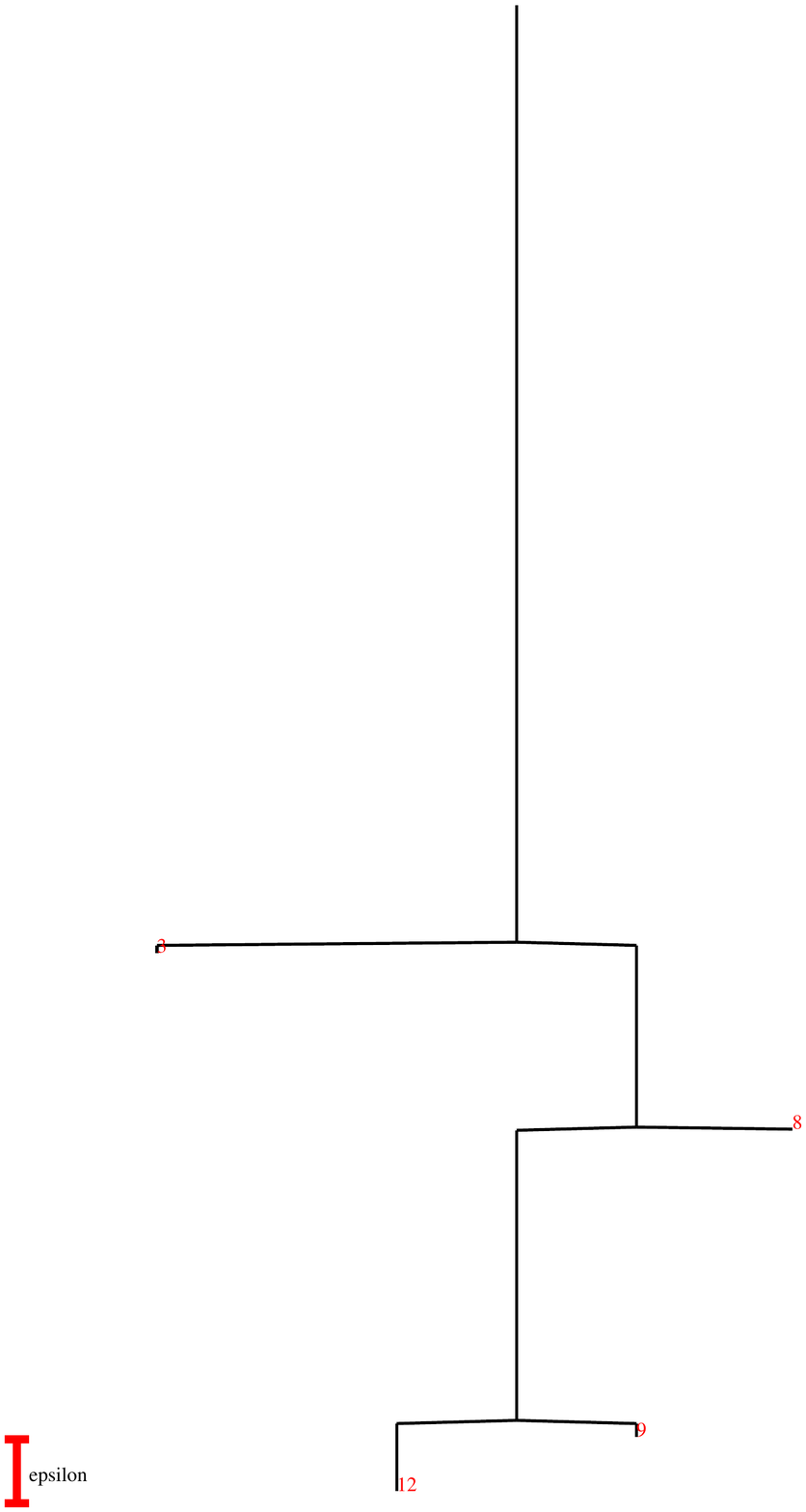}}
	\hfill
	\subfloat[$32$]{\includegraphics[width=0.05\textwidth,trim = 3.7cm -5cm 3.7cm 0cm,clip]{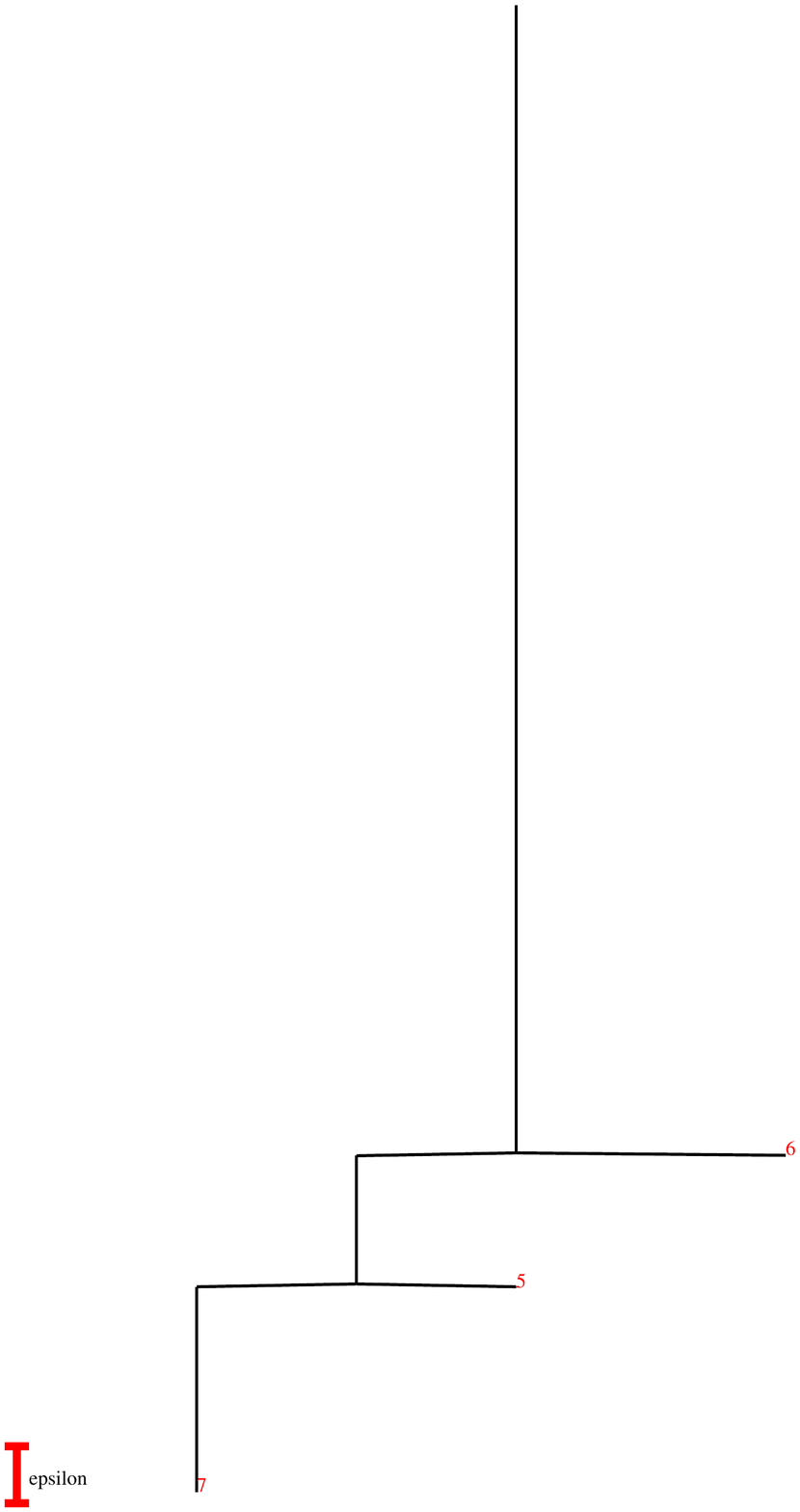}}
	\hfill
\end{figure}

It has to be noted that the effect is not purely an artefact of the data points ordering and it can be replicated by different orderings of the MID to symmMID evolution with sometimes the sharp drop appearing one point earlier (still without completely equal length scales however). This can be considered as a dependence on the importance of the last few data points.  If they are very similar to other points in the training set they might not have a large impact on the landscape and allow the change from first to second model to happen earlier.

\begin{figure}[H]
	\centering
	\subfloat[]{\includegraphics[width=0.22\textwidth]{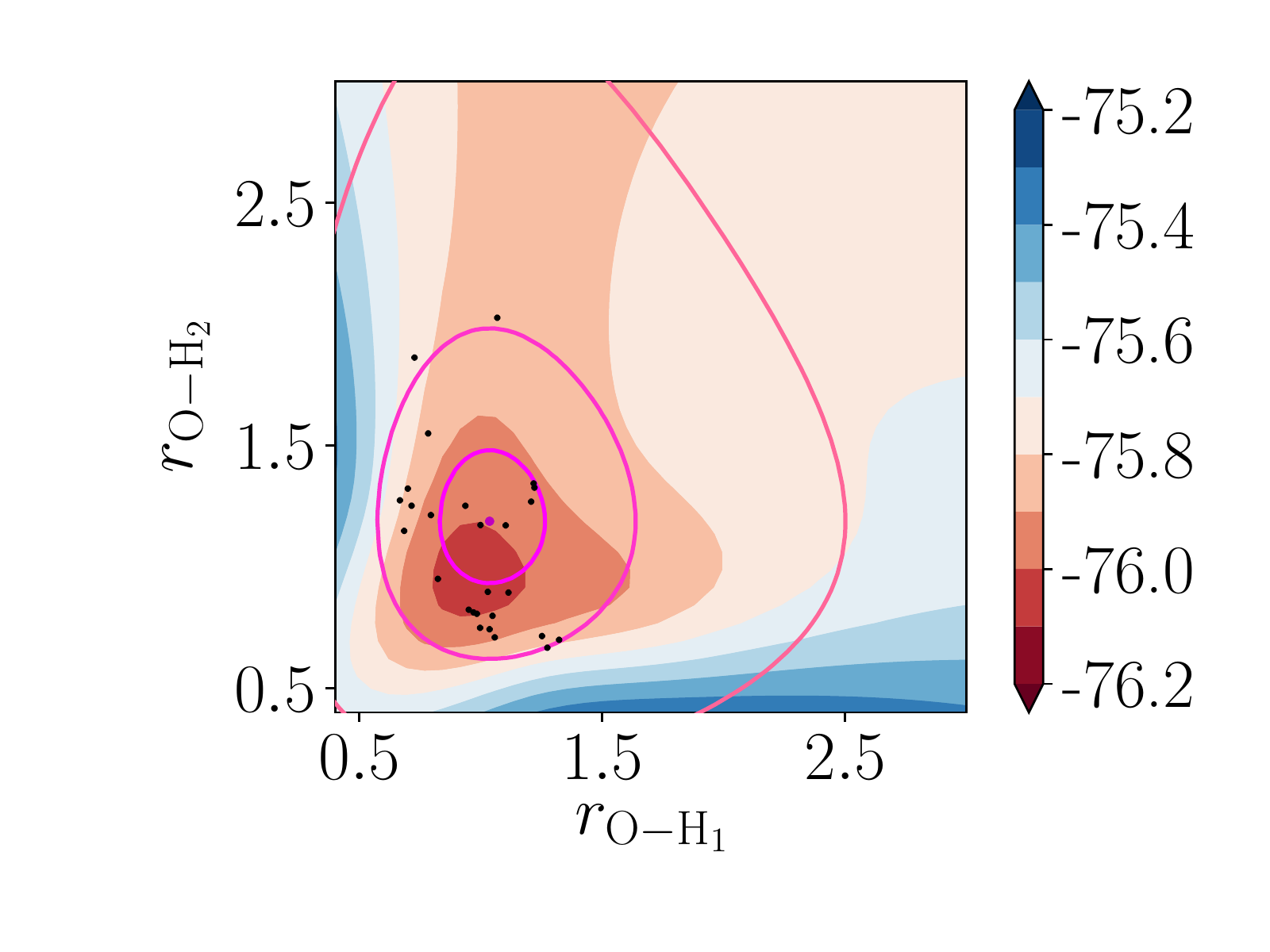}}
	\subfloat[]{\includegraphics[width=0.22\textwidth]{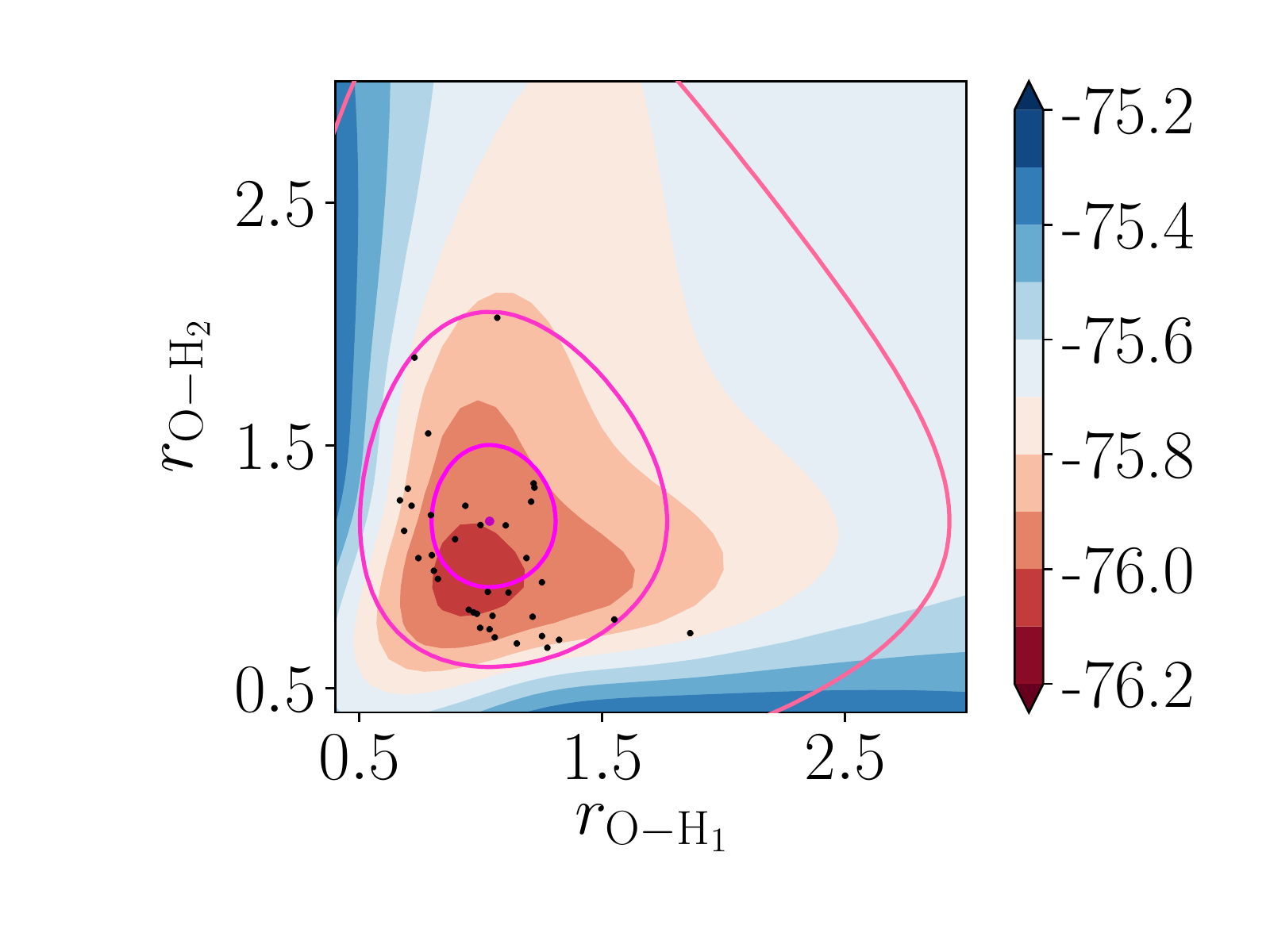}}
	\hfill
	\caption{Two latent function of Gaussian processes trained along the way of the original training set,  MID,  tending to symmMID.  The graphs have an additional 5 data points,  in panel (a),  and 15 data points,  in panel (b).  One can see that the symmetry of the resulting PES is improving as more data is present in the training set but it does remain only partially symmetric.  The slow symmetrisation is also seen in the second panel of figure \ref{fig:symmetry-slow} where $\rho_0$ and $\rho_1$,  after an initial regime where selected models have quite different,  start to converge to the same value. }
	\label{fig:symmetry-slow-cont}
\end{figure}

\subsection{Symmetrising NM-based feature spaces}

Despite normal modes transforming according to the irreducible representation,  since we do differentiate between each hydrogen in the water molecule,  a hydrogen permutation yields very different displacements,  $\Delta \mathbf{q}$.  These displacements do not project in a meaningful way on the normal modes coefficients since the labels are exchanged\footnote{If one looks at figure \ref{fig:h2o-truesurface} these are the H$_1$ and H$_2$ labels.}.  However,  by switching the hydrogens in the equilibrium geometry as well,  one can obtain useful projections of the displacements and thus,  unlike the FI feature space,  still increase the size of the training set without added computational cost.  This correspond to symmetrising training data by permuting atoms and then apply a C$_2$ rotation (or a $\sigma_v$ reflection) of the geometry.
\par
The last point seems to imply that hydrogens are swapped twice but this is only due to our computational perspective and need for labelling.  In a physical sense all we need is a C$_2$ operation on the molecule which allows us to understand how normal mode coefficients are affected through the C$_{2v}$ character table (since we use a C$_{2v}$ equilibrium geometry).  Both the bending and symmetrical stretching are permutationally invariant normal modes (A$_1$ in the character table) that do not have different coefficients upon swapping of hydrogens.  On the other hand,  the asymmetrical stretching is B$_1$ meaning that the permutation of hydrogens leads to an inversion of its coefficients. The resulting symmetrised normal modes coefficients for the data point in the normal mode coefficients feature space $[ v_1, v_2, v_3 ]$ is simply given by  $[ v_1, v_2,  -v_3  ]$. 

\begin{figure}[H]
\vspace{0.4cm}
\centering
	\begin{tikzpicture}[scale=0.5]
	\node[rotate=90] (a) at (-2,-1.5) {\footnotesize MAE:  7.5/22.5 mHa};
	\node[inner sep=0pt] (graph1) at (2,-2) {\includegraphics[width=0.22\textwidth]{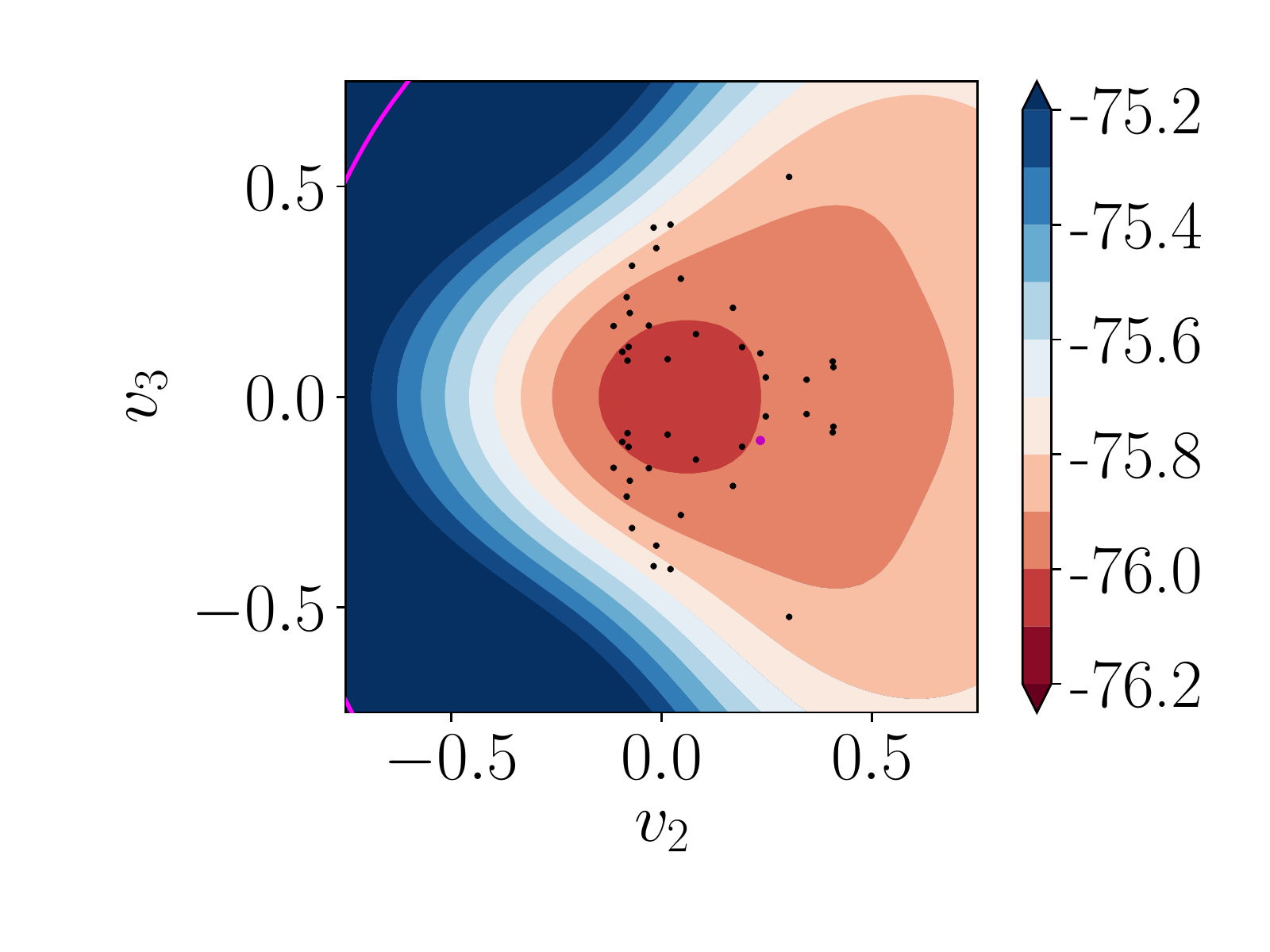}};
	\node[inner sep=0pt] (graph2) at (9,-2) {\includegraphics[width=0.22\textwidth]{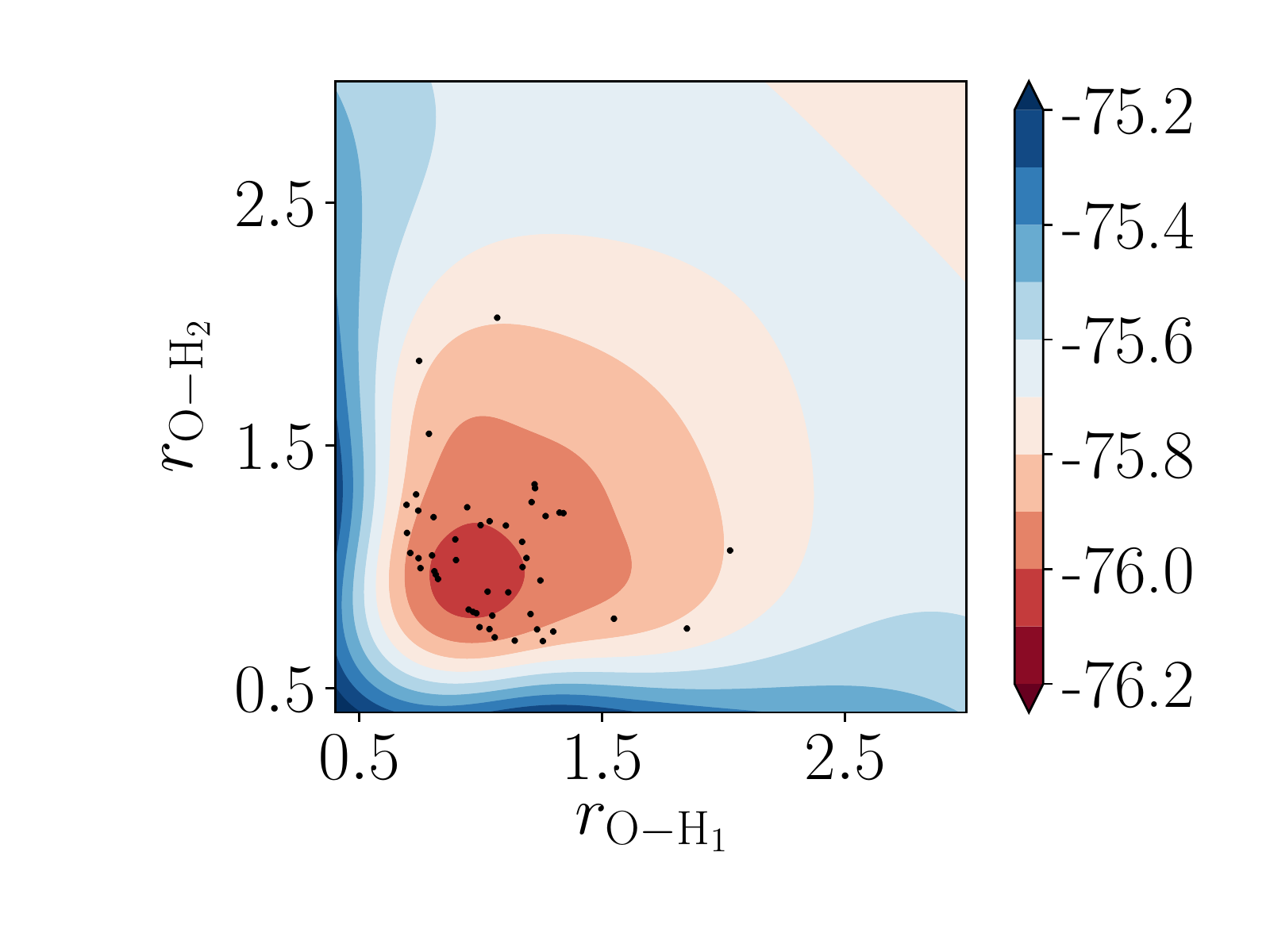}};
	\end{tikzpicture}
	\vspace{0.4cm}
	\caption[Predicted PES for the symmetrised NM feature space.]{Latent functions of GP trained on the NM feature space with the symmetrised data.  One can see that compared to panel (b) in figure \ref{fig:nm-pes},  the symmetrised set restores the symmetry of the PES.  Moreover,  one can see the first magenta isovalue contour corresponding to 0.75$\sigma^2$ covariance extending much further away.}
	\label{fig:nm-pes-symm}
\end{figure}

The resulting GP model is surprisingly still much worse than other symmetrised GP models.  Since the NM feature space was improved by the NLNM description,  one expects the same to happen with symmetrised training data for the latter. The same reasoning is applied to the NLNM feature space with the feature dimensions $[ x, w_2,  w_3 ]$ being equivalent to $[ x, w_2,  -w_3 ]$ under permutational invariance.

\begin{figure}[H]
\vspace{0.4cm}
\centering
	\begin{tikzpicture}[scale=0.5]
	\node[rotate=90] (a) at (-2,-1.5) {\footnotesize MAE:  1.9/13.3 mHa};
	\node[inner sep=0pt] (graph1) at (2,-2) {\includegraphics[width=0.22\textwidth]{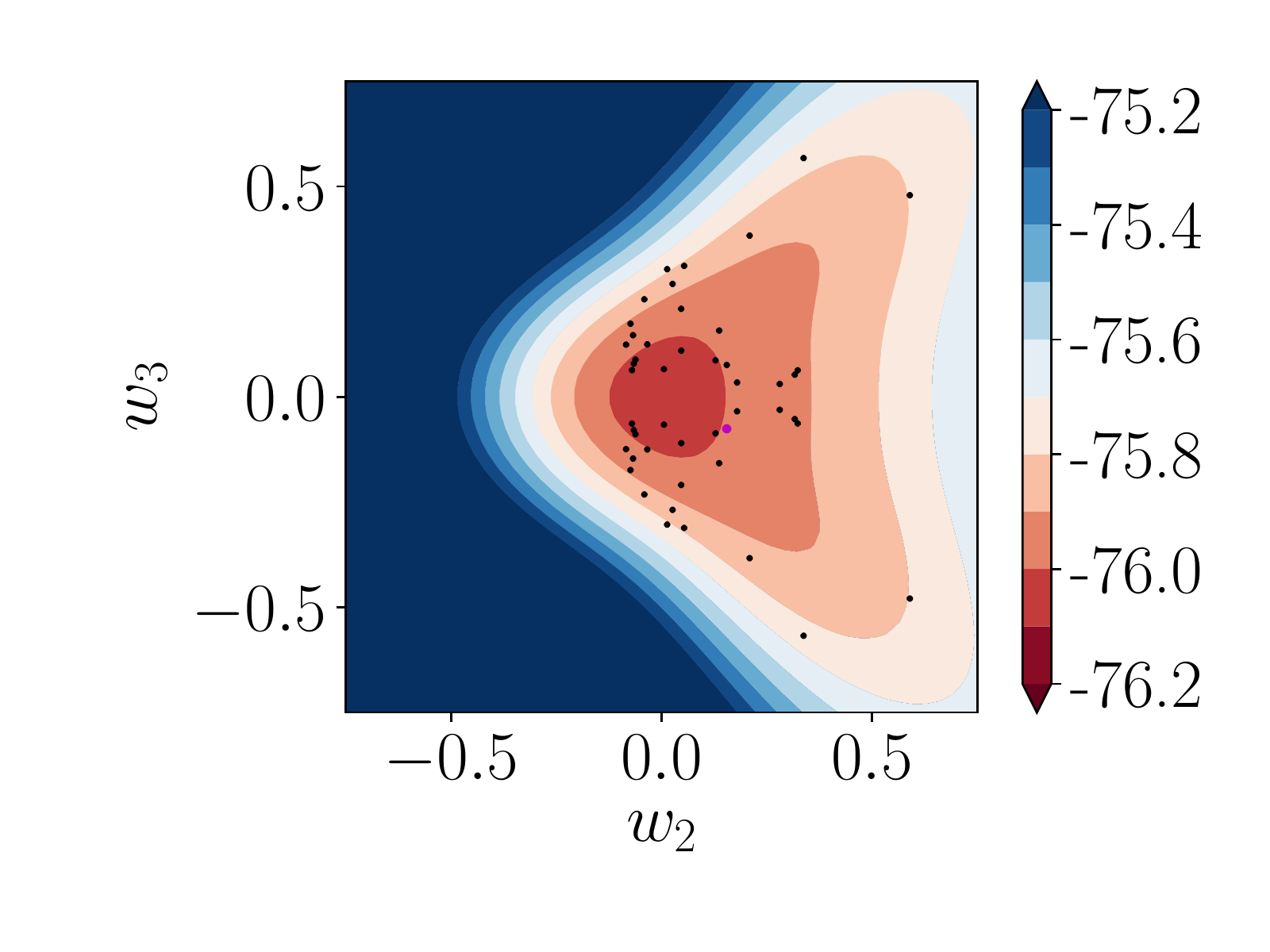}};
	\node[inner sep=0pt] (graph2) at (9,-2) {\includegraphics[width=0.22\textwidth]{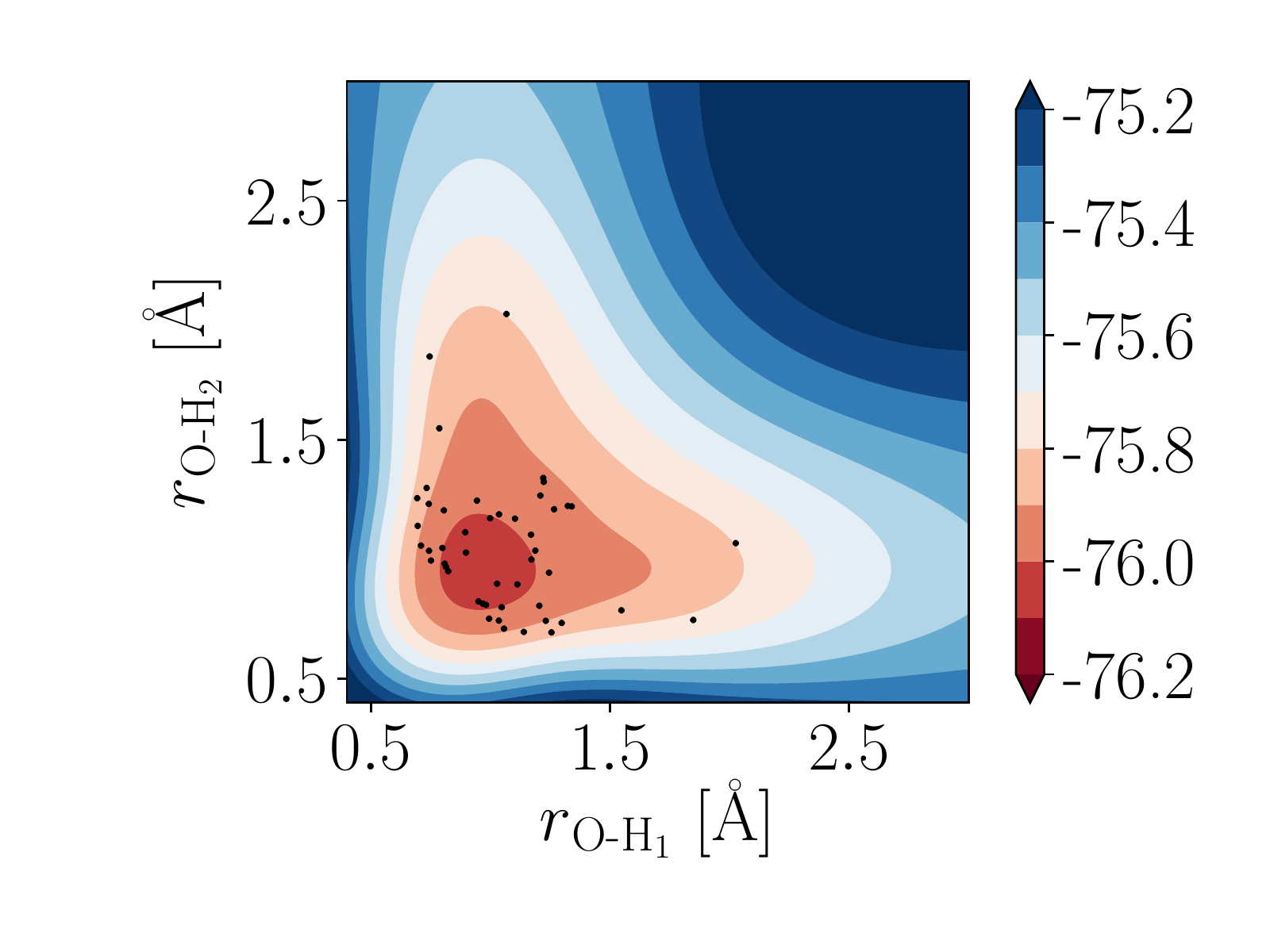}};
	\end{tikzpicture}
	\vspace{0.4cm}
	\caption[Predicted PES for the symmetrised NLNM feature space.]{Latent functions trained on non-local normal mode coefficients with the symmetrised training data.  The usual isovalue contours of the kernel function are not seen as they extend past the plotted area.}
	\label{fig:nlnm-pes-symm}
\end{figure}

The NLNM feature space is a large improvement on the local description of the PES and the MAE does become similar to the symmetrised training data projected onto the MID feature space.  However,  the ``high energy'' testing set,  is not described as well.

\section{Conclusion}
Feature spaces onto which GPs are trained can considerably change the ability of the latter to learn complex surfaces when using sparse  training data.  For the specific case of surfaces with known properties,  such as invariance w.r.t.  explicit coordinate change,  the coordinate transformation that produces spaces which are themselves invariant can be extremely powerful for large training sets but can also fail to accurately reproduce the true target surface as the training data is projected onto heavily distorted feature space.  Moreover,  these invariants spaces grow exponentially larger in complexity with system size impeding considerably the ability to use ML models. 
\par
We also notice that when optimising GPs though LML maximisation,  one often finds multiple minima on the surface and it is worth exploring models produced by each minima and not the global one only.  Bayesian ``scoring'' is not always equivalent to GP model performance.  It is preferable to use the LML to find stable hyperparameter which define a model that can scored with a different metric.  
\par
Increasing the size of the training data,  especially if it can be done without additional calculations,  is very efficient.  For PES models,  only molecules where same atom permutations exist can produce ``new data'' when training on feature spaces without explicit permutational invariance.  Although this is not a big restriction as one often finds equivalent atoms in molecules.  Moreover,  for centro-symmetric AB$_n$ molecules,  GPs with ``symmetrised'' training sets are guaranteed to produce symmetrical models on feature spaces without explicit invariance when optimised using the Bayesian LML maximisation. 
\par
It is thus often appropriate to consider feature spaces for regression problems with care and not treat the latter as a ``black box method''.  In terms of kernels,  one expects the Mat\'{e}rn class of kernels to be the most suited to PES modelling.  However,  one could consider more complex kernels that include convolutions or compositions that have summed kernels other than noise,  for example two with different $\nu$ parameters.  Moreover,  one could have explored more exotic noise kernels than a white noise that assumes homogenous noise in the data.  We have seen that one can improve GP models with regularisation induced by weighted white kernels\autocite{Fabiochap6}.  A step further,  one could consider to leave the choice of kernel to the ML model itself using method that are able to explore a space of kernels and pick the most suited one\autocite{Duvenaud2013}.

\section*{Acknowledgments}
F.  E.  A.  Albertani would like to thank the Royal Society for funding as well as the Wales group of the University of Cambridge for providing access to the GMIN suite\autocite{GMIN}. 

\printbibliography
\end{multicols}
\end{document}